\documentclass[traditabstract]{aa} 
\usepackage{graphicx}
\usepackage{mathtools}
\def\la{\lower.5ex\hbox{$\; \buildrel < \over \sim \;$}}
\def\ga{\lower.5ex\hbox{$\; \buildrel > \over \sim \;$}}
\begin{document}
   \title{Deciphering the radio-star formation correlation on kpc-scales}

   \subtitle{III. Radio-dim and bright regions in spiral galaxies}

   \author{B.~Vollmer\inst{1}, M.~Soida\inst{2}, R.~Beck\inst{3}, \and J.D.P.~Kenney\inst{4}}

   \institute{Universit\'e de Strasbourg, CNRS, Observatoire Astronomique de Strasbourg, UMR 7550, 67000 Strasbourg, France \and
          Astronomical Observatory, Jagiellonian University, ul. Orla 171, 30-244 Krak\'ow, Poland \and
	  Max-Planck-Institut f\"{u}r Radioastronomie, Auf dem H\"{u}gel 69, 53121 Bonn, Germany \and
	  Yale University Astronomy Department, P.O. Box 208101, New Haven, CT 06520-8101, USA}

   \date{Received ; accepted }


  \abstract
{
The relation between the resolved star formation rate per unit area and the non-thermal radio continuum emission is studied in
$21$ Virgo cluster galaxies and the two nearby spiral galaxies, NGC~6946 and M~51.
For the interpretation and understanding of our results we used a 3D model where star formation, 2D cosmic ray (CR) propagation, 
and the physics of synchrotron emission are included. Based on the linear correlation between the star formation rate per unit area
and the synchrotron emission and its scatter radio-bright and radio-dim regions can be robustly defined for our sample of spiral galaxies.
We identified CR diffusion or streaming as the physical causes of radio-bright regions of unperturbed symmetric spiral galaxies as NGC~6946.
The enhanced magnetic field in the region of ISM compression via ram pressure is responsible for the southwestern radio-bright region in NGC~4501.
We identified the probable causes of radio-bright regions in several galaxies as CR transport, via either gravitational tides (M~51) or 
galactic winds (NGC~4532) or ram pressure stripping (NGC~4330 and NGC~4522).
Three galaxies are overall radio-dim: NGC~4298, NGC~4535, and NGC~4567.
Based on our model of synchrotron-emitting disks we suggest that the overall radio-dim galaxies have a significantly lower
magnetic field than expected by equipartition between the magnetic and turbulent energy densities.
We suggest that this is linked to difference between the timescales of the variation in SFR and the small-scale dynamo.
In NGC~4535 shear motions increase the total magnetic field strength via the induction equation, which leads to an enhanced 
synchrotron emission with respect to the star formation rate in an otherwise radio-dim galactic disk.
Radio-bright regions frequently coincide with asymmetric ridges of polarized radio continuum emission, and we found a clear albeit 
moderate correlation between the polarized radio continuum emission and the radio/SFR ratio.
When compression or shear motions of the interstellar medium (ISM) are present in the galactic disk, the radio-bright regions are linked 
to the commonly observed asymmetric ridges of polarized radio continuum emission and represent a useful tool for the interaction diagnostics.
The magnetic field is enhanced (as observed in NGC~4535 and NGC~4501) and ordered by these ISM compression and shear motions. 
Wheres the enhancement of the magnetic field is rather modest and does not significantly 
influence the radio/SFR correlation, the main effect of ISM compression and shear motions is the ordering of the magnetic field,
which significantly affects the CR transport.
CR energy losses and transport also affect the spectral index, which we measure between $4.85$ and $1.4$~GHz.
The influence of CR losses and transport on the spectral index distribution with respect to the synchrotron/SFR ratio is discussed
with the help of model calculations.
Based on our results, we propose a scenario for the interplay between star formation, CR electrons, and magnetic fields in spiral galaxies.
}


   \authorrunning{Vollmer et al.}

   \maketitle
%

\section{Introduction\label{sec:introduction}}

Relativistic particles, or cosmic rays, and magnetic fields are important components of the interstellar medium (ISM) in galaxies. 
Cosmic rays are dynamically coupled to the thermal gas, can drive outflows and modify the structure of shocks. 
They are the dominant source of ionization and a major source of heating in regions of high surface densities. 
Moreover, cosmic rays and magnetic fields provide pressure support to the ISM.
Cosmic ray electrons spiralling around magnetic field lines emit synchrotron emission, which can easily be detected in the classical radio domain.

One of the tightest correlations in astronomy is the relation between the integrated radio continuum (synchrotron) and the far-infrared emission
(Helou et al. 1985; Condon  1992; Mauch \& Sadler  2007; Yun et al. 2001; Bell 2003; Appleton et al. 2004; Kovacs et al. 2006;
Murphy et al. 2009; Sargent et al. 2010, Li et al. 2016). It holds over five orders of magnitude in various types of galaxies including starbursts.
The common interpretation of the correlation is that both emission types are proportional to star formation:
the radio emission via (i) the cosmic ray source term which is due to supernova explosions and the turbulent amplification of
the small-scale magnetic field (small-scale dynamo, see, e.g., Schleicher \& Beck 2013) and (ii) the far-infrared emission via
the dust heating mainly through massive stars. Within nearby galaxies, variations in the radio-FIR correlation have been observed
by Gordon et al. (2004), Hughes et al. (2006), Murphy et al. (2006, 2008), Dumas et al. (2011), and Tabatabaei et al. (2013a,b).

Gordon et al. (2004) found variations of $q=\log (FIR/3.75 \times 10^{-12}~{\rm Hz})-\log (S_{1.49~{\rm GHz}})$ of the order of $1$~dex in M~81.
The variation in $q$ was found to be coherent with structures related to the spiral arms, confirming that the radio-infrared
correlation is more complex inside individual galaxies than between galaxies.

Hughes et al. (2006) found that the
slope of the relation between the radio and FIR emission is non-linear in the LMC. In bright star-forming
regions, the radio emission increases faster than linearly with respect to the FIR emission
(power-law slope of $\sim 1.2$), whereas a flatter slope of $\sim 0.6$--$0.9$ applies more generally across
the LMC.

Dumas et al. (2011) also found a large scatter in the resolved radio--$24~\mu$m correlation of the galactic disk of M~51. 
The radio--$24~\mu$m ratio presents a complex behavior with local extrema
corresponding to various galactic structures, such as complexes of H{\sc ii} regions, spiral arms, and interarm filaments,
indicating that the contribution of the thermal and non-thermal radio emission is a strong function of environment.
In particular, the relation of the resolved $24$~$\mu$m and $20$~cm emission presents a linear relation within the spiral arms and
globally over the galaxy, while it deviates from linearity in the interarm and outer regions ($R > 7.5$~kpc) as well in the inner region ($R < 1.6$~kpc).

Tabatabaei et al. (2013a) investigated the correlation between
the IR and free-free/synchrotron radio continuum emission at 20~cm from the two local group galaxies M~31 and M~33 on spatial
scales between 0.4 and 10~kpc. They found that the synchrotron-IR correlation is stronger in M~33 than in M~31 on small 
scales ($<1$~kpc), but it is weaker than in M~31 on larger scales. They argued that the difference on small scales can be 
explained by the smaller CR propagation length in M~33 than in M~31 and the difference on large scales is due to the thick 
disk/halo in M~33, which is absent in M~31.
Moreover, Tabatabaei et al. (2013b) found that the slope of the radio-FIR correlation across NGC~6946 varies as a 
function of the star formation rate (SFR) and the magnetic field strength. 

For the integrated correlation between the non-thermal 20~cm continuum emission (RC) and the SFR in 17 nearby spiral galaxies Heesen et al. (2014) found 
$RC \propto SFR^{1.11 \pm 0.08}$. The locally averaged correlation between the RC emission and SFR
tracers in pixels of 1.2 and 0.7~kpc resolution yielded a slope $0.63 \pm 0.25$. These authors argued that diffusion of cosmic-ray (CR) electrons 
is responsible for flattening the local radio--SFR relation, resulting in a sub-linear relation. 

CR electrons are produced in supernova remnants.
Before losing their energy by synchrotron radiation CR electrons propagate significantly farther via diffusion or 
streaming than the mean free path of dust-heating photons.
Murphy et al. (2006)  tested a phenomenological model that describes the radio continuum image as a smeared version of the 
far-infrared image. They found that galaxies with higher infrared
surface brightnesses have substantially shorter best-fit smoothing scale lengths than those for lower surface brightness galaxies.
Murphy et al. (2008) decomposed the IR images into one component containing the star-forming structures and a second
one for the diffuse disk. The components were then smoothed separately. They found that the disk component dominates for 
galaxies having low star formation activity, whereas the structure component dominates at high star formation activity. 

Berkhuijsen et al. (2013) studied the radio--SFR correlation in M~31 and M~33 and found sub-linear
slopes. They convolved the thermal radio continuum map, assuming it to be a representation of the star formation surface density, 
with a Gaussian kernel, so that the non-thermal radio--SFR relation becomes linear, and use this as an estimate for the cosmic-ray 
diffusion length (see also Heesen et al. 2014).

Vollmer et al. (2020) obtained predicted radio continuum maps of Virgo cluster spiral galaxies by convolving the maps of
cosmic-ray electron sources, represented by that of the star formation, with adaptive Gaussian and exponential kernels. The ratio
between the smoothing lengthscales at 6~cm and 20~cm could be used to determine, between diffusion and streaming, which is the
dominant transport mechanism: whereas the diffusion lengthscale is proportional to $\nu^{-0.25}$, the streaming lengthscale is
proportional to $\nu^{-0.5}$, where $\nu$ is the frequency.
The dependence of the smoothing lengthscale on the SFR contains information on the
dependence of the magnetic field strength on the SFR, or the ratio between the ordered and turbulent magnetic field strengths on the SFR. 
These authors found that perturbed spiral galaxies tend to have smaller lengthscales, which is a natural consequence 
of the enhancement of the magnetic field caused by the interaction probably via large-scale ISM compression.

In the present effort we take an approach to investigate the physics governing the resolved radio--SFR correlation, which
is different from our previous work (Vollmer et al. 2020). 
Rather than convolving the SFR maps with Gaussian or exponential kernels, as we did in our previous work, in this work we directly calculate the ratio
between the resolved non-thermal radio continuum emission and the SFR to identify radio-bright and radio-dim regions within the galaxies.
In this way we avoid the artificial smoothing of sharp edges in the radio continuum distribution, which are naturally produced by 
ram pressure in cluster galaxies (e.g., Vollmer et al. 2007).
In addition, we compare the radio/SFR ratios to the spectral index to identify the dominant energy loss and the influence of
CR propagation. In a second step the analytical model for non-thermal synchrotron emission developed by Vollmer et al. (2022)
is applied to dynamical simulations, which can be directly compared to our observations.

The structure of this article is the following: the radio, infrared (IR), and UV observations and the calculation of the
SFR maps are described in Sect.~\ref{sec:observations}. Maps of the radio/SFR ratio in connection with the radio continuum spectral index
are presented in Sect.~\ref{sec:results}. Our model for galactic synchrotron emission is introduced and its results are
compared to our observations in Sect.~\ref{sec:model}. Our findings are discussed in Sect.~\ref{sec:discussion} and we give
our conclusions in Sect.~\ref{sec:conclusions}.

\section{Observations\label{sec:observations}}

For the radio continuum maps we used published VLA data at $4.85$ and $1.4$~GHz at $15$-$22''$ resolution for all galaxies.
The Virgo spiral galaxies were observed at 4.85 GHz between November 8, 2005 and January 10, 2006 and 
between October 12, 2009 and December 23, 2009 with the Very Large
Array (VLA) of the National Radio Astronomy Observatory (NRAO) in the D array configuration. The band passes were
$2 \times 50$~MHz. We used 3C286 as the flux calibrator and 1254+116 as the phase calibrator, the latter of which was observed 
every 40~min. Maps were made for both wavelengths using the AIPS task IMAGR with ROBUST = 3. The final cleaned maps
were convolved to a beam size of $18'' \times 18''$ (Vollmer et al. 2010) or $22'' \times 22''$ (Vollmer et al. 2013). 
In addition, we observed the galaxies at $1.4$~GHz on August 15, 2005  and on March 21, 2008
in the C array configuration. The band passes were $2 \times 50$~MHz. We used the same calibrators
as for the 4.85~GHz observations. The final cleaned maps were convolved to a beam size of $20'' \times 20''$
(Vollmer et al. 2010) or $22'' \times 22''$ (Vollmer et al. 2013)

In addition, we used VLA 4.85 and 1.4~GHz images of NGC~4254 (Chy\.zy et al. 2007), NGC~6946 (Beck 2007), 
and M~51 (Fletcher et al. 2011) with spatial resolutions of $15'' \times 15''$.
NGC~6946 and M~51 are nearby nearly face-on spiral galaxies with clear spiral structure (plus a tidal interaction for M~51), 
that show clear trends between the radio-SFR ratio and galaxy morphology/ISM substructure (Tabatabaei et al. 2013a, Dumas et al. 2011).
We subtracted strong point sources and the thermal free-free radio emission according to the recipe of Murphy et al. (2008)
\begin{equation}
\big(\frac{S_{\rm therm}}{\rm Jy}\big) = 7.9 \times 10^{-3} \big(\frac{T}{10^4~{\rm K}}\big)^{0.45} \big(\frac{\nu}{\rm GHz}\big)^{-0.1}
\big(\frac{I_\nu(24~\mu{\rm m})}{\rm Jy}\big)\ ,
\end{equation}
where $T$ is the electron temperature and $I_\nu(24~\mu{\rm m})$ is the flux density at a wavelength of $24~\mu$m.
We note that there are other alternative methods to
account for the thermal free-free emission using, e.g., the extinction-corrected H$\alpha$ emission (Tabatabaei et al. 2007, Heesen et al. 2014).
We define the spectral index SI as $I_{\nu} \propto \nu^{-{\rm SI}}$. In the following, the term radio continuum emission denotes
the non-thermal or synchrotron component of the resolved radio continuum emission.

The star formation rate was calculated from the FUV luminosities corrected by the total infrared to FUV luminosity ratio (Hao et al. 2011).
This method takes into account the UV photons from young massive stars which escape the galaxy and those which
are absorbed by dust and re-radiated in the far infrared:
\begin{equation} 
\label{eq:sfr}
\dot{\Sigma}_{*} = 8.1 \times 10^{-2}\ (I({\rm FUV}) + 0.46\ I({\rm TIR}))\ ,
\end{equation}
where $I({\rm FUV})$ is the GALEX far ultraviolet and $I({\rm TIR})$ the total infrared intensity based on Spitzer IRAC and MIPS data
in units of MJy\,sr$^{-1}$. $\dot{\Sigma}_{*}$ has the units of M$_{\odot}$kpc$^{-2}$yr$^{-1}$.
This prescription only holds for a constant star formation rate over the last few $100$~Myr.
The recipes for the calculation of the TIR emission together with comments on the validity of different star formation estimators are 
given in Appendix~\ref{app:sfr}.

The integrated SFR, the radio continuum flux at $4.85$~GHz from Vollmer et al. (2010, 2013), and the stellar mass
from Boselli et al. (2015) of each galaxy are presented in Table~\ref{tab:sample}.
\begin{table*}
      \caption{Galaxy sample}
         \label{tab:sample}
      \[
       \begin{tabular}{lcccccc}
        \hline
        name & D & i & $S_{4.85~{\rm GHz}}$ & $S_{1.4~{\rm GHz}}$ & SFR & log($M_*$) \\
         & (Mpc) & (deg) & (mJy) & (mJy) & (M$_{\odot}$yr$^{-1}$) & (M$_{\odot}$) \\
        \hline
        NGC6946 & 7.7 & 38 & 480 & 1400 &  3.5 & 10.5 \\
        M51 & 8.6 & 20 & 400 & 1300 &  2.9 & 10.6 \\
        NGC4254 & 17 & 30 & 162.0 & 510.0 & 3.6 & 10.4 \\
        NGC4294 & 17 & 70 & 11.5 & 32.3 & 0.3 & 9.2 \\
        NGC4298 & 17 & 57 & 8.0 & 24.8 & 0.4 & 10.1 \\
        NGC4299 & 17 & 22 & 7.7 & 17.6 & 0.3 & 9.5 \\
        NGC4302 & 17 & 90 & 13.2 & 43.2 & 0.5 & 10.4 \\
        NGC4303 & 17 & 25 & 135.1 & 398.5 & 3.6 & 10.5 \\
        NGC4321 & 17 & 27 & 99.1 & 256.0 & 3.2 & 10.7 \\
        NGC4330 & 17 & 90 & 4.7 & 13.3 & 0.1 & 9.5 \\
        NGC4396 & 17 & 72 & 8.3 & 15.8 & 0.2 & 9.3 \\
        NGC4402 & 17 & 74 & 26.3 & 68.0 & 0.5 & 10.0 \\
        NGC4419 & 17 & 74 & 9.2 & 17.7 & 0.6 & 10.2 \\
        NGC4457 & 17 & 33 & 16.2 & 34.1 & 0.4 & 10.4 \\
        NGC4501 & 17 & 57 & 96.2 & 331.0 & 2.3 & 11.0 \\
        NGC4522 & 17 & 79 & 7.6 & 24.0 & 0.1 & 9.4 \\
        NGC4532 & 17 & 70 & 51.2 & 119.9 & 0.8 & 9.2 \\
        NGC4535 & 17 & 43 & 28.7 & 66.0 & 1.6 & 10.5 \\
        NGC4567/68 & 17 & 49/66 & 46.8 & 128.8 & 1.8 & 9.92/10.33 \\
        NGC4579 & 17 & 38 & 43.3 & 88.4 & 0.8 & 10.9 \\
        NGC4654 & 17 & 51 & 49.9 & 134.6 & 1.7 & 10.1 \\
        NGC4808 & 17 & 68 & 19.4 & 62.6 & 0.7 & 9.5 \\
        \hline
        \end{tabular}
      \]
\end{table*}

\begin{figure}
  \centering
  \resizebox{\hsize}{!}{\includegraphics{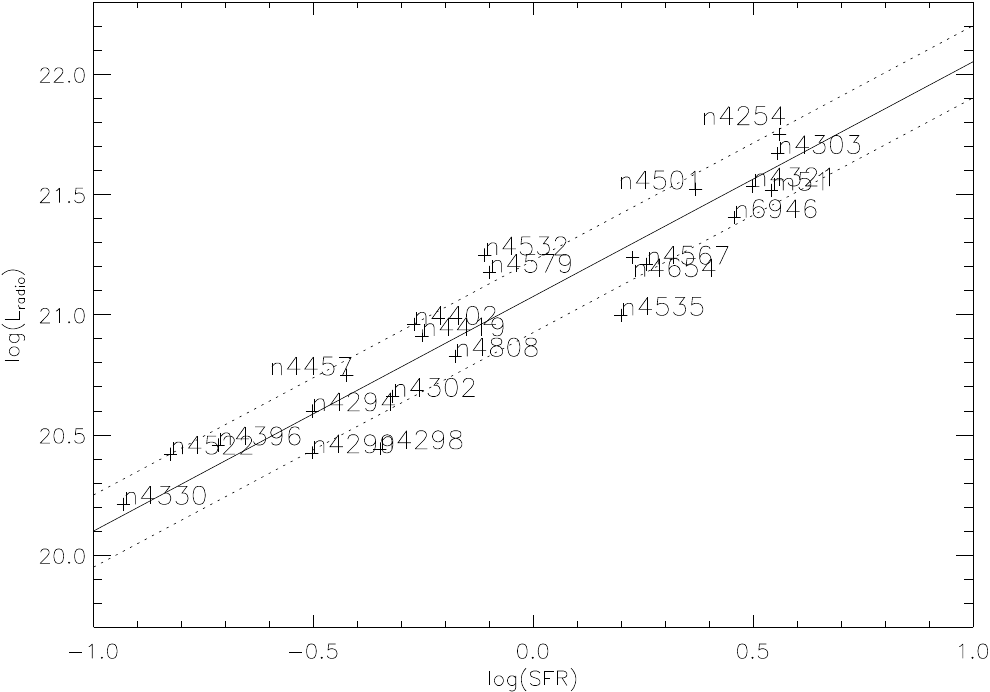}}
  \caption{Radio luminosity at $4.85$~GHz in W\,Hz$^{-1}$ as a function of the star formation rate in M$_{\odot}$yr$^{-1}$.
    The black line represents the result of a outlier-resistant linear regression with a slope of $1.0$, the dotted lines the scatter of $0.15$~dex.
  \label{fig:sfrradc_int2}}
\end{figure}

The integrated radio luminosities at $4.85$~GHz as a function of the integrated star formation rates for our sample galaxies are
shown in Fig.~\ref{fig:sfrradc_int2}. 
The slope of the log(L$_{\rm rad}$)--log(SFR) relation is $1.00$ with a scatter of $0.15$~dex. Many of the perturbed Virgo cluster galaxies
(NGC~4522, NGC~4457, NGC~4402, NGC~4532, NGC~4501, and NGC~4254) are located at the upper end of the observed range of radio luminosities
with respect to the SFR. On the other hand, NGC~4298 and NGC~4535 have radio luminosities about $2\,\sigma$ or a factor of about two lower 
than expected based on their SFR.

\section{Results\label{sec:results}}

For the spatially resolved correlations we show one data point per resolution element ($15-22''$).
The resolved correlation between the $100$~$\mu$m surface brightness $I_{100\mu{\rm m}}$ and the $4.85$~GHz radio continuum surface brightness
$I_{4.85{\rm GHz}}$ is presented in Fig.~\ref{fig:sfrrad_all_fir}. Only galaxies for which Herschel $100$~$\mu$m data are available are shown.
The radio surface brightness distributions of NGC~4303, NGC~4535, and NGC~4579 
significantly deviate from the mean relation over the whole disk regions and were excluded from the outlier-resistant linear regression.
NGC~4535 is overall radio-dim (see Sect.~\ref{sec:radiodimgd}) and the radio continuum emission within the disk of NGC~4579 probably has a strong AGN contribution 
(see Sect.~\ref{sec:radiobright}).
The slope of the log($I_{100\mu{\rm m}}$)--log($I_{4.85{\rm GHz}}$) relation is $1.02$ with a scatter of $0.17$~dex. 

\begin{figure}
  \centering
  \resizebox{\hsize}{!}{\includegraphics{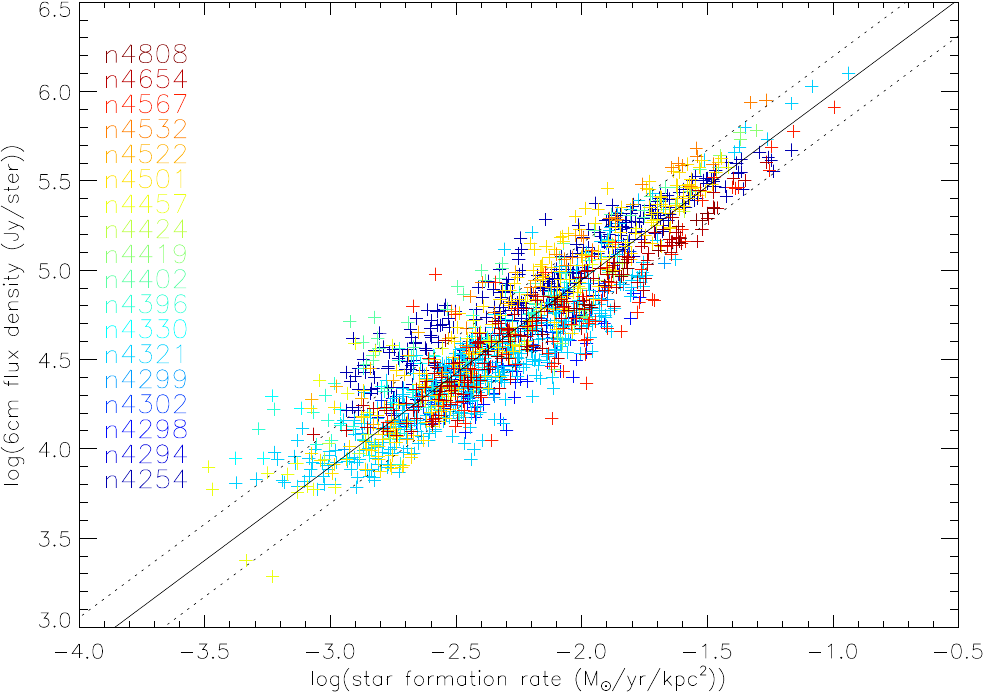}}
  \caption{Resolved properties measured in $15-22''$ apertures. Radio continuum surface brightness at $4.85$~GHz as a function of the star formation
    rate per unit area. The black line represents the result of a outlier-resistant linear regression with a slope of $1.05$, the dotted lines the scatter of $0.20$~dex.
  \label{fig:sfrrad_all}}
\end{figure}

The resolved correlation between the SFR per unit area $\dot{\Sigma}_*$ and the $4.85$~GHz radio continuum surface brightness $I_{4.85{\rm GHz}}$ is
presented in Fig.~\ref{fig:sfrrad_all}. As for the $100$~$\mu$m data, galaxies whose radio continuum surface brightness significantly deviates from 
the mean were excluded (NGC~4303, NGC~4535, and NGC~4579). The slope of the log($\dot{\Sigma}_*$)--log($I_{4.85{\rm GHz}}$) relation is $1.05$ with a scatter of $0.20$~dex. 
This result implies that $\dot{\Sigma}_*$ and $I_{100\mu {\rm m}}$ show a very strong correlation.
The slopes of the log($\dot{\Sigma}_*$)--log($I_{4.85{\rm GHz}}$) relations of the individual galaxies are given in Table~\ref{tab:coeffs2}.

\subsection{Radio-bright and radio-dim regions \label{sec:radiobright}}

\begin{figure*}
  \centering
  \resizebox{16cm}{!}{\includegraphics{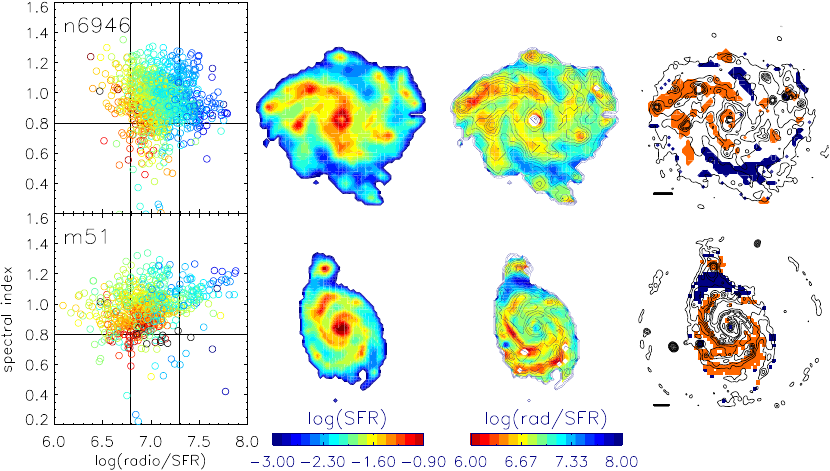}}
  \caption{From left to right: (i) spectral index as a function of the ratio between the radio surface brightness
    and the star formation rate per unit area, the colors correspond to the star formation rate per unit area (red corresponds to high values) 
    (ii) star formation rate per unit area (M$_{\odot}$kpc$^{-2}$yr$^{-1}$); 
    (iii) star formation rate per unit area
    (contours) on the  ratio between the radio continuum surface brightness (Jy/sr) and the star formation rate per unit area (color);
    (iv) radio continuum surface brightness (contours) on the radio-dim (red) and radio-bright (blue) regions.
    The radio continuum contour levels are $(1,2,4,6,8,10,20,30,40,50) \times 200~\mu$Jy/beam.
    The horizontal lines correspond to a size of $1'$. 
  \label{fig:rcfir_spixx1c1_nice_5}}
\end{figure*}

We used the results of the linear regression of Fig.~\ref{fig:sfrrad_all} to define radio-bright ($\log(I_{4.85{\rm GHz}}) > \log(I_{\rm exp})+0.25$) 
and radio-dim regions ($\log(I_{4.85{\rm GHz}}) < \log(I_{\rm exp})-0.25$). 
$I_{\rm exp}$ is the expected surface brightness from the linear fit of Fig.~\ref{fig:sfrrad_all}.
Radio-bright and radio-dim regions are thus defined on an absolute scale, which is justified by the tightness of the resolved radio--SFR correlation.
The second, third and fourth columns of Fig.~\ref{fig:rcfir_spixx1c1_nice_5} show the star formation rate per unit area, the star formation rate per unit area
(contours) on the  ratio between the radio continuum surface brightness and the star formation rate per unit area (color), and the
radio continuum surface brightness (contours) on the radio-dim (red) and radio-bright (blue) regions.

The ratio between the resolved radio continuum emission and the SFR depends on the fraction of thermal electrons,
CR losses, CR propagation (diffusion, streaming, advection), and radio active galactic nucleus (AGN) components.
CRs can move relative to the ISM by streaming along magnetic field lines down a CR density gradient.
Advection means the common transport of CRs together with the ISM and the associated magnetic field.
Vertical advection of CR electrons can lead to CR escape and/or thick radio disks or halos
in disk galaxies (e.g., Krause et al. 2018). Moreover, ram pressure can push CR electrons and magnetic fields out of the galactic plane
(e.g., Vollmer et al. 2021a).
The ratio between the resolved radio continuum emission and the SFR also depends on the frequency of the radio continuum observations 
via the synchrotron lifetime of the CR electrons, which increases for decreasing frequency:
\begin{equation}
\label{eq:synch}
t_{\rm sync} \simeq 4.5 \times 10^7 \big( \frac{B}{10~\mu{\rm G}} \big)^{-3/2} \big(\frac{\nu}{\rm GHz}\big)^{-1/2}~{\rm yr}\ .
\end{equation}
In addition, CR electrons can lose their energy via bremsstrahlung, inverse Compton (IC) scattering, ion, and pion production.  

To estimate these losses we assumed a stationary CR electron density distribution ($\partial n/\partial t$=0; Eq.~\ref{eq:diffeq}) 
and that the source term of CR electrons is
proportional to the star formation rate per unit volume $\dot{\rho}_*$. For the energy distribution of the cosmic
ray electrons the standard assumption is a power law with index $-q$, which leads to a power law of the radio continuum spectrum
with index $(1-q)/2$ (e.g., Beck 2015). 
Under these assumptions, the synchrotron emissivity is given by the density per unit energy interval of the primary CR 
electrons, where $E$ is the energy, $n_0 \propto \dot{\rho}_* t_{\rm eff}$, and
\begin{equation}
\label{eq:emissivity}
\epsilon_\nu {\rm d}\nu \propto \dot{\rho}_* t_{\rm eff} E^{-q} \frac{E}{t_{\rm sync}} {\rm d}E
\end{equation}
(Vollmer et al. 2022).
The effective lifetime of synchrotron-emitting CR electrons $t_{\rm eff}$ is set by by the synchrotron $t_{\rm sync}$, bremsstrahlung $t_{\rm brems}$, IC $t_{\rm IC}$,
ion $t_{\rm ion}$, and pion $t_{\pi}$ loss timescales. Moreover, CR electrons can escape from the galactic disk via vertical diffusion or outflows $t_{\rm escp}$:
\begin{equation}
\label{eq:emts}
\frac{1}{t_{\rm eff}}=\frac{1}{t_{\rm sync}}+\frac{1}{t_{\rm escp}}+\frac{1}{t_{\rm brems}}+\frac{1}{t_{\rm IC}}+\frac{1}{t_{\rm ion}}+\frac{1}{t_{\pi}}\ .
\end{equation}
Whereas a small synchrotron timescale (Eq.~\ref{eq:synch}) leads to an increase of the synchrotron emission, smaller 
timescales of bremsstrahlung, IC, ionization, or pion losses lead to a decrease of synchrotron emission.
In addition, CR electrons are transported toward regions of low SFR by diffusion or streaming. This decreases the synchrotron emission
in regions of high SFR and increases the synchrotron emission in regions of low SFR.

All these effects affect the synchrotron spectral index, which we measure between $4.85$ and $1.4$~GHz. 
The comparison between the radio/SFR to the spectral index thus contains information on the physical processes affecting the CR electrons 
(first row of Fig.~\ref{fig:rcfir_spixx1c1_nice_5} where we only show our diagnostic plots for the nearby spiral galaxies NGC~6946 and M~51.
The corresponding plots for the Virgo cluster galaxy sample are presented in Appendix~\ref{sec:virgoplots}; Figs.~\ref{fig:rcfir_spixx1c1_nice_1}
to \ref{fig:rcfir_spixx1c1_nice_4}).
In Sect.~\ref{sec:SI} we discuss the link between the spectral index distribution and the radio-bright/radio-dim regions.
We also produced the diagnostic plots for the $1.4$~GHz data (Appendix~\ref{sec:1.4GHzdata}; Figs.~\ref{fig:rcfir_spixx1c1_nice_20_5} 
to \ref{fig:rcfir_spixx1c1_nice_20_4}). The results are qualitatively
similar to those for the $4.85$~GHz data, albeit the radio/SFR--SI correlation is significantly stronger 
at $1.4$~GHz compared to the correlation at $4.85$~GHz. Since our $4.85$~GHz observations have a much better UV coverage because of the much longer 
observation times, we preferred to use and discuss these data. 

The observed local radio/SFR values range from $0.4$~dex to $1.4$~dex, which corresponds to about $5$ times the
scatter of the overall correlation (Fig.~\ref{fig:sfrrad_all}).
Even if there is an imbalance between radio-dim or radio-bright regions, the integrated radio/SFR ratio is not
strongly affected (Fig.~\ref{fig:sfrradc_int2}) because these regions do not prevail.
Even if NGC~4501 and NGC~4522 show radio-bright regions and lie above the integrated radio--SFR correlation,
their radio flux densities are still within the scatter ($0.15$~dex) of the correlation.
The radio-dim and radio-bright regions thus represent a second order effect to the radio--SFR correlation.
As shown by Vollmer et al. (2022) the losses of Eq.~\ref{eq:emts} are most relevant for the integrated radio--SFR correlation.

The radio-bright regions in NGC~6946 (Fig.~\ref{fig:rcfir_spixx1c1_nice_5}) correspond to the interarm regions and are thus most probably caused by CR diffusion or streaming.
The radio-bright region in M~51 (Fig.~\ref{fig:rcfir_spixx1c1_nice_5}) is located toward NGC~5195 and is thus most probably due to CR advection via gravitational tides. Indeed, the atomic hydrogen has a relatively high surface density in this region (Walter et al. 2008).
On the other hand, the H{\sc i} velocity dispersion outside the spiral arm is quite high in the radio-bright region, which might have led to an
increase of the magnetic field.   
The one-sided radio-bright regions in NGC~4330 (Fig.~\ref{fig:rcfir_spixx1c1_nice_2}) and NGC~4522 (Fig.~\ref{fig:rcfir_spixx1c1_nice_3}) coincide with
stripped extraplanar gas (Vollmer et al. 2006, 2012, 2021) and are thus due to CR transport together with the interstellar 
medium (ISM) via ram pressure. The one-sided radio-bright region in NGC~4402 (Fig.~\ref{fig:rcfir_spixx1c1_nice_2}) might correspond to stripped gas  (Crowl et al. 2005)
or a radio halo or a thick radio disk (Krause et al. 2018), 
which has been pushed toward the galactic disk in the south and/or be due to CR transport together with the ISM via ram pressure. 
In the former scenario one would expect a radio-bright southern edge of the galactic disk due to compression. Since this is not observed, we prefer the latter scenario.
The southern radio-bright region in NGC~4501 (Fig.~\ref{fig:rcfir_spixx1c1_nice_3}) coincides with a region of ISM compression caused by ram pressure 
(Vollmer et al. 2008). This might also be the case for the southern radio-bright region in NGC~4254 (Fig.~\ref{fig:rcfir_spixx1c1_nice_1}; Chyzy et al. 2007), 
the other radio-bright regions being probably caused by shear motions that were revealed by Phookun et al. (1993) or by CR diffusion or streaming. 
The radio-bright region in NGC~4532  (Fig.~\ref{fig:rcfir_spixx1c1_nice_3}) is extended in the polar direction and shows a strong poloidal magnetic field
component visible in polarized emission (Vollmer et al. 2013). Therefore, this region is most probably due to a nuclear starburst outflow.

The radio-bright region in NGC~4579 (Fig.~\ref{fig:rcfir_spixx1c1_nice_4}) is enigmatic. 
Since a CR source (the AGN), which is not connected to star formation, is probably involved in NGC~4579 (Appendix~\ref{sec:n4579}), 
we will discard this galaxy when we calculate sample-average 
values.

The radio-dim regions in all galaxies can be due to (i) CR diffusion or streaming out of regions of high SFR or 
(ii) bremsstrahlung, IC, ionization, pion, or escape losses. We refer to Sect.~\ref{sec:SI} for further discussion with respect to the spectral index.
Three galaxies are overall radio-dim: NGC~4298 (Fig.~\ref{fig:rcfir_spixx1c1_nice_1}), NGC~4535, and NGC~4567 (Fig.~\ref{fig:rcfir_spixx1c1_nice_4}).
NGC~4567 is not present in Fig.~\ref{fig:sfrradc_int2} because we did not separate NGC~4567 from NGC~4568 for the integrated data. 
It is remarkable that two out of three overall radio-dim galaxies are gravitationally interacting (NGC~4298 and NGC~4567).

NGC~4535 has an asymmetric distribution of polarized radio continuum emission in the western half of its disk (Vollmer et al. 2007), 
which coincides with strong shear motions
detected in the H{\sc i} velocity field (Chung et al. 2009). These shear motions are thus responsible for the region of enhanced polarized
radio continuum emission. Within this region the radio/SFR is normal, i.e. it is enhanced with respect to the otherwise radio-dim 
galactic disk. We argue that the shear motions increase the ordered magnetic field strength via the induction equation and hence also the total magnetic field strength.

\subsection{Correlations \label{sec:correlations}}

The synchrotron radio/SFR ratio can be modified by CR transport or a change in the magnetic field strength.
The magnetic field strength might be enhanced by ISM shear or compression motions, which are traced by
the polarized radio continuum emission (Vollmer et al. 2007, 2010, 2013).

To identify the causes of the variations of the radio/SFR ratio, we investigated the relations between the SFR and (i) radio continuum, (ii) 
polarized emission (PI), and (iii) fractional polarization (FP) and (iv) the radio continuum emission and the polarized emission.
For each of these four relations we calculated the Spearman rank correlation coefficients (Table~\ref{tab:coeffs1}) and the slopes (Table~\ref{tab:coeffs2})
of the log-log relations via  a Bayesian approach to linear regression with errors in both directions (Kelly 2007).
The detailed results on these correlations are presented in Appendix~\ref{app:correlations}.
  
The three basic correlations are SFR--radio, SFR--PI, and radio--PI with slopes of $1.11 \pm 0.02$, $0.43 \pm 0.03$, and $0.41 \pm 0.02$, 
respectively\footnote{The slopes were determined for the combined data of all galaxies except NGC~4579.}.
The SFR--radio correlation is much steeper and tighter than the SFR--PI correlation resulting in an SFR--FP
anticorrelation. The difference of the SFR--radio, SFR--PI slopes is $-0.83 \pm 0.04$, which corresponds to the measured SFR--FP slope.
The radio--FP slope follows directly from the radio--PI slope.

\subsection{The SFR--radio/SFR and PI--radio/SFR relations \label{sec:piradiosfr}}

\begin{figure}
  \centering
  \resizebox{\hsize}{!}{\includegraphics{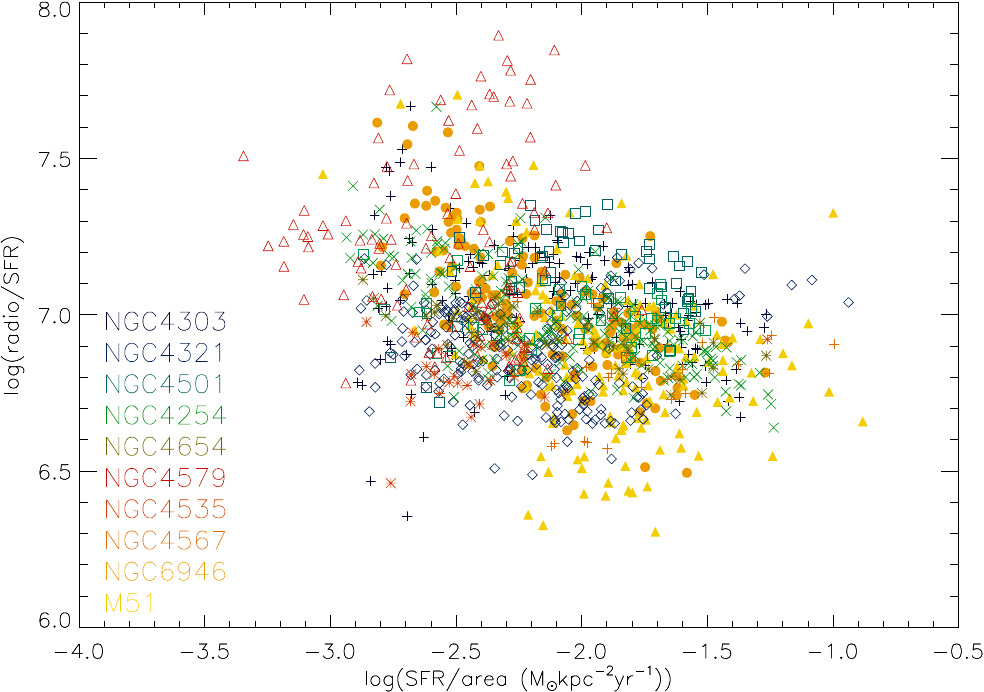}}
  \caption{Ratio between the radio surface brightness (Jy/sr) and the star formation rate per unit area (M$_{\odot}$pc$^{-2}$yr$^{-1}$) 
    as a function of the star formation rate per unit area. 
  \label{fig:sfrrad_all_ratio}}
\end{figure}

To further identify the causes of the variations of the radio/SFR ratio, we investigated if the local radio/SFR ratio depends on the SFR per unit area.
Only a weak anticorrelation was found between these two quantities for the whole sample 
(Spearman rank correlation coefficient $\rho=-0.29$; Fig.~\ref{fig:sfrrad_all_ratio}). Only NGC~6946, NGC~4254, and NGC~4654 show strong 
anticorrelations ($\rho < -0.6$).
 
\begin{figure*}
  \centering
  \resizebox{\hsize}{!}{\includegraphics{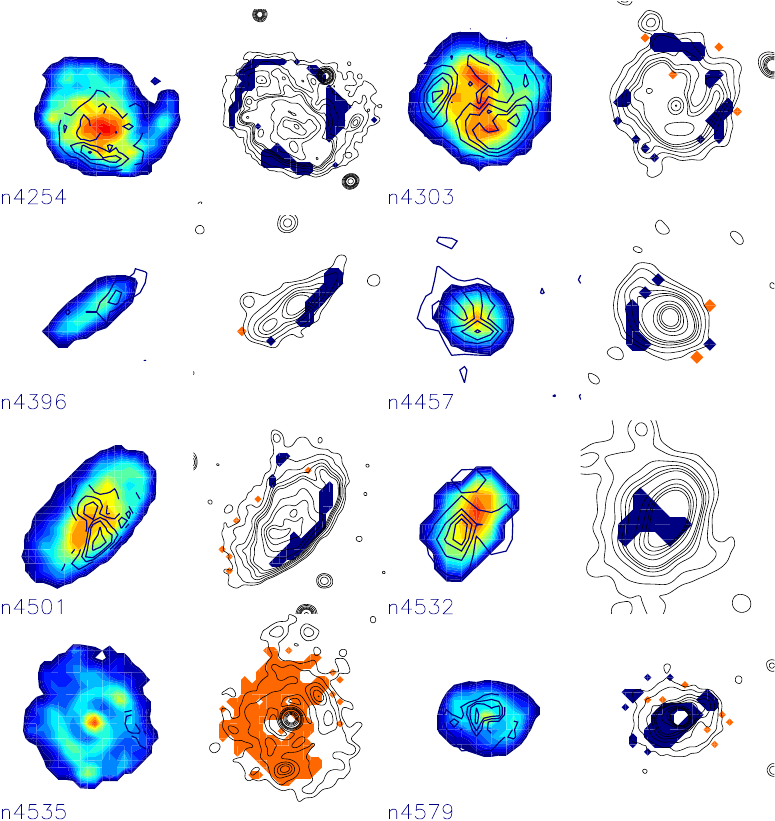}}
  \caption{For each galaxy: left panel: polarized intensity (contours) on the star formation rate per unit area (color);
    right panel: radio continuum surface brightness (contours) on the radio-dim (red) and radio-bright (blue) regions.
    For the color and contour levels we refer to Appendix~\ref{sec:virgoplots}.
  \label{fig:rcfir_spixx1c1_nice_pol}}
\end{figure*}

We realized that the radio-bright regions frequently coincide with the asymmetric ridges of polarized radio continuum emission
presented in Vollmer et al. (2010, 2013) and regions of enhanced polarized radio continuum emission, which does not follow the spiral structure. 
These regions are clearly different from classical interarm regions with high polarization.
In the eight galaxies presented in Fig.~\ref{fig:rcfir_spixx1c1_nice_pol} 
regions of high polarized radio continuum emission coincide at least partly with radio-bright regions.
Since the inverse is not always true, other mechanisms than enhanced polarized emission can lead to radio-bright regions.

\begin{figure}
  \centering
  \resizebox{\hsize}{!}{\includegraphics{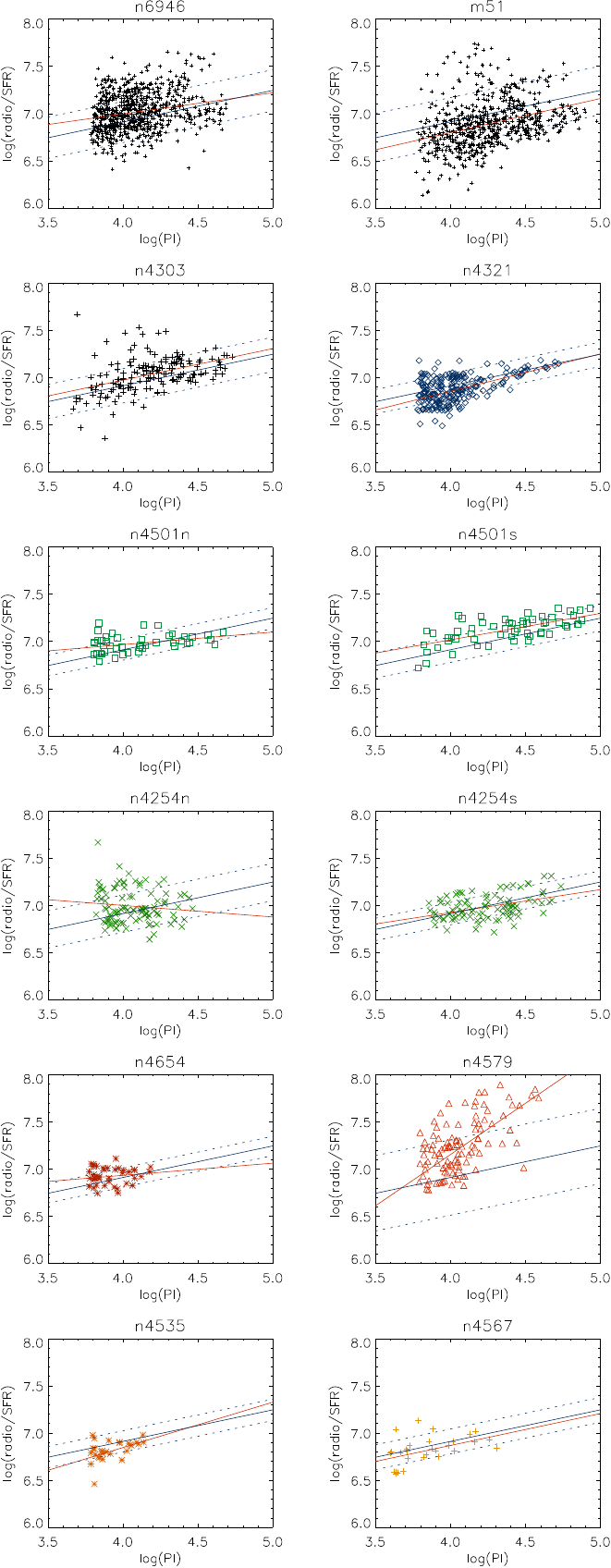}}
  \caption{Ratio between the radio surface brightness and the star formation rate per unit area as a function of the polarized 
    intensity. The blue lines show the common linear fit and scatter for the relation of all Virgo galaxies except NGC~4579,
    the red lines show the individual linear fits for each galaxy.
  \label{fig:radsfr_pi}}
\end{figure}

The asymmetric regions of enhanced polarized emission are located in the outer disks of NGC~4254 (south), NGC~4303 (ring-like),
NGC~4396 (west), NGC~4457 (east), NGC~4501 (southwest), and NGC~4535 (west); they are located in the inner parts of the disks in NGC~4532
and NGC~4579. In the two latter galaxies, the enhanced total power and polarized radio continuum emission is due to
an outflow (starburst and AGN, respectively).
To investigate the relationship between polarization and radio-brightness, we correlated the local radio/SFR ratios with the surface 
brightness of polarized radio continuum emission for the $11$ individual galaxies (Fig.~\ref{fig:radsfr_pi}) and for the sample of $11$ 
galaxies (lower right panel of Fig.~\ref{fig:radsfr_piall}).

The corresponding Spearman rank correlation coefficients $\rho$ are presented in the seventh column of Table~\ref{tab:coeffs1}.
Moderate correlations ($0.4 \la \rho \la 0.6$) are found in most galaxies, except for NGC~6946, NGC~4254, and NGC~4654.
The strongest correlation is found in the southern compressed region of NGC~4501 ($\rho=0.62$). The slopes of the subset of
moderately correlating log(PI)--log(radio/SFR) relations (seventh column of Table~\ref{tab:coeffs2}) vary between $0.33$ and $0.50$. 
Only the log(PI)--log(radio/SFR) 
relation of NGC~4579 has an exceptional slope of $1.08$. 
The  Spearman rank correlation coefficient for the whole sample is $\rho=0.41$ and the slope is $0.31$.
If we remove NGC~6946, NGC~4254, and NGC~4654 from the sample, the correlation coefficient and slope become $0.51$ and $0.36 \pm 0.03$, respectively.
We thus observe a clear albeit moderate correlation between the surface brightness of polarized radio continuum emission and
the radio/SFR ratio(see Appendix~\ref{sec:sfrpiapp} for a detailed discussion of the correlation). 
This correlation is consistent with an enhancement and ordering of the magnetic field in regions of ISM compression and shear motions.

The two perturbed galaxies NGC~4254 and NGC~4654, which show a weak or absent correlation between the surface brightness of polarized 
radio continuum emission and the radio/SFR ratio, present a moderate or strong $(\rho > 0.5)$ correlation between the fractional polarization (FP) 
and the radio/SFR ratio (second column of Table~\ref{tab:coeffs2}). The slopes of these correlations (Fig.~\ref{fig:radsfr_dp}) range between $0.12$ and $0.42$.
The FP--radio/SFR correlation of NGC~6946 is very weak ($\rho=0.3$) but significantly higher than the PI-radio/SFR correlation.
It is remarkable that, except for the isolated galaxy NGC~6946, the radio/SFR ratio is correlated with either FP or PI.
In the latter galaxy, the FP--radio/SFR relation is significantly stronger than that of the PI--radio/SFR relation.
The two correlations thus seem to be exclusive.

\subsection{Interpretation}

Polarized synchrotron intensity at high radio frequencies (to avoid Faraday depolarization) measures 
the ordered magnetic field projected on the plane of the sky within the telescope beam. 
A regular (or coherent) field has a well-defined direction within the beam of the telescope.
Turbulent magnetic fields can be isotropic (i.e., the same dispersion in all three spatial dimensions) or anisotropic (i.e., different dispersions; 
Beck et al. 2019). In the latter case local field fluctuations are not isotropic but rather prefer certain orientations (not directions; Jaffe 2019). 
The ordered field is made of the regular and anisotropic field components. In general, the ordered magnetic field component is dominated 
by the anisotropic field component (Table~3 of Beck et al. 2019).

For the interpretation of our findings we assume the following framework:
CR electrons are created in supernova shocks, which are located in starforming region within the galactic disk.
These CR electrons undergo mainly synchrotron losses, bremsstrahlung, and IC losses in regions of high SFR and diffuse or stream from
regions of high SFR to regions of low SFR. Since there are no starburst galaxies\footnote{NGC~4532, which is located slightly above 
the main-sequence of starforming galaxies (Renzini \& Peng 2015), has an outflow that is radio-bright.} in our sample, escape losses 
via galactic winds are not important. 
The isotropic random magnetic field is created by a small-scale dynamo.
It is enhanced (as observed in NGC~4535 and NGC~4501) and becomes anisotropic by ISM compression and shear motions in perturbed 
regions of galactic disks. In addition, the large-scale dynamo creates a regular field
component in regions of low SFR, i.e. the interarm regions. This regular field might dominate the total field
in the so-called magnetic arms (e.g., Beck et al. 2019). The regular field is tangled in regions of high star 
formation rates where the turbulent energy density is high.
 
In the simulations of Seta \& Federrath (2020) the small-scale, random field (with zero mean and non-zero rms) is generated by 
the tangling of the non-zero mean field due to turbulence. Energetically, the input turbulent kinetic energy is converted into small-scale 
magnetic energy by shredding of the mean field (the second term on the right hand side of their Eq.~8). This mean field is maintained 
by the large-scale dynamo and held constant due to periodic boundary conditions in these simulations.

A discussion on the log(SFR)--log(radio) and log(SFR)--log(FP) correlations can be found in Appendix~\ref{app:fpradsfr}.
We suggest that the SFR--FP correlation in all galaxies is driven by the fact that the large-scale dynamo operates
preferably in regions with low SFR (see e.g. Beck et al. 2019). Moreover, the steeper slopes of the SFR--FP correlation are
caused by compression and shear of isotropic random fields (which lead to anisotropy via field ordering),
which mainly occur in regions with low SFR where the gas energy density is also low. 

Shear affects the magnetic field geometry if the kinetic energy density due to shear is larger than that of the turbulent gas
\begin{equation}
\rho v_{\rm turb} l_{\rm driv} R \frac{\partial \Omega}{\partial R} > \rho v_{\rm turb}^2 \ ,
\end{equation}
where $v_{\rm turb}$ is the turbulent velocity dispersion, $l_{\rm driv}$ the turbulent driving length scale, $R$ the galactic radius,
$\Omega$ the angular rotation velocity, and $\rho$ the gas density.
For a constant rotation curve this leads to
\begin{equation}
\label{eq:shearr}
t_{\rm shear} = \big( R \frac{\partial \Omega}{\partial R} \big) ^{-1} \sim \Omega^{-1} < t_{\rm turb} = \frac{l_{\rm driv}}{v_{\rm turb}} \ . 
\end{equation}
For typical values of $l_{\rm driv} \sim 100$~pc and $v_{\rm turb} \sim 10$~km\,s$^{-1}$ the turbulent timescale is $t_{\rm turb} \sim 10$~Myr.
For a spiral arm width of $500$~pc (Velusamy et al. 2015) a velocity difference of about $50$~km\,s$^{-1}$ is needed for shear leading to 
field ordering. Such high shear velocities are not observed in symmetric spiral arms.
On the other hand, such a velocity difference is observed in the western gas arm of NGC~4535 (Chung et al. 2009).

Similarly, compression leads to magnetic field ordering if the compression timescale is smaller than the turbulent timescale
\begin{equation}
\label{eq:compr}
t_{\rm comp}=\rho \big(\frac{\partial \rho}{\partial t}\big)^{-1} <   t_{\rm turb} \ .
\end{equation}
We note that the compression timescale is smaller in regions with low gas densities.
The bridge region of the Taffy galaxies (Vollmer et al. 2021b) or the large-scale shock region in Stephan's Quintet (Nikiel-Wroczynski et al. 2013)
might be examples of large-scale ICM compression with associated radio continuum polarisation.
The compression timescale cannot be derived from observations but from simulations (Vollmer et al. 2021b).
A future project will be devoted to this issue. Here we suggest that Eq.~\ref{eq:compr} is fulfilled in parts of the
southwestern compressed gas of NGC~4501.

We conclude that in NGC~4535 and NGC~4501 the radio continuum emission in the sheared and compressed regions is enhanced with respect to the
radio continuum emission of the rest of the disk. Thus shear and compression lead to field ordering and a locally enhanced total magnetic field 
via the induction equation. We argue that the observed PI--radio/SFR correlation is caused by magnetic field enhancement and ordering via 
ISM shear and compression motions in the interacting galaxies (Sect.~\ref{sec:piradiosfr}). 

Finally, we ascribe the higher Spearman coefficients and steeper slopes of the radio--PI relation 
with respect to the SFR--PI relation to the influence of (i) the CR bremsstrahlung and IC losses (Eq.~\ref{eq:emts}), which
are higher in regions of high SFR, and (ii) diffusion or streaming of CR electrons: CR electrons diffuse or stream out of the regions of 
high SFR and low PI into regions of low SFR and high PI.

\subsection{Tying in the spectral index \label{sec:SI}}

As stated in Sect.~\ref{sec:radiobright},  the CR energy losses of Eq.~\ref{eq:emts} affect the synchrotron spectral index, 
which we measure between $4.85$ and $1.4$~GHz.
Pure synchrotron losses lead to a synchrotron spectral index of $-q/2$.
Whereas bremsstrahlung, ionization, and pion losses lead to a flatter spectrum, IC losses lead to a steeper spectrum (see Fig.~\ref{fig:model3}).
CR transport removes CR electrons from regions of high SFR without changing the spectrum. On the other hand, the CR electrons 
that were transported into regions of low SFR will age. Since CR electrons of higher energies have smaller synchrotron energy loss times than
CR electrons of lower energies, aging leads to a steeper spectrum.
The spectral index distributions of NGC~6946 and M~51 are shown in the left panels of Fig.~\ref{fig:rcfir_spixx1c1_nice_5}, those
of the Virgo cluster galaxies in the corresponding figures of Appendix~\ref{sec:virgoplots}).
Within these plots, the horizontal line corresponds a spectral index (SI) of $0.8$ and the vertical lines to the limits of radio-bright and
radio-dim regions (Sect.~\ref{sec:radiobright}). The colors correspond to the SFR per unit area with red being high values of SFR/area.

In all galaxies but NGC~4501 and NGC~4532, the resolution elements with the highest SFR per unit area ($\ga 3 \times 10^{-2}$~M$_{\odot}$kpc$^{-2}$yr$^{-1}$)
have SI around a value of $0.8$. Most frequently, the resolution elements with lower SFR per unit area have steeper SI.
The vast majority of resolution elements with the lowest SFR per unit area ($\dot{\Sigma}_* \la 3 \times 10^{-3}$~M$_{\odot}$kpc$^{-2}$yr$^{-1}$)
have SI$> 0.8$. In NGC~4396 all resolution elements, in NGC~4321, NGC~4522, NGC~4535, NGC~4579, and NGC~4654 some resolution elements with 
$\dot{\Sigma}_* \la 3 \times 10^{-3}$~M$_{\odot}$kpc$^{-2}$yr$^{-1}$ have SI$<0.8$.
In particular, we observe a clear gradient in the radio/SFR--SI space in NGC~6946, M~51, NGC~4254, NGC~4303, NGC~4396, NGC~4501, NGC~4522, and NGC~4567/68: 
regions with a low radio/SFR ratio have the shallowest SI ($\sim 0.8$). The other 13 galaxies show a vertical distribution of points in the radio/SFR--SI space. 
The SI distributions in the radio-dim regions of the overall radio-dim galaxies show different behaviors:
whereas these regions show a steep SI in NGC~4298, they have shallow SI in NGC~4567 and a broad range of SI in NGC~4535.
Lastly, the two nearest galaxies, NGC~6946 and M~51, show quite different log(radio/SFR)--SI distributions: 
whereas the resolution elements with $\dot{\Sigma}_* \ga 3 \times 10^{-2}$~M$_{\odot}$kpc$^{-2}$yr$^{-1}$ are distributed more vertically in NGC~6946,
they are distributed more horizontally in M~51. Furthermore, the radio-bright regions in M~51 
have significantly steeper SI ($\ga 1.0$) than those in NGC~6946 (SI$\la 1.0$; left panels of Fig.~\ref{fig:rcfir_spixx1c1_nice_5}). 

NGC~4330, NGC~4402, NGC~4522, NGC~4532, and M~51 are clear cases of CR particle advection via ICM ram pressure, galactic winds,
or gravitational tides (Sect.~\ref{sec:radiobright}).
The CR electrons, which were transported out of the galactic disk will age leading to a steep radio continuum spectrum with SI$>0.8$. 
Indeed, this effect is observed in these five galaxies, which therefore do not need further discussion.

For the other galaxies the interplay between the SFR and radio/SFR ratio on the one hand and the spectral index on the 
other hand is complex. To understand this interplay, modelling of the synchrotron emission in disk galaxies is needed. 
A first approach to such a modelling is presented in Sect.~\ref{sec:model}.

\section{Model \label{sec:model}}

For the interpretation and understanding of our results a 3D model, where star formation and the physics of synchrotron 
emission are included, is highly valuable, as presented in this Section. The CR propagation by diffusion within the disk plane is modelled by
the convolution of the model radio continuum image with a symmetric Gaussian kernel, the vertical CR diffusion by an
escape timescale.
For the calculation of the galaxy dynamics at scales of about $\sim 1$~kpc we used the 3D dynamical model
introduced by Vollmer et al. (2001) and applied to NGC~4501 by Vollmer et al. (2008) and Nehlig et al. (2016).
For the calculation of the radio continuum emission we use the analytical formalism introduced by Vollmer et al. (2022).

\subsection{Dynamical model}

The 3D N-body code consists of two components: a non-collisional component that simulates the stellar bulge/disk and
the dark halo, and a collisional component that simulates the ISM. A scheme for star formation was implemented where stars
were formed during cloud collisions and then evolved as non-collisional particles (see Vollmer et al. 2012).
The non-collisional component consists of $81920$ particles that simulate the galactic halo, bulge, and disk. The 
characteristics of the different galactic components are adapted to the observed properties. 
We adopted a model where the ISM is simulated as a collisional component, i.e., as discrete particles that possess a 
mass and a radius and can have inelastic collisions (sticky particles). The $20000$ particles of the collisional
component represent gas cloud complexes that evolve in the gravitational potential of the galaxy. During the disk evolution,
the particles can have inelastic collisions, the outcome of which (coalescence, mass exchange, or fragmentation) is simplified
following Wiegel (1994). This results in an effective gas viscosity in the disk.

As the galaxy moves through the ICM, its clouds are accelerated by ram pressure. Within the galaxy’s inertial system, its
clouds are exposed to a wind coming from a direction opposite to that of the galaxy’s motion through the ICM. The temporal ram
pressure profile has the form of a Lorentzian, which is realistic for galaxies on highly eccentric orbits within the Virgo 
cluster (Vollmer et al. 2001). The effect of ram pressure on the clouds is simulated by an additional force on the clouds 
in the wind direction. Only clouds that are not protected by other clouds against the wind are affected. 
Since the gas cannot develop instabilities, the influence of turbulence on the stripped gas is not included in
the model. The mixing of the intracluster medium into the ISM is very crudely approximated by a finite penetration length of
the intracluster medium into the ISM, i.e., up to this penetration length the clouds undergo an additional acceleration 
caused by ram pressure. A scheme for star formation was implemented where stars are formed during cloud collisions and 
then evolve as non-collisional particles. These newly formed star particles carry their time of formation. 
This star formation scheme reproduces the Schmidt-Kennicutt law (Fig.A.1 of Vollmer et al. 2012).

\subsection{Radio continuum emission model}

The diffusion--advection--loss equation for the CR electron density $n$ reads as
\begin{equation}
\label{eq:diffeq}
\begin{multlined}
\frac{\partial n}{\partial t} = D {\nabla}^2 n + \frac{\partial}{\partial E} \big( b(E) n(E) \big) - \\
(\vec{u}+\vec{v}) \nabla n + \frac{p}{3} \frac{\partial n}{\partial p} \nabla \vec{u} + Q(E) - \frac{n}{t_{\rm loss}}\ ,
\end{multlined}
\end{equation}
where $D$ is the diffusion coefficient, $E$ the CR electron energy, $\vec{u}$ the advective flow velocity, 
$\vec{v}$ the streaming velocity, $p$ the CR electron pressure,
$Q$ the source term, and $b(E)$  the  rate of energy loss.
The first part of the right-hand side of Eq.~6 is the diffusion term, followed by the synchrotron loss, advection, streaming, adiabatic energy gain or loss,
and the source terms. The advection and adiabatic terms are only important for the transport in a vertical direction. 
Energy can be lost via inverse Compton (IC) radiation, bremsstrahlung, pion or ionization energy loss, and most importantly synchrotron emission
(e.g., Murphy 2009, Lacki et al. 2010). Within our basic model we use a constant diffusion coefficient.

Observations of beryllium isotope ratios at the Solar Circle have shown that the confinement timescale for particles $\ge 3$~GeV 
is dependent on energy (Webber \& Higbie 2003). Lacki et al. (2010) use $D(E) \propto \sqrt{E/(3~{\rm GeV})}$ for $E > 3$~GeV.
Following Mulcahy et al. (2016), we also tested a constant diffusion coefficient $D=D_0$ for $E \le 3$~GeV and 
$E=D_0 \big(E/(3~{\rm GeV})\big)^\kappa$ for $E > 3$~GeV. Theoretically predicted values of $\kappa$ range between $0.3$ and $0.6$ 
(Schlickeiser 2002, Shalchi \& Schlickeiser 2005, Trotta et al. 2011).

We assumed a stationary CR electron density distribution ($\partial n/\partial t$=0; Eq.~\ref{eq:diffeq}). 
The CR electrons are transported into the halo through diffusion or advection
where they lose their energy via adiabatic losses or where the energy loss through synchrotron emission is so small that the
emitted radio continuum emission cannot be detected. Furthermore, we assumed that the source term of CR electrons is
proportional to the SFR per unit volume $\dot{\rho}_*$. For the energy distribution of the CR electrons, the standard assumption is a power law with index $q$, which leads to a power law of the radio continuum spectrum
with index $-(q-1)/2$ (e.g., Beck 2015). For the synchrotron emissivity we used Eq.~\ref{eq:emissivity} with the effective timescale given by Eq.~\ref{eq:emts}.

For the characteristic timescales, we follow the prescriptions of Lacki et al. (2010).
The diffusive escape timescale based on observations of beryllium isotope ratios at the solar circle (Connell 1998, Webber et al. 2003) is
\begin{equation}
\label{eq:tdiffe}
t_{\rm esc}=26/\sqrt{E/3~{\rm GeV}}~{\rm Myr}
\end{equation}
for $E > 3$~GeV and $t_{\rm diff}=26$~Myr otherwise.
The mean energy $E$ is calculated via the mean synchrotron frequency 
\begin{equation}
\label{eq:meanfreq}
\nu_{\rm s}=1.3 \times 10^{-1} \big(\frac{B}{10~\mu {\rm G}} \big) \big(\frac{E}{\rm GeV}  \big)^2 ~{\rm GHz}\ .
\end{equation}
The characteristic time for bremsstrahlung is
\begin{equation} 
t_{\rm brems}=37\, (\frac{n}{{\rm cm}^{-3}})^{-1}~{\rm Myr},
\end{equation}
and that for IC energy losses is
\begin{equation}
t_{\rm IC}=180\, (\frac{B}{10~\mu{\rm G}})^{\frac{1}{2}}(\nu_{\rm GHz})^{-\frac{1}{2}}(\frac{U}{10^{-12}~{\rm erg\,cm}^{-3}})^{-1}~{\rm Myr}
,\end{equation}
where $U$ is the interstellar radiation field. The timescale of ionization-energy loss is
\begin{equation}
t_{\rm ion}=210\, (\frac{B}{10~\mu{\rm G}})^{-\frac{1}{2}}(\nu_{\rm GHz})^{\frac{1}{2}}(\frac{n}{{\rm cm}^{-3}})^{-1}~{\rm Myr}\ .
\end{equation}
The magnetic field strength $B$ is calculated under the assumption of energy equipartition between the turbulent kinetic energy of the gas
and the magnetic field:
\begin{equation}
\label{eq:Bmag2}
\frac{B^2}{8 \pi}=\frac{1}{2}\rho v_{\rm turb}^2\ ,
\end{equation}
where $\rho$ is the total midplane density of the gas and $v_{\rm turb}$ its turbulent velocity dispersion.
Thus, within the framework of the symmetric analytical model the magnetic field is characterized only by its energy density.

With $\nu=C B E^2$, the synchrotron emissivity of Eq.~\ref{eq:emissivity} becomes
\begin{equation}
\label{eq:emfinal}
\epsilon_{\nu}= \xi \dot{\rho}_* \frac{t_{\rm eff}(\nu)}{t_{\rm sync}(\nu)} B^{\frac{q}{2}-1} \nu^{-\frac{q}{2}}\ .
\end{equation}
The constant is $C=e/(2 \pi m_{\rm e}^2 c^2)$.
With the CR electron density $n_0 \propto \dot{\rho}_* t_{\rm eff}$ and Eq.~\ref{eq:synch}, the classical expression 
$\epsilon_{\nu} \propto n_0 B^{(q+1)/2} \nu^{(1-q)/2}$ is recovered.
In Vollmer et al. (2022) the factor $\xi$ was chosen such that the radio--IR correlations measured by Yun et al. (2001) 
and Molnar et al. (2021) are reproduced within $2 \sigma$ (their Fig.~8).
We assume $q=2.3$ as our fiducial model but also investigated the case of $q=2.6$.
The gas density $\rho$, turbulent gas velocity dispersion $v_{\rm turb}$, and interstellar radiation field $U$ are 
directly taken from the dynamical model.
 
The radio continuum emission map was calculated in the following way:
\begin{enumerate}
\item
for all gas particles the gas density and velocity dispersion are calculated with the $50$ nearest neighbors;
\item 
the magnetic field strength $B$ is calculated for each gas particle using Eq.~\ref{eq:Bmag2};
\item
the radiation field $U$ is calculated for each star particle using all other star particles that are younger than $10$~Myr;
\item
the synchrotron and loss timescales are calculated for each star particle;
\item
Eq.~\ref{eq:emfinal} is evaluated for each star particle and added to the model radio continuum map;
\item
the model map is convolved with Gaussian kernels of size $l=2\,\sqrt{D\,t_{\rm eff}}$ with $D=10^{28}$~cm$^2$s$^{-1}$;
in this way vertical diffusion within the galactic disk was neglected;
\item
the projection angles are applied to the model map.
\end{enumerate}
In the following the escape timescale is neglected. Its influence is described in Appendix~\ref{sec:additional}.
For the synchrotron emission, bremsstrahlung and IC losses are most important. Ionization and pion losses are less
important but cannot be neglected (Fig.~\ref{fig:model3}).
We intentionally did not calibrate the units of the SFR per unit area and the radio continuum surface brightness
because we mainly use their ratio and the spectral index, which does not depend on units.

\subsection{Large-scale magnetic field model \label{sec:largescaleBfield}}

To calculate the large-scale regular magnetic field for our simulations, we follow the same procedure as Soida et al. (2006). Time-
dependent gas-velocity fields provided by the 3D dynamical simulations were used as input for solving the dynamo
(induction) equation using a second-order Godunov scheme with second-order upstream partial derivatives together with a
second-order Runge-Kutta scheme for the time evolution. To avoid non-vanishing $\nabla \cdot \vec{B}$, we evolved the dynamo equation
expressed by the magnetic potential $A$, where $\vec{B} = \nabla \times \vec{A}$, ($\vec{B}$ is the magnetic induction):
\begin{equation}
\frac{\partial \vec{B}}{\partial t}=\nabla \times (\vec{v} \times \vec{B})  - \nabla \times (\eta \nabla \vec{B})\,,
\end{equation} 
where $\vec{v}$ is the large-scale velocity of the gas, and $\eta$ the coefficient of a turbulent diffusion. We assume the magnetic field to
be partially coupled to the gas via the turbulent diffusion process (Elstner et al. 2000) whilst assuming the magnetic diffusion
coefficient to be $\eta = 3 \times 10^{25}$~cm$^2$s$^{-1}$. We do not implement any explicit dynamo process. 
The resulting polarized emission is calculated by assuming a density of relativistic electrons that is proportional to the model
gas density $\rho$. This rather crude approximation is motivated by the fact that in quiescent galaxies, the density of relativistic electrons 
is approximately proportional to the star formation density which depends on $\rho^{1-1.7}$. These calculations permit to determine
the geometry of the large-scale magnetic field but not its strength.

\subsection{Model results}

For the comparison with observations three characteristic models were calculated: a symmetric model, a perturbed model with one-sided compressed 
ISM, and a model with a globally radio-dim galactic disk.

\subsubsection{Symmetric model \label{sec:symmetricm}} 

For our basic model we used a timestep of the NGC~4654 simulations presented in Lize\'e et al. (2021).
NGC~4654 is perturbed by a rapid flyby of a spherical galaxy with a mass of $9.2 \times 10^{10}$~M$_{\odot}$.
At the time of interest the perturber is located at a distance of $18.9$~kpc. It will reach its minimum distance of $18.3$~kpc within $\sim 20$~Myr.
We chose this particular timestep because the galaxy developed marked spiral arms being still symmetric.

\begin{figure*}
  \centering
  \resizebox{\hsize}{!}{\includegraphics{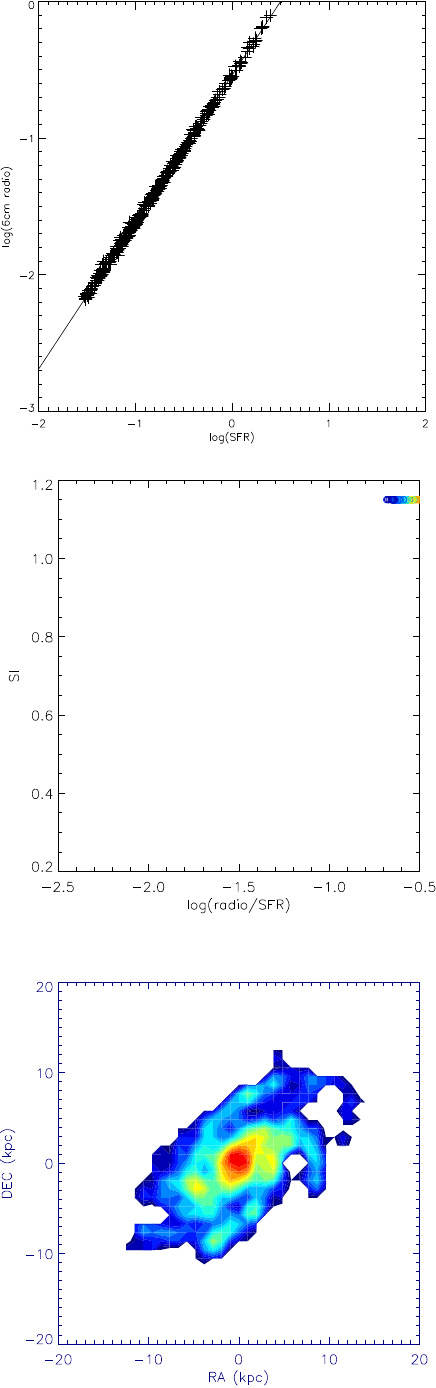}\includegraphics{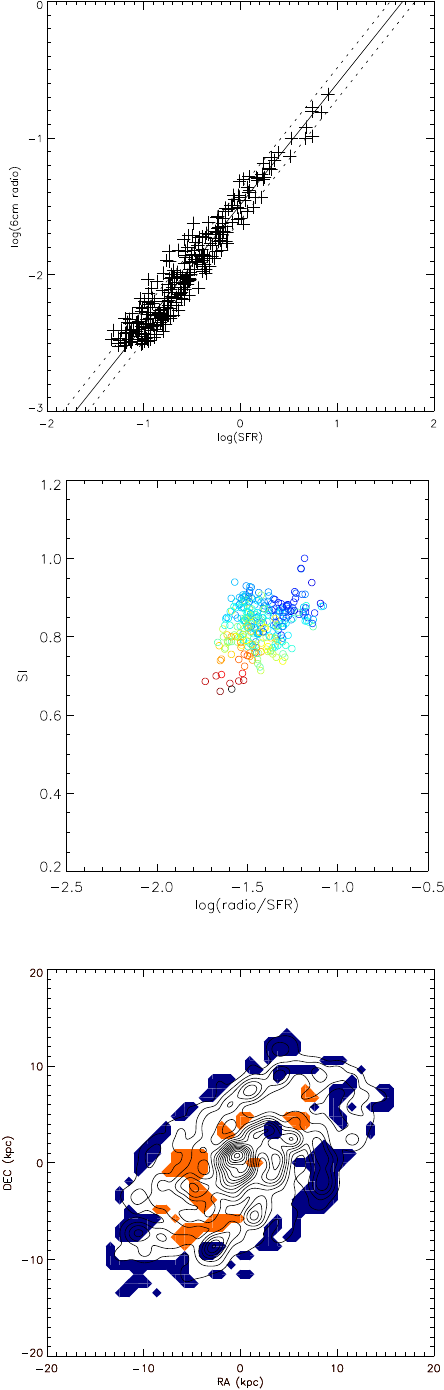}\includegraphics{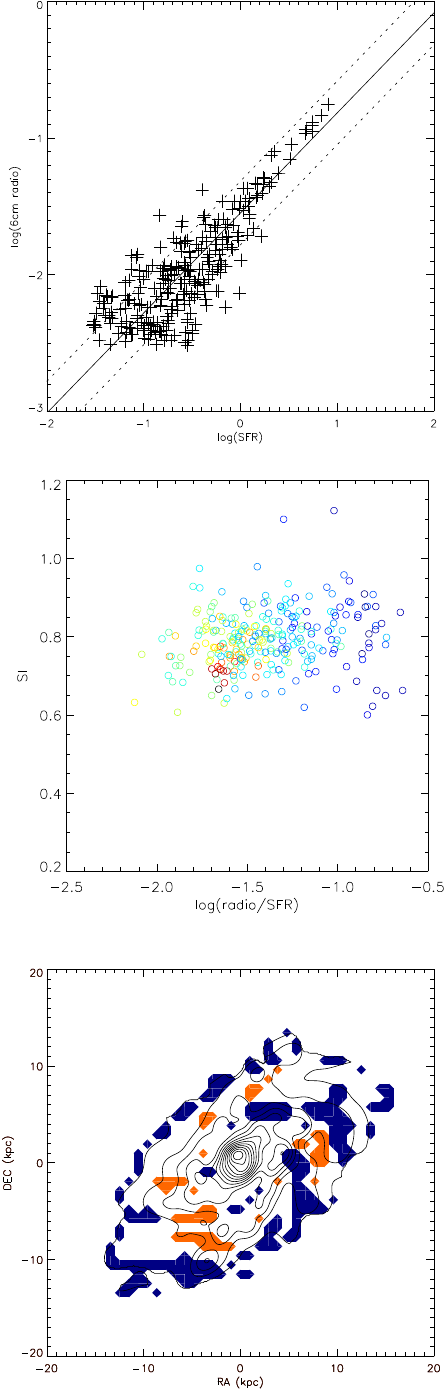}}
  \put(-310,220){\Large without diffusion}
  \put(-130,220){\Large with diffusion}
  \put(-480,220){\Large $t_{\rm eff}=t_{\rm sync}$}
  \put(-480,25){\Large SFR}
  \put(-310,25){\Large blue: radio-bright}
  \put(-130,25){\Large red: radio-dim}
  \caption{Model calculations in arbitrary units. Left column: $t_{\rm eff}=t_{\rm sync}$ without diffusion; middle column: 
    $t_{\rm eff}^{-1}=(t_{\rm sync}^{-1}+t_{\rm brems}^{-1}+t_{\rm IC}^{-1}+t_{\rm ion}^{-1}+t_{\pi}^{-1})^{-1}$ without diffusion;
    right column: with diffusion. Upper row: radio continuum surface brightness at $4.85$~GHz as a function of the
    star formation rate per unit area; middle row: spectral index as a function of the ratio between 
    the radio continuum surface brightness and the star formation rate per unit area; lower row:
    left: star formation rate per unit area; middle and right: radio continuum surface brightness at $4.85$~GHz (contours)
    on radio-dim (red) and radio-bright (blue) regions.
  \label{fig:model1}}
\end{figure*}
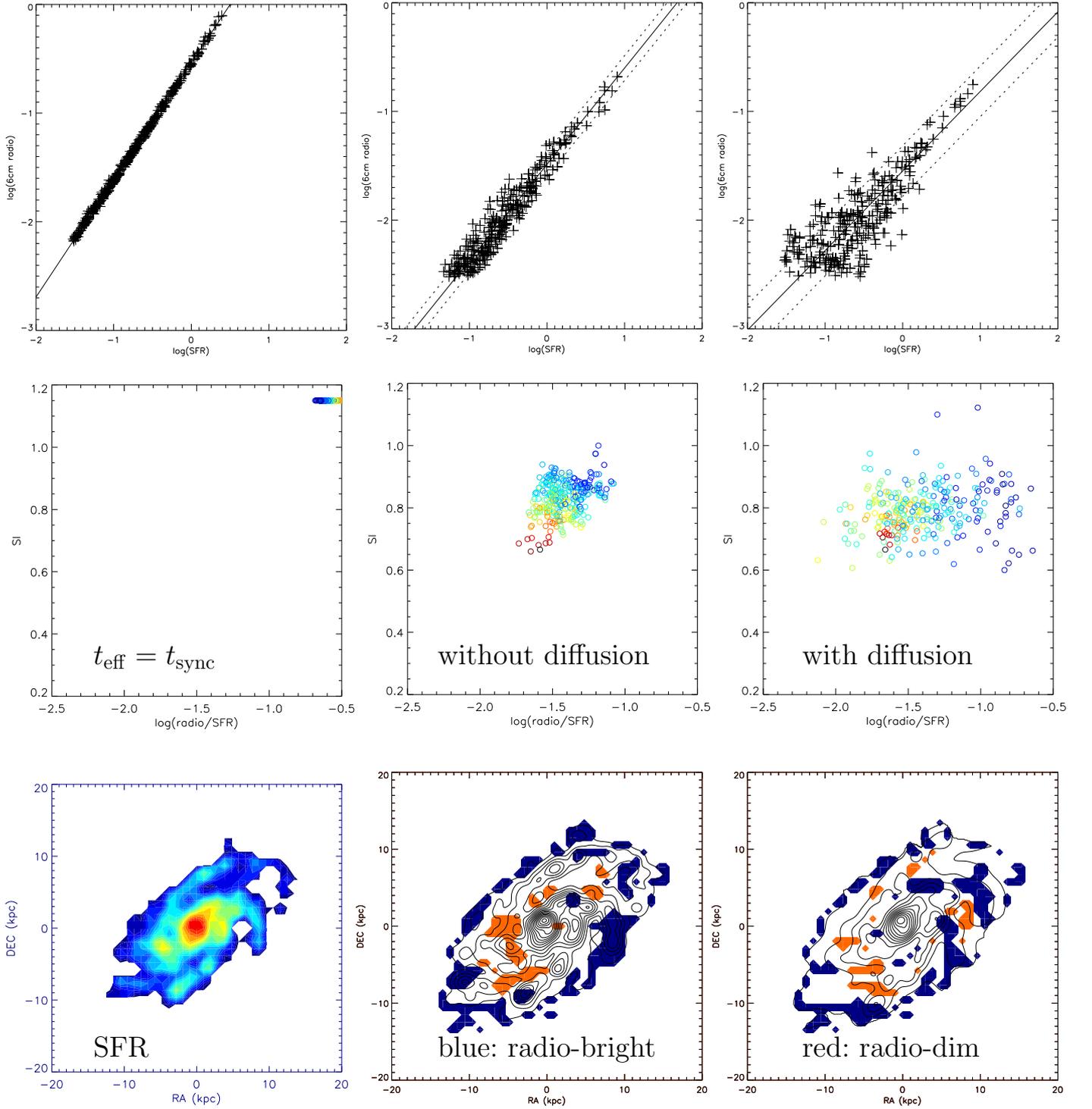

For the calculation of the synchrotron emission we set the exponent of the energy distribution of CR particles to $q=2.3$ (Eq.~\ref{eq:emissivity}).
As a first model we set $t_{\rm eff}=t_{\rm sync}$, i.e. there are no other than synchrotron losses. CR diffusion is absent.
The resulting radio--SFR relation and SI as a function of the radio/SFR ratio are shown in the left column of Fig.~\ref{fig:model1}.
The log(radio)--log(SFR) correlation has a slope of $1.08$ and a scatter of $0.02$~dex. As expected, the spectral index is constant SI$=1.15$.

We then set $t_{\rm eff}^{-1}=t_{\rm brems}^{-1}+t_{\rm IC}^{-1}+t_{\rm ion}^{-1}+t_{\pi}^{-1}$ (middle column of Fig.~\ref{fig:model1}).
The log(radio)--log(SFR) correlation has a slope of $0.89$ and the scatter increases to $0.11$~dex.
We divided the galaxies into radio-bright and radio-dim regions according to the mean and scatter. There is no clear correlation
between the these regions and the SFR per unit area.
It has to be noted that all regions are radio normal if one applies the observed scatter of $0.20$~dex.

In the third model we allow for CR diffusion within the disk plane (right column of  Fig.~\ref{fig:model1}). 
To do so we set the diffusion length to 
\begin{equation}
\label{eq:diffl}
l={\rm diff}=2\,\sqrt{D\,t_{\rm eff}}\ , 
\end{equation}
where we assume a diffusion coefficient of $D=D_0=10^{28}$~cm$^{2}$s$^{-1}$. 
This procedure corresponds to an instantaneous diffusion approximation.

Heesen et al. (2019) found $D = (0.13 - 1.5) \times 10^{28}$~cm$^2$s$^{-1}$ at $1$~GeV within three
local spiral galaxies. Mulcahy et al. (2016) derived a value of $D = 6.6 \times 10^{28}$~cm$^2$s$^{-1}$ for $E < 3$~GeV in M~51.
Moreover, the Milky Way diffusion coefficient is 
$D = 3 \times 10^{28}$~cm$^2$s$^{-1}$ (Strong et al. 2007). Vollmer et al. (2020) derived a value of 
$D = (1.8 \pm 0.6) \times 10^{28}$~cm$^2$s$^{-1}$ at $\sim 5$~GeV from $6$~cm radio continuum data. Our model diffusion coefficient is thus
at the lower end of the observationally derived range. Models with a two times higher and an energy-dependent
diffusion coefficient ($\kappa=0.3$) are presented in Fig.~\ref{fig:modelDE}. Models with diffusion coefficients 
$\ga 2 \times 10^{28}$~cm$^{2}$s$^{-1}$ or $\kappa > 0.3$ lead to model radio/SFR--SI relations which are not consistent 
with our observations.

The log(radio)--log(SFR) correlation has a slope of $0.73$ and the scatter increases to $0.23$~dex, which is comparable to the observed value of $0.20$~dex.
We note that the log(radio)--log(SFR) correlation of NGC~6946 has a slope of $0.76$ (Table~\ref{tab:coeffs2}).
CR diffusion therefore creates radio-dim and radio-bright regions. As observed, the radio-bright regions mainly coincide with interarm regions.
We caution the reader that we only modelled isotropic diffusion. Anisotropic diffusion along the large-scale magnetic field
lines will still lead to radio-bright and radio-dim regions but decrease the contrast between these regions.
CR diffusion does not increase the spectral index in our model. The only way to significantly increase SI is to increase $q$.
The inclusion of CR streaming where $l=l_{\rm str}$ if $l_{\rm str} > l_{\rm diff}$ (see Vollmer et al. 2020) with $l_{\rm str}=v_{\rm str}t_{\rm eff}$ and
$v_{\rm str}=100$~km\,s$^{-1}$ leads to a somewhat steeper slope of $0.80$.
The inclusion of diffusive CR escape increases the slope of the log(radio)--log(SFR) correlation to $0.95$ (Fig.~\ref{fig:model1a}).

We conclude that we are able to reproduce the resolved radio/SFR and SI values of a symmetric spiral galaxy as, e.g., NGC~6946.
Whereas the different CR electron losses lead to the overall level of radio continuum emission and are thus relevant for
the integrated radio--SFR correlation (Vollmer et al. 2022), diffusion or streaming lead to the observed radio-bright and radio-dim regions
in symmetric spiral galaxies.

\subsubsection{Perturbed model \label{sec:perturbedm}}

\begin{figure}
  \centering
  \resizebox{\hsize}{!}{\includegraphics{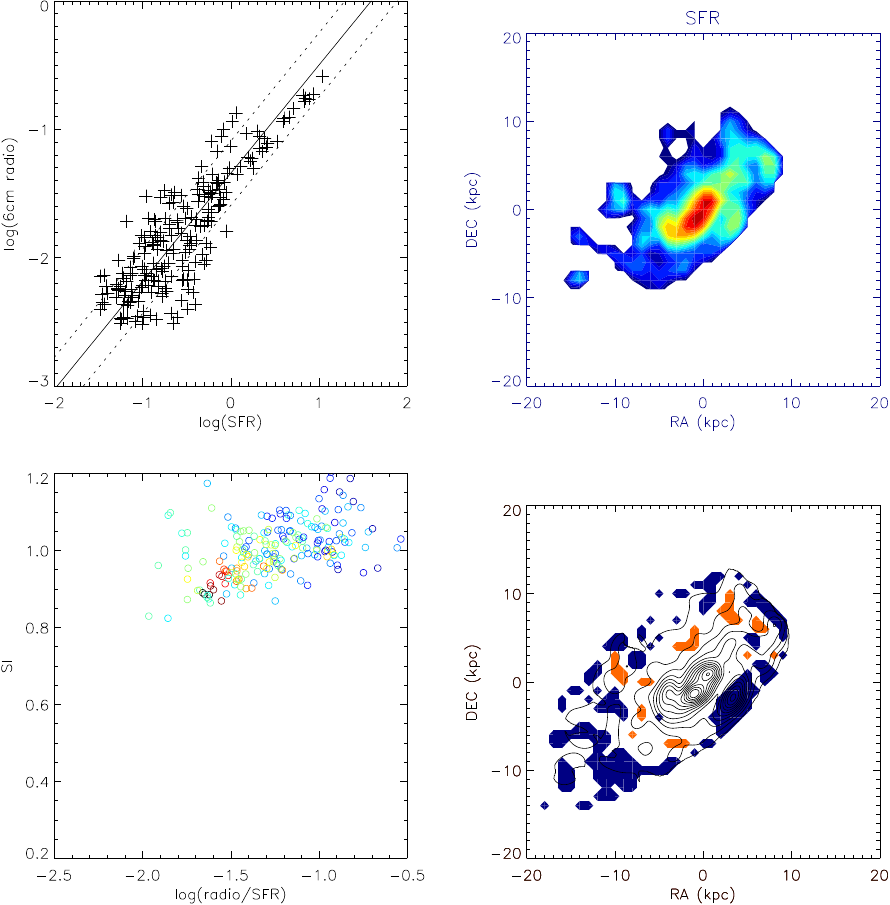}}
  \caption{NGC~4501 model (arbitrary units). Upper left: radio continuum surface brightness at $4.85$~GHz as a function of the
    star formation rate per unit area; upper right:  star formation rate per unit area; lower left:
    spectral index as a function of the ratio between the radio continuum surface brightness and the 
    star formation rate per unit area; lower right: radio continuum surface brightness at $4.85$~GHz (contours)
    on radio-dim (red) and radio-bright (blue) regions.
  \label{fig:model2}}
\end{figure}

NGC~4501 is a well-studied showcase for almost edge-on ram pressure stripping (Vollmer et al. 2008).
The ISM in the southern half of the galactic disk is compressed by ram pressure and shows a high molecular fraction
(Nehlig et al. 2016) and a bright asymmetric ridge of polarized radio continuum emission (Vollmer et al. 2007).
This region is clearly radio-bright (Fig.~\ref{fig:rcfir_spixx1c1_nice_3}).
For the comparison with observations we used the dynamical simulations presented in Nehlig et al. (2016).
Since this galaxy shows particularly high spectral indices, we set $q=2.7$ instead of $q=2.3$.
The model results including CR diffusion are presented in Fig.~\ref{fig:model2}.

Even without CR diffusion, the southwestern region of compressed ISM is radio-bright ($\Delta$ log(radio/SFR) $> 0.25$). This is due to the
enhanced magnetic field strength in the compressed gas (Eq.~\ref{eq:Bmag2}). Including CR diffusion steepens 
the log(SFR)--log(radio) and flattens the log(radio/SFR)--SI distribution. Our diffusion lengths (Eq.~\ref{eq:diffl}) seem to be somewhat overestimated.

The log(SFR)--log(radio) correlation of the model with and without CR diffusion have a slope of $0.84/0.96$ and a scatter of $0.25/0.16$~dex, respectively. 
The southwestern compressed region is radio-bright as it is observed (Fig.~\ref{fig:rcfir_spixx1c1_nice_3}). However, it is less extended than 
the observed radio-bright region. The radio-bright region in the southeast is absent in the $4.85$~GHz data but present in the $1.4$~GHz data.
The model distribution of the spectral index with respect to log(radio/SFR) (lower left panel of Fig.~\ref{fig:model2})
is well comparable to the observed distribution (Fig.~\ref{fig:rcfir_spixx1c1_nice_3}).

\begin{figure}
  \centering 
  \resizebox{4.4cm}{!}{\includegraphics{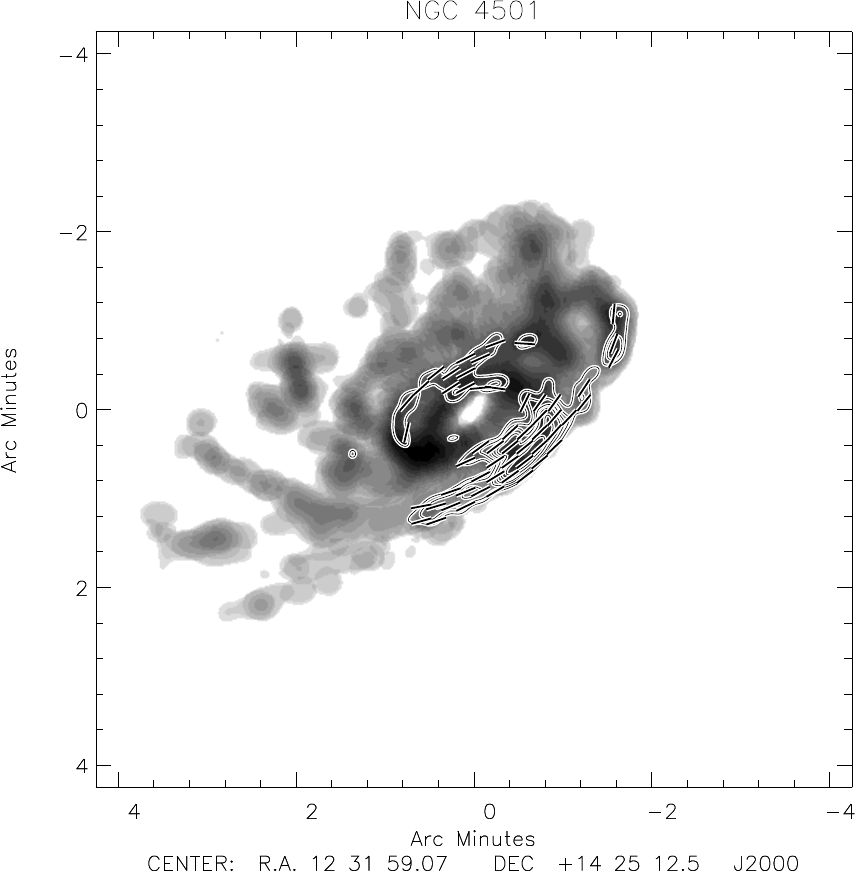}}
  \resizebox{4.4cm}{!}{\includegraphics{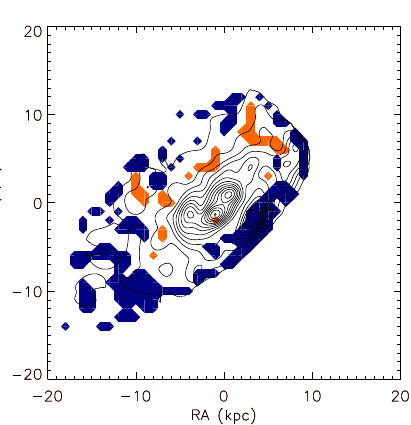}}
  \caption{NGC~4501 model. Left panel: polarized radio continuum emission (contours) with magnetic field vectors on H{\sc i}
    emission (greyscale). Right panel: same model as in Fig.~\ref{fig:model2} but with an additional regular field component $B_{\rm reg}=12$~$\mu$G in regions
    of enhanced model polarized intensity. Radio continuum surface brightness at $4.85$~GHz (contours)
    on radio-dim (red) and radio-bright (blue) regions.
 \label{fig:model2a}}
\end{figure}

Based on our conclusion that shear motions in NGC~4535 enhance the magnetic field strength via the induction equation (Sect.~\ref{sec:results}),
we increase the magnetic field strength by $12$~$\mu$G in regions of high model polarization (see Sect.~\ref{sec:largescaleBfield};
left panel of Fig.~\ref{fig:model2a}).
This field strength is comparable to the ordered field in the southern PI ridge of NGC~4501 assuming energy equipartition
between the magnetic field and the CR particles and a pathlength along the line-of-sight of $2$~kpc.
In this way we obtained a  more extended radio-bright region in the southern half of the galaxy (right panel of Fig.~\ref{fig:model2a}).
As before, the southwestern region of compressed ISM is radio-bright even without CR diffusion.

For the basic model with CR diffusion the Spearman rank correlation coefficient of the log(FP)--log(radio/SFR) relation and the log(PI)--log(radio/SFR)
are $0.08$ and $0.55$, respectively (upper panels of Fig.~\ref{fig:radsfr_pi_0}). This means that only the polarized radio continuum 
emission is correlated with the radio/SFR ratio. The slope of the log(PI)--log(radio/SFR) relation is $0.47$.
For the model with increased magnetic field strength in regions of high polarized emission the Spearman rank correlation coefficient the 
log(FP)--log(radio/SFR) relation increases to $0.37$, that of the  log(PI)--log(radio/SFR)
relation decreases to $0.51$ (lower panels of Fig.~\ref{fig:radsfr_pi_0}).
The slope of the log(PI)--log(radio/SFR) relation decreases to $0.31$, which is close to the average of the observed slopes of $0.33$.
However, we caution against the absence in our model of depolarization effects due to the tangling of the ordered magnetic field by
the turbulent motions of the ISM. 

We conclude that we can reproduce the observed main characteristics of the perturbed Virgo spiral galaxy NGC~4501: the radio-bright
region of compressed ISM and the correlation between the polarized radio continuum emission and the radio/SFR ratio.
The radio-bright region is mainly caused by the high magnetic field strength in the compressed ISM. CR diffusion or streaming
enhance the radio-brightness and somewhat flatten the radio continuum spectrum within the radio-bright regions.

\subsection{Radio-dim galactic disk model \label{sec:radiodimgd}}

\begin{figure}
  \centering
  \resizebox{\hsize}{!}{\includegraphics{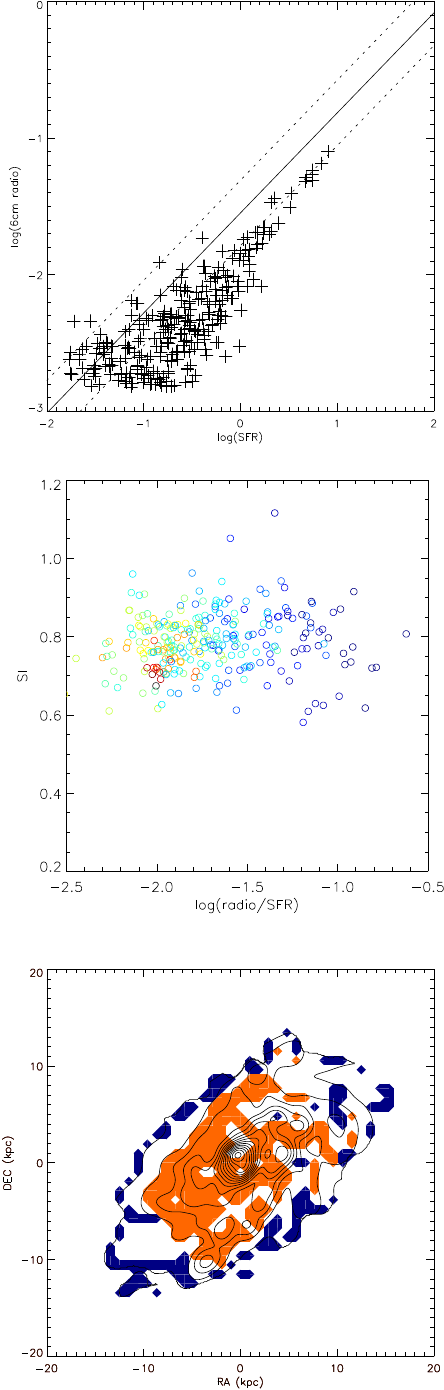}\includegraphics{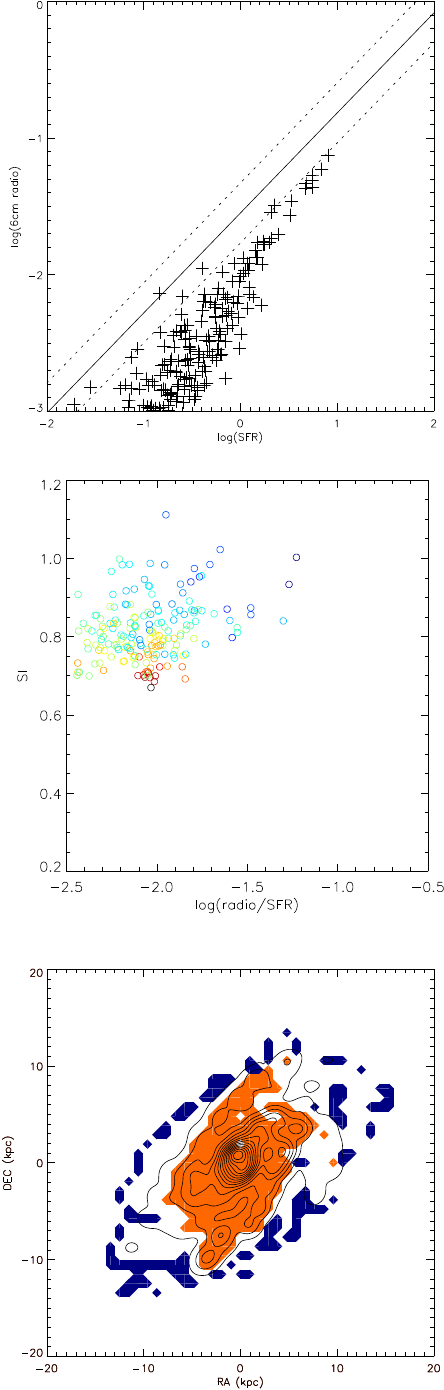}}
  \caption{Model as in Fig.~\ref{fig:model1} with $B=0.6 \times \sqrt{4\,\pi\,\rho\,v_{\rm turb}^2}$. 
    This model reproduces the globally low radio/SFR ratios in NGC~4298, NGC~4535, and NGC~4567.
    Left panels: without diffusive escape. Right panels: with diffusive escape.
  \label{fig:compositemove_n4501uv_TP_indQ_paper_n4535}}
\end{figure}

There are three overall radio-dim galaxies in our sample: NGC~4298, NGC~4535, and NGC~4567.
Radio-dim model galactic disks can be obtained by a decrease of the magnetic field strength with respect to its equipartition value
(Eq.~\ref{eq:Bmag2}). As an example, we set $B=0.6 \times \sqrt{4\,\pi\,\rho\,v_{\rm turb}^2}$ in the symmetric model (left panels of 
Fig.~\ref{fig:compositemove_n4501uv_TP_indQ_paper_n4535}). Indeed, the bulk of the disk becomes radio-dim.
In a second step we added a vertical escape of CR electron by diffusion (Eq.~\ref{eq:tdiffe}).
As expected, the disk becomes even more radio-dim (right panels of Fig.~\ref{fig:compositemove_n4501uv_TP_indQ_paper_n4535}).
Remarkably, the log(radio/SFR)--SI distribution becomes vertical as it is observed in NGC~4298 and NGC~4535.
In general, vertical CR escape makes the log(radio/SFR)--SI distribution more vertical. Vertical CR escape might thus
also be important in many other galaxies and most notably in NGC~4321 (Fig.~\ref{fig:rcfir_spixx1c1_nice_2}).

Including diffusive escape together with equipartition between the energy densities of the turbulent gas and the magnetic field
leads to a galactic disk, which is only radio-dim in the outer parts ($R \ga 5$~kpc).
Therefore, within our model framework diffusive escape does not seem to be sufficient for a globally radio-dim galactic disk as observed 
in NGC~4298, NGC~4535, and NGC~4567.
Diffusive escape is not important in the perturbed spiral galaxies because of their high magnetic field strength.

Within the framework of our model a globally radio-dim galactic disk is obtained by a decrease ($\sim 60$\,\%) of the
magnetic field with respect to its equipartition value. Diffusive escape can enhance the radio-dimness in the outer disk and leads 
to a vertical log(radio/SFR)--SI distribution. In addition, in galaxies with high SFRs CR advection via a galactic wind is expected to play a role.


\section{Discussion\label{sec:discussion}}

\subsection{Radio-deficit regions}

In undisturbed galaxies the radio emission extends beyond the associated FIR emission/SF region, due to the diffusion and streaming of the CR electrons.
However in many Virgo galaxies the radio emission does not extend
beyond the associated FIR emission on the leading sides of ram pressure interactions, apparently because the diffusion and streaming of the CR electrons has been 
inhibited in this direction. This inhibition is presumably due to a sharp edge of the ISM distribution and an associated magnetic field configuration, 
in which the magnetic field lines are largely aligned along 
the leading edge of the gas disk as witnessed by the asymmetric ridges of polarized radio continuum emission (see Sect.~\ref{sec:piradiosfr}).

The presence of these regions lacking the expected radio emission caused by CR diffusion and streaming can be shown by subtracting the radio map from a 
model made by smoothing the FIR/SFR distribution by an amount that matches the typical radio behavior in undisturbed galaxies.
For many Virgo galaxies, these difference maps reveal a deficit with respect to the expected radio emission along the side of the outer edge of the gas disk known from 
other indicators to be the leading side of the ram pressure interaction (Murphy et al. 2009).

We have seven galaxies in common with Murphy et al. (2009): NGC~4254, NGC~4321, NGC~4330, NGC~4396, NGC~4402, NGC~4522, and NGC~4579. 
These authors found significant radio-deficit regions in NGC~4254, NGC~4330, NGC~4402, and NGC~4522.
Since we do not smooth the SF maps in this work, we cannot evaluate these deficit regions.
Interestingly, we observe low radio/SFR values, which we classified as radio-normal though, 
in the radio-deficit regions of NGC~4330, NGC~4402, and NGC~4522 (Figs.~\ref{fig:rcfir_spixx1c1_nice_1} to \ref{fig:rcfir_spixx1c1_nice_3}
and Figs.~\ref{fig:rcfir_spixx1c1_nice_20_1} to \ref{fig:rcfir_spixx1c1_nice_20_3}).

\subsection{FP--PI dichotomy \label{sec:dichotomy}}

Within sample of interacting galaxies the radio/SFR ratio is correlated either with the fractional polarization (FP) or polarized radio continuum 
emission (PI) (Table~\ref{tab:coeffs1}). We call this the FP--PI dichotomy.
This is contrary to expectation that radio/SFR should be correlated to PI if it is correlated with FP.
In most galaxies of our sample the radio/SFR ratio correlates with PI.
On the other hand, the radio/SFR ratio is only correlated with the SFR ($|\rho| > 0.5$) in
NGC~6946, NGC~4254, and NGC~4654 and these are the galaxies where the radio/SFR ratio is best correlated with FP.
The SFR--radio/SFR correlation is driven by (i) synchrotron, bremsstrahlung, and IC losses in regions of high SFR and (ii) diffusion or streaming from
regions of high SFR to regions of low SFR (see Sect.~\ref{sec:symmetricm}). Mechanisms (i) and (ii) decrease the radio continuum emission
in regions of high SFR. In addition, mechanism (ii) can increase the radio continuum emission in regions of low SFR.
Whether the radio continuum emission of low-SFR regions is enhanced depends on the timescale of CR energy losses.
The FP--radio/SFR correlation arises because the polarized emission stems from the magnetic field, which is ordered 
by shear and compression regions related to galactic structure, i.e. spiral arms. 

Once the galaxies are significantly perturbed the picture changes: 
the SFR--radio/SFR correlation is significantly weakened in interacting galaxies where externally triggered shear and compression motions
locally enhance the magnetic field and thus the radio continuum emission without significantly affecting the SFR.
Within these regions the radio continuum and the polarized emission are enhanced leading to the PI--radio/SFR correlation.
As expected, the observed correlation is driven by regions of strong polarized emission.

\subsection{Overall radio-dim spiral galaxies}

\begin{figure}
  \centering
  \resizebox{\hsize}{!}{\includegraphics{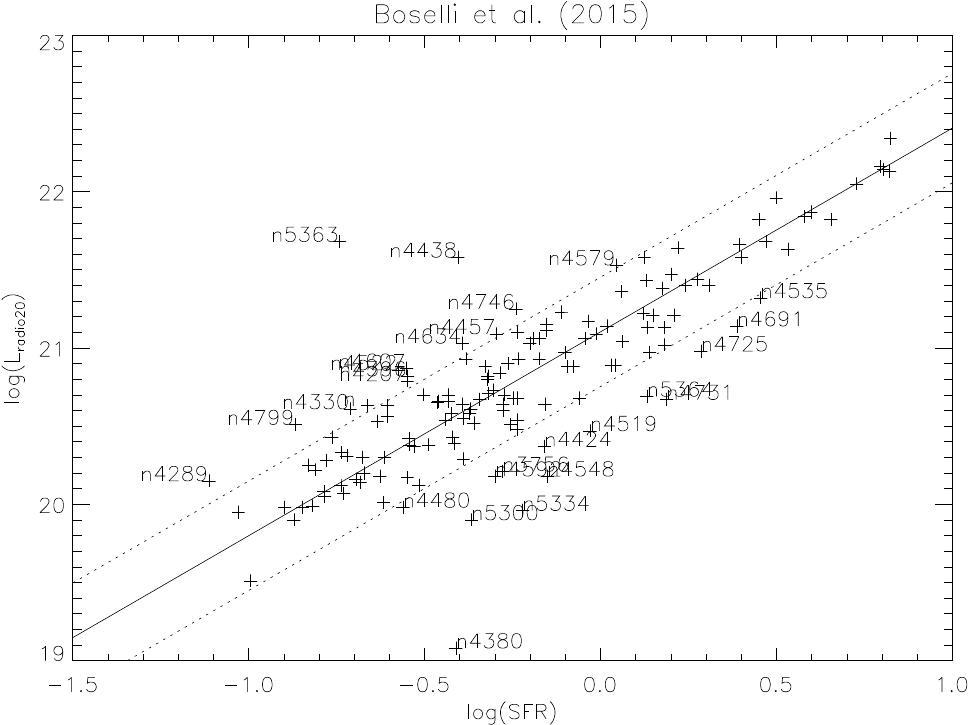}}
  \caption{Radio continuum luminosity at $1.4$~GHz as a function of star formation rate
    for the galaxy sample of Boselli et al. (2015). The solid line represent an outlier-resistant linear
    regression. The dotted line correspond to to the $1\,\sigma$ scatter.
  \label{fig:boselli}}
\end{figure}

To further investigate the physical reason for a global radio-dimness of spiral galaxies, we
show the integrated SFR--radio continuum relation at $1.4$~GHz for the sample of Boselli et al. (2015),
which is mostly comprised of Virgo spiral galaxies, in Fig.~\ref{fig:boselli}.
The slope of the log(SFR)--log(radio) relation is $1.3$ with a scatter of $0.27$.
We identify galaxies with ${\mid}{\Delta}$log(radio/SFR)${\mid} > 0.35$ as radio-dim and radio-bright.

By comparing VLA FIRST to NVSS flux densities we found that nine out of $14$ radio-bright galaxies seem to be overall 
radio-bright. Eight out of these nine galaxies are Virgo galaxies.
Four out of these eight galaxies are interacting with the intracluster medium (NGC~4330, NGC~4396, NGC~4438, and NGC~4522).
NGC~4457 might have had a minor merger in the recent past (Vollmer et al. 2013).
As already stated, NGC~4579 is enigmatic: it is an anemic galaxy as NGC~4548, which is radio-dim, but it is radio-bright
probably due to additional radio continuum emission from an AGN-driven outflow (Sect.~\ref{sec:radiobright}).
We are thus left with three radio-bright galaxies: NGC~4207, NGC~4289,  and NGC~4746.
All these galaxies are observed edge-on and are of relatively low mass ($M_* < 5 \times 10^9$~M$_{\odot}$) and have a range 
of specific star formation rates (sSFR) from normal ($10^{-10}$yr$^{-1}$) to low ($3 \times 10^{-11}$~yr$^{-1}$).

Out of the $14$ radio-dim galaxies six are Virgo galaxies. Thus radio-dimness is not related to the cluster environment. 
There are two merger remnants (NGC~4691, NGC~4424) and one interacting galaxy (NGC~4731). This is quite surprising 
because one would expect that gravitational interactions enhance the magnetic fields (Drzazga et al. 2011), but they
also enhance the SFR. 

We conclude that ongoing hydrodynamic interactions (ram pressure) do whereas ongoing gravitational interaction can
enhance the radio emission of disk galaxies. Evolved gravitational interactions can decrease the radio/SFR ratio by
up to a factor of about three.

NGC~4548 is a radio-dim anemic gas-poor galaxy with a low sSFR. However, there is a similar galaxy with a normal 
radio/SFR ratio, NGC~2841 (Heesen et al. 2014). Moreover, the nearby spiral galaxy NGC~3198
is overall radio-dim (Heesen et al. 2014) but otherwise normal (gas content, SFR, sSFR). It thus seems that overall
radio-dimness cannot be predicted by any known physical property of a galaxy.

We speculate that the global radio/SFR differences stem from difference between the timescales of (i) the 
variation in SFR ($t_*$) and (ii) the small-scale dynamo ($t_{\rm dynamo}$). If $t_* > t_{\rm dynamo}$, then the galaxy is radio-normal.
If the SFR increases and  $t_* < t_{\rm dynamo}$, the galaxy is radio-deficient (see also Roussel et al. 2003).
This is consistent with the fact that two (NGC~4298 and NGC~4567) out of the 
three overall radio-dim galaxies from Table~\ref{tab:sample} are parts of interacting pairs. If we interpret the western gas arm
of NGC~4535 as sign of an interaction that occurred in the recent past, all three cases are consistent with a scenario where the
interaction enhanced the SFR of the galaxies within less time than the timescale of the small-scale dynamo, which is
about the turbulent turnover time $\Omega^{-1}=R/v_{\rm rot} \sim 100$~Myr.
Alternatively, different saturation levels of the galactic small-scale dynamo might also play a role.

\section{Summary and conclusions\label{sec:conclusions}}

In this work we investigated the relation between the resolved SFR per unit area and the non-thermal radio continuum emission.
We used the radio continuum observations of $20$ Virgo spiral galaxies presented in Vollmer et al. (2010, 2013).
For the Virgo spiral galaxy NGC~4254 and the two nearby spiral galaxies NGC~6946 and M~51 we used the radio continuum data from Chyzy et al. (2007),
Beck (2007) and Fletcher et al. (2011). In addition, we created SFR maps using archival Spitzer, Herschel, and GALEX data 
(Sect.~\ref{sec:observations}).

For the interpretation and understanding of our results we used a 3D model where star formation, CR propagation, and the physics of synchrotron 
emission are included (Sect.~\ref{sec:model}).
For the calculation of the galaxy dynamics at scales of about $\sim 1$~kpc we used the 3D dynamical model
introduced by Vollmer et al. (2001) and applied to NGC~4501 by Vollmer et al. (2008) and Nehlig et al. (2016).
For the calculation of the radio continuum emission we use the analytical formalism introduced in Vollmer et al. (2022).

The slope of the log($I_{100\mu{\rm m}}$)--log($I_{4.85{\rm GHz}}$) relation is $1.02$ with a scatter of $0.17$~dex, that of the 
log($\dot{\Sigma}_*$)--log($I_{4.85{\rm GHz}}$) relation is $1.05$ with a scatter of $0.20$~dex.
We used the results of the linear regression of Fig.~\ref{fig:sfrrad_all} to define radio-bright ($\log(I_{4.85{\rm GHz}}) > \log(I_{\rm exp})+0.25$) 
and radio-dim regions ($\log(I_{4.85{\rm GHz}}) < \log(I_{\rm exp})-0.25$), where $I_{\rm exp}$ is the expected surface brightness.

The radio/SFR ratio depends on CR energy losses and CR transport. Energy losses decrease the radio continuum emission, whereas CR transport
decreases/increases the radio continuum emission in regions of high/low SFR.
Based on our model of synchrotron-emitting disks we identified CR diffusion or streaming as the physical causes of radio-bright regions
of unperturbed symmetric spiral galaxies as NGC~6946 (Sect.~\ref{sec:symmetricm}). 
The enhanced magnetic field in the region of ISM compression via ram pressure is responsible for the southwestern radio-bright region in NGC~4501
(Sect.~\ref{sec:perturbedm}). CR diffusion or streaming
enhance the radio-brightness and somewhat flatten the radio continuum spectrum within the radio-bright region.
Based on our knowledge of the Virgo spiral galaxies, we identified the physical causes of radio-bright regions:
CR advection transport via gravitational tides in M~51, CR transport via a galactic wind in NGC~4532,
CR transport via ram pressure stripping in NGC~4330, and NGC~4522.
Three galaxies are overall radio-dim: NGC~4298 (Fig.~\ref{fig:rcfir_spixx1c1_nice_1}), NGC~4535, and NGC~4567 (Fig.~\ref{fig:rcfir_spixx1c1_nice_4}).
It is remarkable that two out of three overall radio-dim galaxies are gravitationally interacting (NGC~4298 and NGC~4567).

Based on our model of synchrotron-emitting disks (Sect.~\ref{sec:radiodimgd}) we suggest that the overall radio-dim galaxies have a significantly lower
magnetic field than expected by equipartition between the magnetic and turbulent energy densities (Eq.~\ref{eq:Bmag2}). 
We suggest that this is linked to differences between the timescales of the variation in SFR and the small-scale dynamo.
NGC~4535 has an asymmetric distribution of polarized radio continuum emission (Vollmer et al. 2007), which coincides with strong shear motions
detected in the H{\sc i} velocity field (Chung et al. 2009). These shear motions are thus responsible for the region of enhanced polarized
radio continuum emission. Within this region the radio/SFR is normal, i.e. it is enhanced with respect to the otherwise radio-dim 
galactic disk. We argue that the shear motions increase the total magnetic field strength via the induction equation, which leads to an enhanced 
radio continuum emission with respect to the SFR.

Radio-bright and radio-dim regions can be robustly defined for our sample of spiral galaxies. Radio-bright regions are caused by
CR transport out of their acceleration sites and/or the increase of magnetic field strength via ISM compression or shear motions. 
In the two latter cases they are linked to the commonly observed asymmetric ridges of
polarized radio continuum emission and represent a useful tool for the interaction diagnostics.  

To understand the reasons for the radio-bright and radio-dim regions, we calculated the Spearman rank correlation coefficients
and slopes of the resolved SFR--radio, SFR--FP, radio--PI, SFR--PI, PI--radio/SFR, and FP--radio/SFR relations (Sect.~\ref{sec:correlations}).
The three basic correlations are SFR--radio, SFR--PI, and radio--PI with slopes of $1.11$, $0.43$, and $0.41$, respectively.
The SFR--radio correlation is much steeper and tighter than the SFR--PI correlation resulting in an SFR--FP
anticorrelation. The slope of the SFR--FP correlation is significantly steeper for the interacting galaxies
than for NGC~6946. We suggest that the SFR--FP correlation in all galaxies is driven by the action of a large-scale
dynamo together with the tangling of regular magnetic fields in regions of relatively high SFR and field ordering by
compression and shear motions.
The steeper slopes of the SFR--FP correlation in interacting galaxies are caused by compression and shear of isotropic random fields, 
which lead to field ordering, mainly occur in regions with low SFR where the gas energy density is also low.

We realized that the radio-bright regions frequently coincide with the asymmetric ridges of polarized radio continuum emission. 
In the eight galaxies presented in Fig.~\ref{fig:rcfir_spixx1c1_nice_pol} radio-bright regions coincide with regions of high polarized 
radio continuum emission. Indeed, we found a clear albeit moderate correlation between PI and the radio/SFR ratio (Fig.~\ref{fig:radsfr_piall}).
This is consistent with an enhancement and ordering of the magnetic field in regions of ISM compression and shear motions.
In our sample galaxies the radio/SFR ratio is either correlated with PI (in most of the cases) or FP.
In galaxies showing a FP--radio/SFR correlation the radio/SFR ratio is also correlated with the SFR.
This correlation is driven by (i) bremsstrahlung and IC losses in regions of high SFR and (ii) diffusion or streaming from
regions of high SFR to regions of low SFR (see Sect.~\ref{sec:symmetricm}).

CR energy losses and transport also affect the spectral index, which we measure between $4.85$ and $1.4$~GHz.
In 8 galaxies of our sample we observe a gradient in radio/SFR--SI space.  Diffusion or streaming flatten this gradient 
and thus lead to a flatter radio continuum spectrum in the radio-bright regions (Sect.~\ref{sec:symmetricm}).
In all other galaxies of our sample the radio/SFR--SI distribution is vertical.
Based on our model of synchrotron-emitting disks (Sect.~\ref{sec:radiodimgd}), we argue that this vertical distribution is
due to diffusive escape of CR electron in rather weak magnetic fields. The diffusive escape makes the outer parts of the galactic disks
radio-dimmer (Fig.~\ref{fig:compositemove_n4501uv_TP_indQ_paper_n4535}).

We suggest the following scenario for the interplay between star formation, CR electrons and magnetic fields,
which is consistent with our results:
CR electrons are created in supernova shocks, which are located in starforming region within the galactic disk.
These CR electrons undergo mainly bremsstrahlung and IC losses in regions of high SFR and diffuse or stream from
regions of high SFR to regions of low SFR. The isotropic random magnetic field is created by a small-scale dynamo.
In addition, the large-scale dynamo creates a regular field
component in regions of low SFR, i.e. the interarm regions. This regular field might dominate the total field
in the so-called magnetic arms (e.g., Beck et al. 2019). The regular field is tangled and thus destroyed in regions of high star 
formation rates where the turbulent energy density is high.
The magnetic field is enhanced (as observed in NGC~4535 and NGC~4501) and ordered by ISM compression and shear motions in spiral arms or 
perturbed regions of galactic disks. In the perturbed galaxies the radio/SFR ratio increases slowly with the polarized intensity
(radio/SFR$\propto$PI$^{0.3}$). The enhancement of the magnetic field in regions of ISM compression and shear motions
is rather modest and does not significantly influence the radio/SFR correlation. The main effect of compression and shear motions is the ordering of the
magnetic field, which can be observed via polarized radio continuum emission.

\begin{acknowledgements}
We would like to thank the anonymous referee for careful reading of the manuscript and Amit Seta for useful discussions.
\end{acknowledgements}

\clearpage

\begin{appendix}

\section{Total infrared and star formation estimators \label{app:sfr}}

The full width at half-maximum (FWHM) of the point spread functions (PSFs), as stated in the Spitzer Observer's Manual 
(Spitzer Science Centre 2006), are 1.7, 2.0, 6, and 18~arcsec at 3.6, 8.0, 24, and 70~$\mu$m, respectively.
In addition, we used Herschel Virgo Cluster Survey (HeViCS) 100 and 160~$\mu$m images (Davies et al. 2010) which have spatial resolutions 
of 7 and 12~arcsec, respectively.
First, the data are convolved with Gaussian kernels that match the PSFs of the images in the 3.6, 8, 24 70, 100, and 160~$\mu$m bands to the PSF of 
the radio continuum data. 
Next, the data were re-binned to the common pixel size of the radio continuum maps. 
We excluded from the analysis regions not detected at the $3 \sigma$ level in one or more wave bands.
This resulted in a loss of surface of $10$\,\% to $20$\,\% compared to the regions with $24~\mu$m $3\sigma$ detections
with a tendency of less loss for truncated galaxies.
Following Helou et al. (2004), we subtracted the stellar continuum from the 8 and 24~$\mu$m surface brightnesses (in MJy\,sr$^{-1}$) using 
\begin{equation}
I_{\nu}({\rm PAH}\ 8\mu {\rm m})=I_{\nu}(8\mu {\rm m})-0.232\ I_{\nu}(3.6\mu {\rm m})\ {\rm and}
\end{equation}
\begin{equation}
I_{\nu}(24\mu {\rm m})=I_{\nu}(24\mu {\rm m})-0.032\ I_{\nu}(3.6\mu {\rm m})\ .
\end{equation}

According to the availability of Spitzer and Herschel data, we calculated the TIR surface brightness in three ways
based on Table~3 from Galametz et al. (2013). 
\begin{enumerate}
\item
For NGC~4254, NGC~4294, NGC~4298, NGC~4299, NGC~4302, NGC~4303, NGC~4321, NGC~4330, NGC~4402, NGC~4522, and NGC~4579 we used
\begin{eqnarray}
I({\rm TIR})=2.064\, \nu I_{\nu}(24\mu {\rm m}) + 0.539\, \nu I_{\nu}(70\mu {\rm m}) + \nonumber \\
0.277\, \nu I_{\nu}(100\mu {\rm m}) + 0.938\, \nu I_{\nu}(160\mu {\rm m})\ ;
\label{eq:tir1}
\end{eqnarray}
\item
for NGC~6946, M~51, NGC~4396, NGC~4419, NGC~4457, NGC~4532, NGC~4654, and NGC~4808 we used
\begin{eqnarray}
I({\rm TIR})=0.95\,\nu I_{\nu}(8\mu {\rm m}) + 1.15\, \nu I_{\nu}(24\mu {\rm m}) + \nonumber \\
2.3\, \nu I_{\nu}(70\mu {\rm m})\ ;
\label{eq:tir2}
\end{eqnarray}
for NGC~4501, NGC~4535, and NGC~4567/68 we used
\begin{eqnarray}
I({\rm TIR})=2.708\, \nu I_{\nu}(24\mu {\rm m}) + 0.734\, \nu I_{\nu}(100\mu {\rm m}) + \nonumber \\
0.739\, \nu I_{\nu}(160\mu {\rm m})\ .
\label{eq:tir3}
\end{eqnarray}
\end{enumerate}
In most galactic environments the error introduced to $I({\rm TIR})$ due to incomplete spectral coverage is about 50\,\%, based on the Draine \& Li (2007) models.

The FIR+FUV relation only works for relatively high SFRs (like those in these spiral galaxies, galaxies near ‘star-forming main sequence’). 
For some ellipticals with dust heated by old stars or AGN, this relation gives a huge overestimate of the SFR.
Moreover, the SF timescales probed by FIR and GALEX UV might not match -- 
the GALEX UV probes star formation timescales of $\sim 50$-$100$~Myr, whereas the FIR probes timescales that are probably shorter, although this depends 
on the star formation history. 
Leroy et al. (2008) estimated the typical uncertainty on their 24~$\mu$m--FUV star formation rate to be of the
order of $50$\,\% or $0.2$~dex. The uncertainty of our star formation rates is of the same order. For a comparison between different star formation
indicators based on mixed processes (direct stellar light (FUV), dust-processed stellar light (FIR, TIR), ionised gas emission(H$\alpha$)) 
we refer to Calzetti (2010). The total star formation rates obtained with Eq.~\ref{eq:sfr} are about $0.2$~dex smaller than
those given by Boselli et al. (2015) with a scatter of $0.07$~dex. 

\section{The $100~\mu$m--4.85~GHz correlation \label{app:sfrrad_all_fir}}

\begin{figure}
  \centering
  \resizebox{\hsize}{!}{\includegraphics{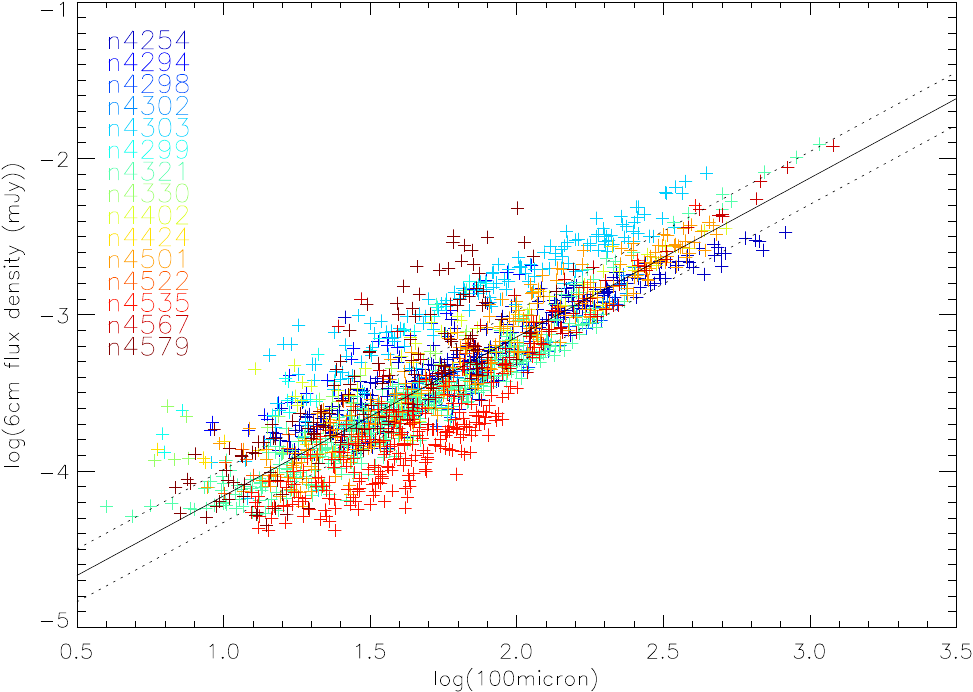}}
  \caption{Resolved properties measured in $15-22''$ apertures. 
    Total radio continuum surface brightness at $4.85$~GHz as a function of the surface brightness at $100$~$\mu$m.
    The black line represents the result of a outlier-resistant linear regression with a slope of $1.02$, the dotted lines the scatter of $0.17$~dex.
    NGC~4303, NGC~4535, and NGC~4579 were excluded from the linear regression.
  \label{fig:sfrrad_all_fir}}
\end{figure}

\section{Diagnostic plots for the Virgo cluster galaxy sample \label{sec:virgoplots}}

\begin{figure*}
  \centering
  \resizebox{16cm}{!}{\includegraphics{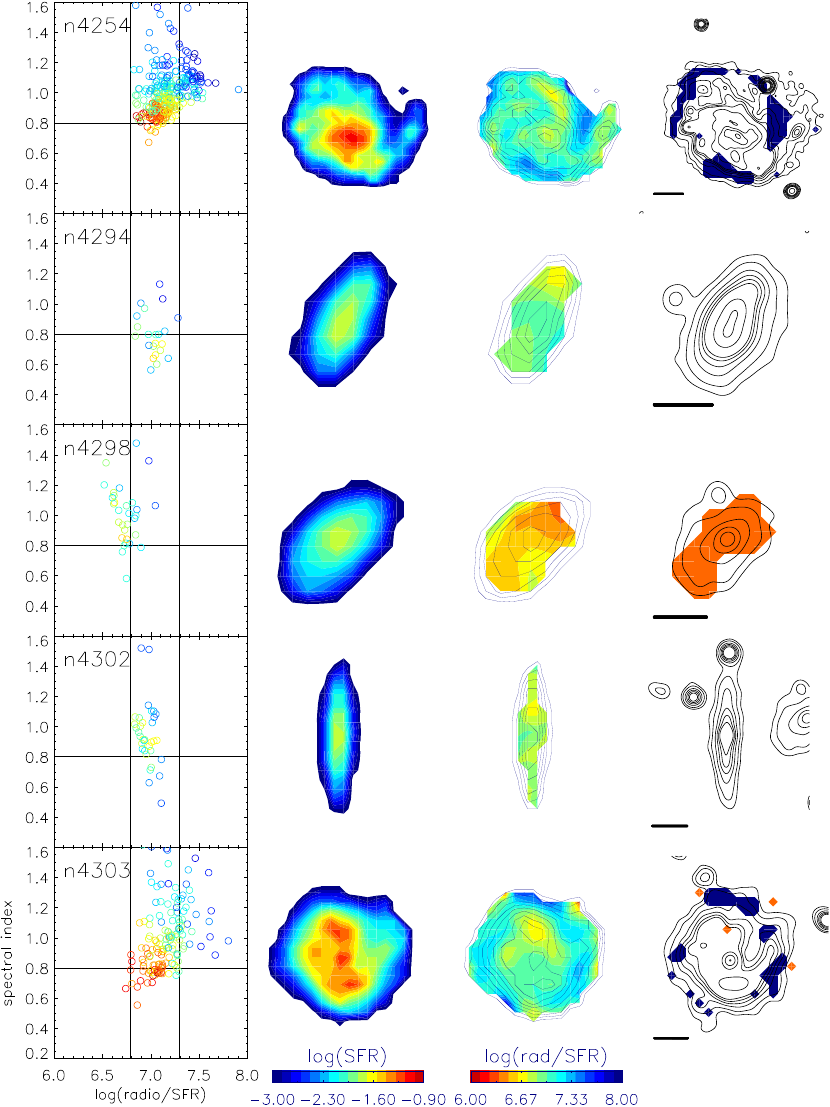}}
  \caption{As Fig.~\ref{fig:rcfir_spixx1c1_nice_5}. The radio continuum contour levels are $(1,2,4,6,8,10,20,30,40,50) \times \xi$ with $\xi=70~\mu$Jy/beam for NGC~4254
    and NGC~4294, $\xi=150~\mu$Jy/beam for NGC~4298 and NGC~4302, and $\xi=250~\mu$Jy/beam for NGC~4303.
  \label{fig:rcfir_spixx1c1_nice_1}}
\end{figure*}

\begin{figure*}
  \centering
  \resizebox{16cm}{!}{\includegraphics{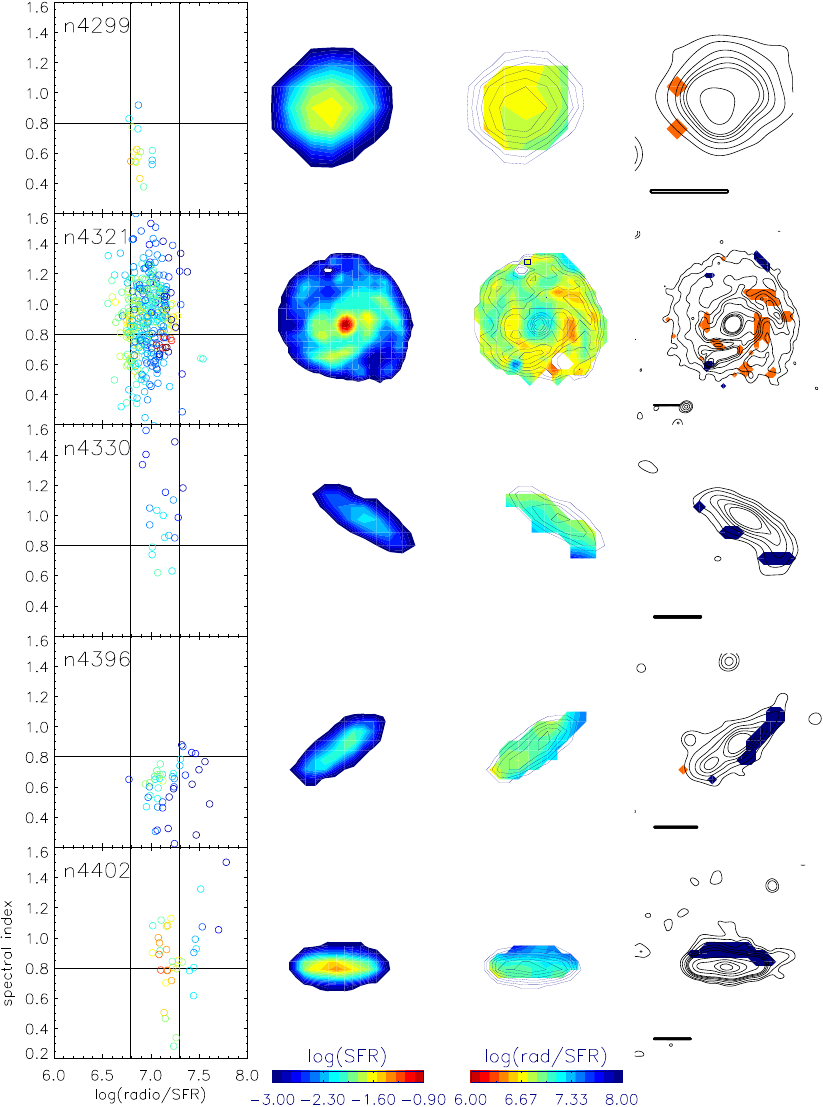}}
  \caption{As Fig.~\ref{fig:rcfir_spixx1c1_nice_5}. The radio continuum contour levels are $(1,2,4,6,8,10,20,30,40,50) \times 70~\mu$Jy/beam.
  \label{fig:rcfir_spixx1c1_nice_2}}
\end{figure*}

\begin{figure*}
  \centering
  \resizebox{16cm}{!}{\includegraphics{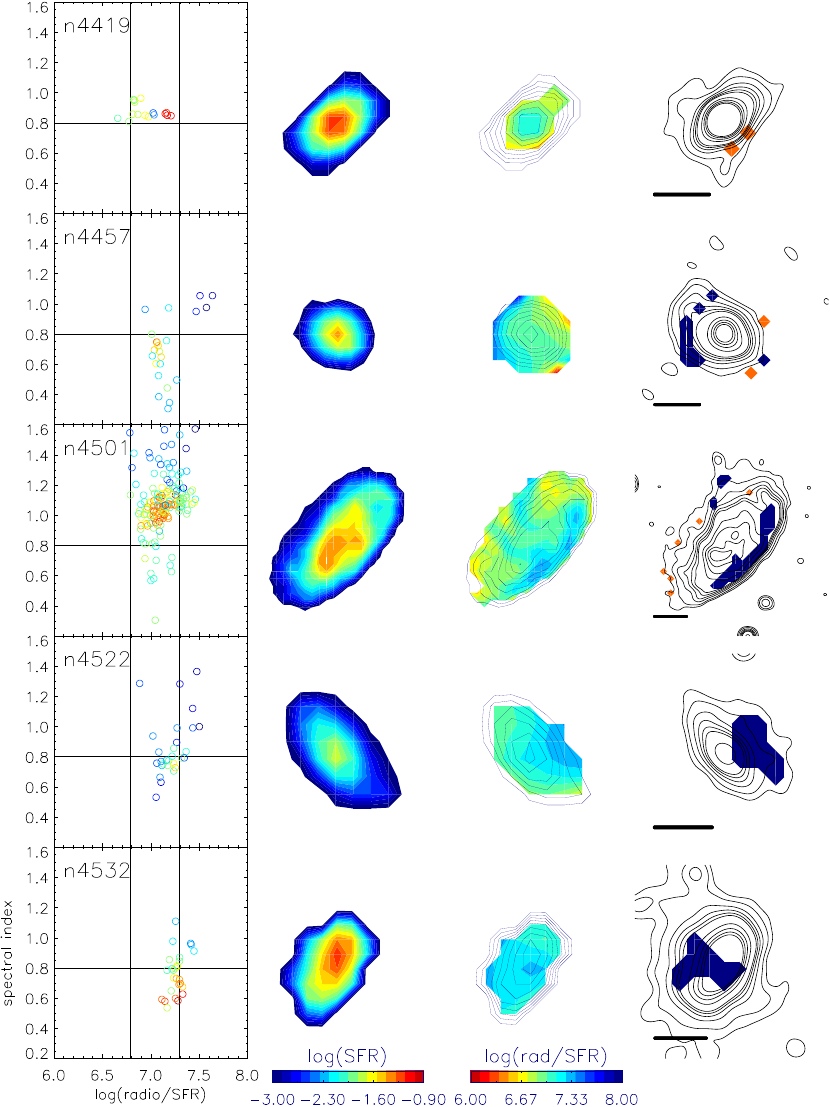}}
  \caption{As Fig.~\ref{fig:rcfir_spixx1c1_nice_5}. The radio continuum contour levels are $(1,2,4,6,8,10,20,30,40,50) \times 70~\mu$Jy/beam.
  \label{fig:rcfir_spixx1c1_nice_3}}
\end{figure*}

\begin{figure*}
  \centering
  \resizebox{16cm}{!}{\includegraphics{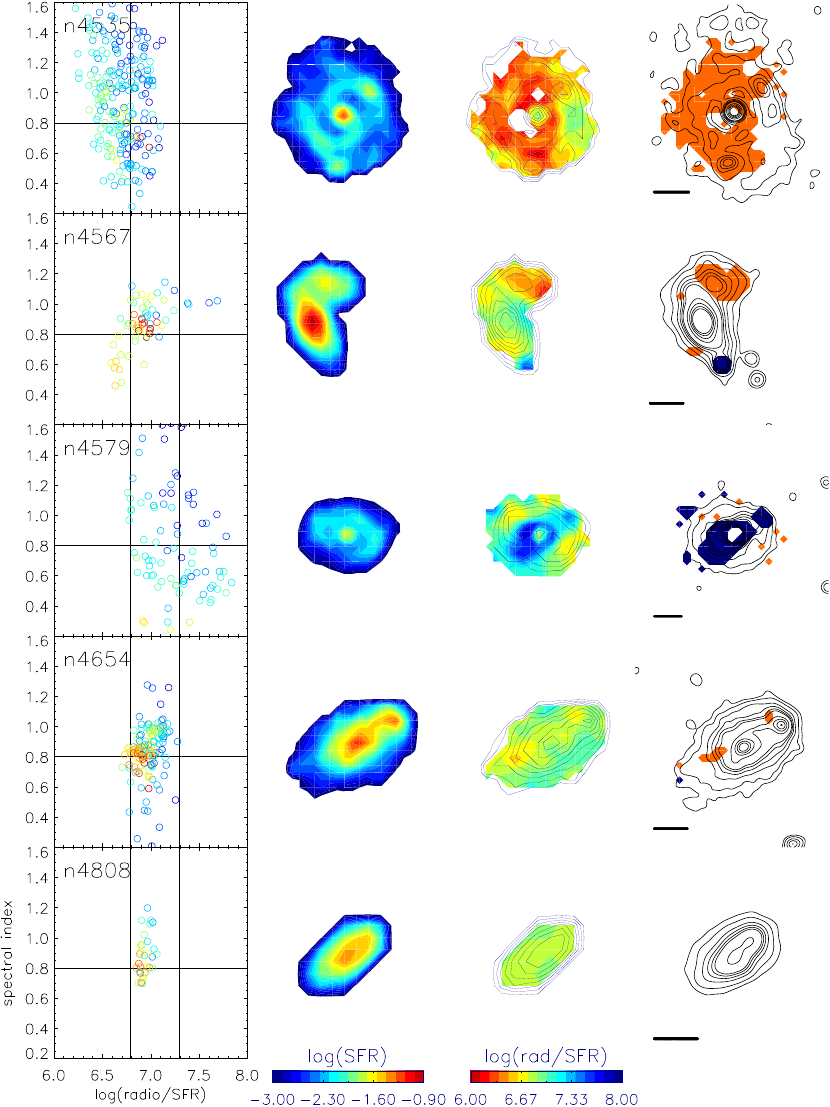}}
  \caption{As Fig.~\ref{fig:rcfir_spixx1c1_nice_5}. The radio continuum contour levels are $(1,2,4,6,8,10,20,30,40,50) \times \xi$ with $\xi=70~\mu$Jy/beam for NGC~4535 and
    NGC~4654 and $\xi=100~\mu$Jy/beam for NGC~4567/68, $\xi=150~\mu$Jy/beam for NGC~4579 and NGC~4808.
  \label{fig:rcfir_spixx1c1_nice_4}}
\end{figure*}

\section{1.4~GHz data \label{sec:1.4GHzdata}}

\begin{figure}
  \centering
  \resizebox{\hsize}{!}{\includegraphics{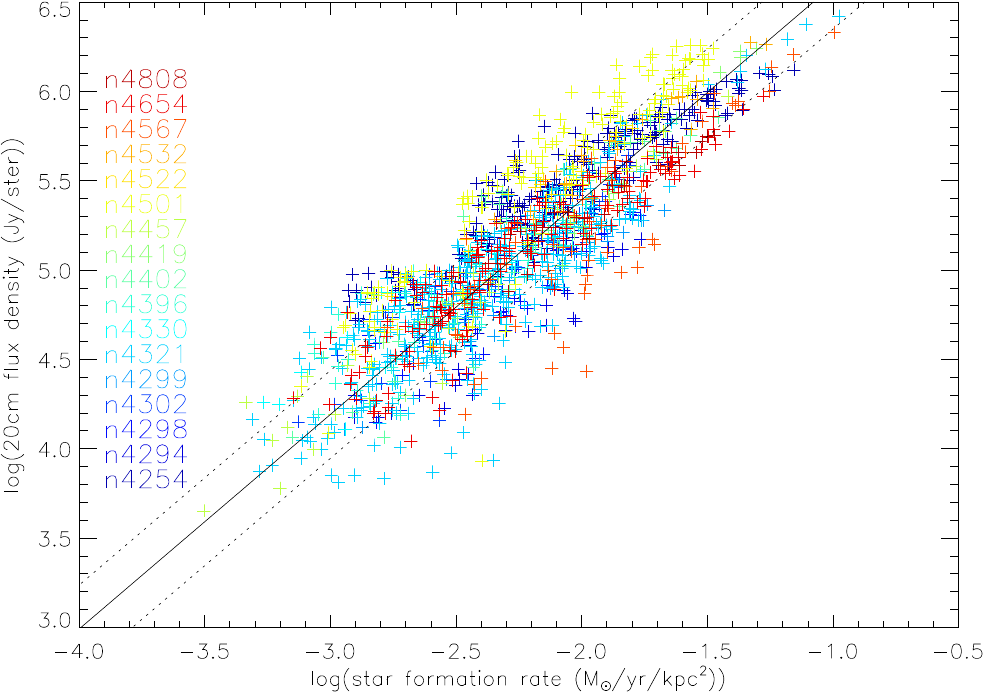}}
  \caption{As Fig.~\ref{fig:sfrrad_all} for the $1.4$~GHz data. 
  \label{fig:radsfr_pi20}}
\end{figure}

\begin{figure*}
  \centering
  \resizebox{16cm}{!}{\includegraphics{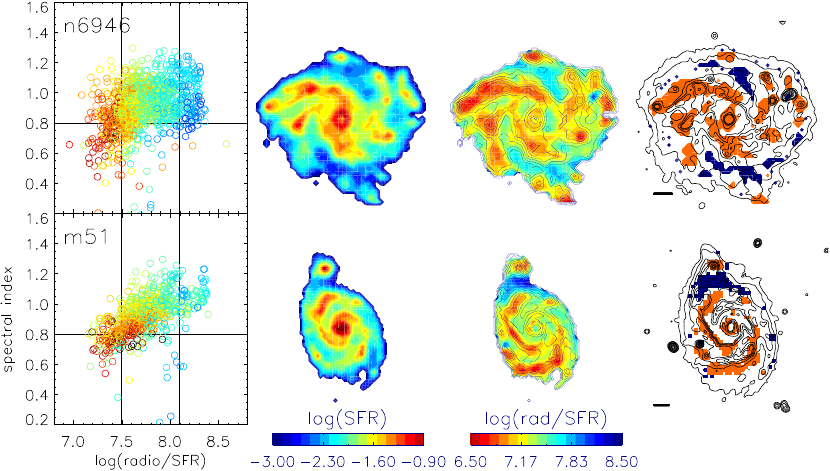}}
  \caption{As Fig.~\ref{fig:rcfir_spixx1c1_nice_5} for the $1.4$~GHz data. The radio continuum contour levels are $(1,2,4,6,8,10,20,30,40,50) \times 1200~\mu$Jy/beam.
  \label{fig:rcfir_spixx1c1_nice_20_5}}
\end{figure*}

\begin{figure*}
  \centering
  \resizebox{16cm}{!}{\includegraphics{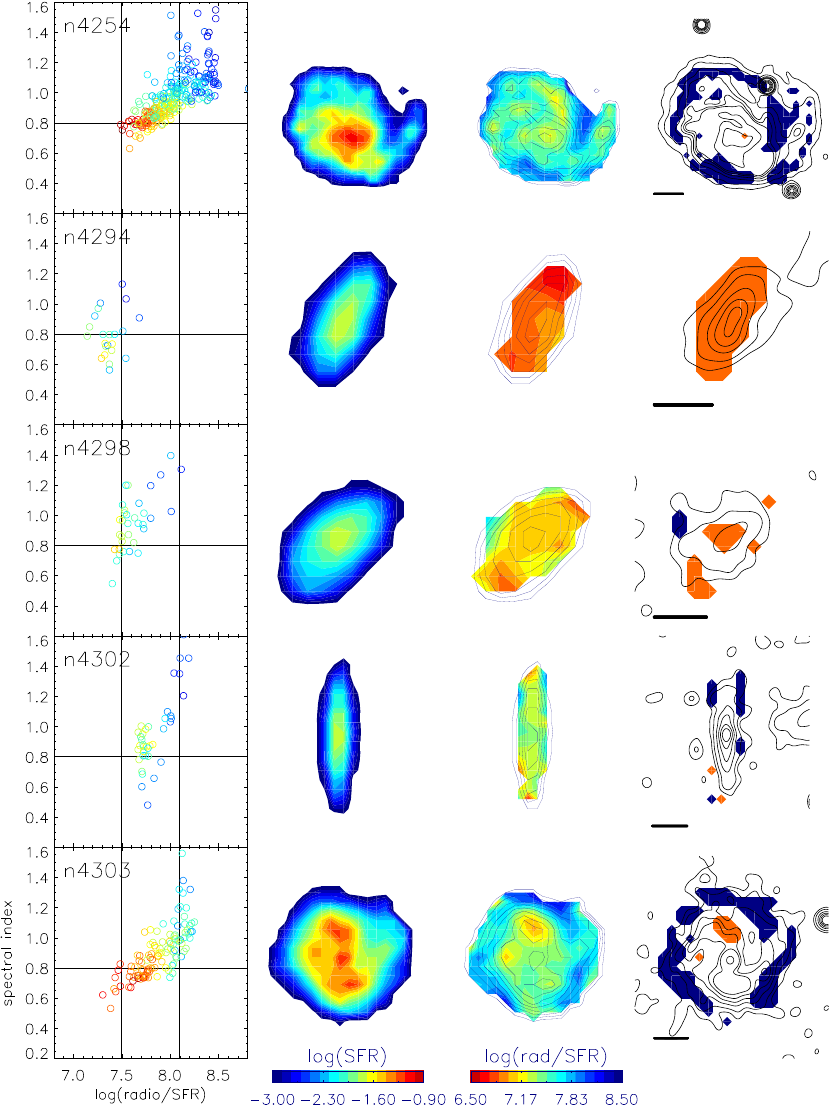}}
  \caption{As Fig.~\ref{fig:rcfir_spixx1c1_nice_5} for the $1.4$~GHz data. The radio continuum contour levels are $(1,2,4,6,8,10,20,30,40,50) \times \xi$ with 
    $\xi=280~\mu$Jy/beam for NGC~4254 and NGC~4294, $\xi=600~\mu$Jy/beam for NGC~4298 and NGC~4302, and $\xi=1000~\mu$Jy/beam for NGC~4303.
  \label{fig:rcfir_spixx1c1_nice_20_1}}
\end{figure*}

\begin{figure*}
  \centering
  \resizebox{16cm}{!}{\includegraphics{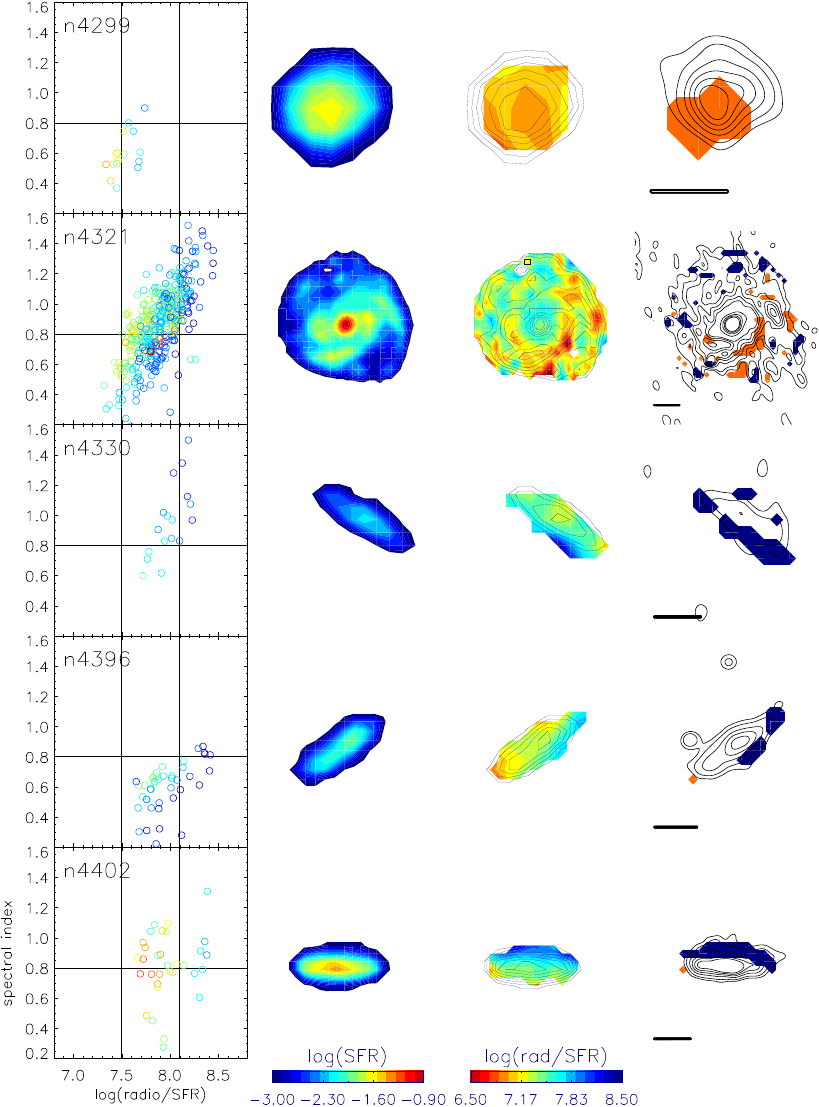}}
  \caption{As Fig.~\ref{fig:rcfir_spixx1c1_nice_5} for the $1.4$~GHz data. The radio continuum contour levels are $(1,2,4,6,8,10,20,30,40,50) \times 280~\mu$Jy/beam.
  \label{fig:rcfir_spixx1c1_nice_20_2}}
\end{figure*}

\begin{figure*}
  \centering
  \resizebox{16cm}{!}{\includegraphics{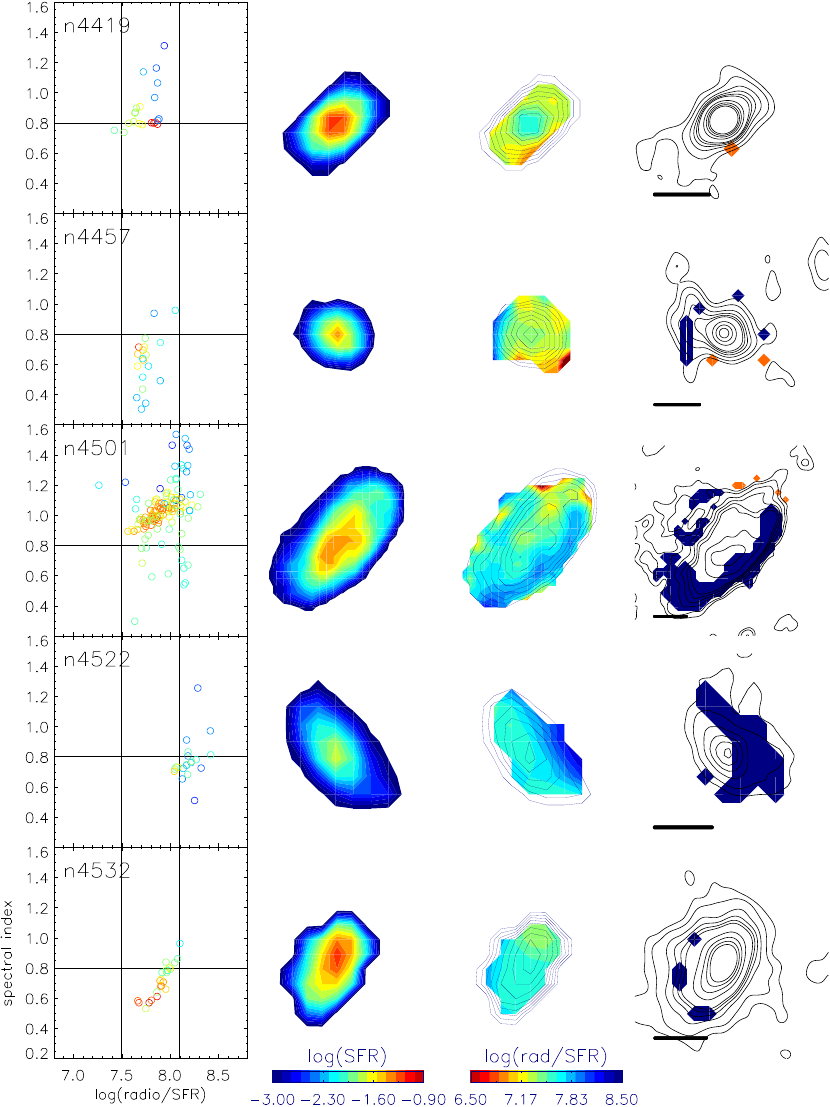}}
  \caption{As Fig.~\ref{fig:rcfir_spixx1c1_nice_5} for the $1.4$~GHz data. The radio continuum contour levels are $(1,2,4,6,8,10,20,30,40,50) \times 280~\mu$Jy/beam.
  \label{fig:rcfir_spixx1c1_nice_20_3}}
\end{figure*}

\begin{figure*}
  \centering
  \resizebox{16cm}{!}{\includegraphics{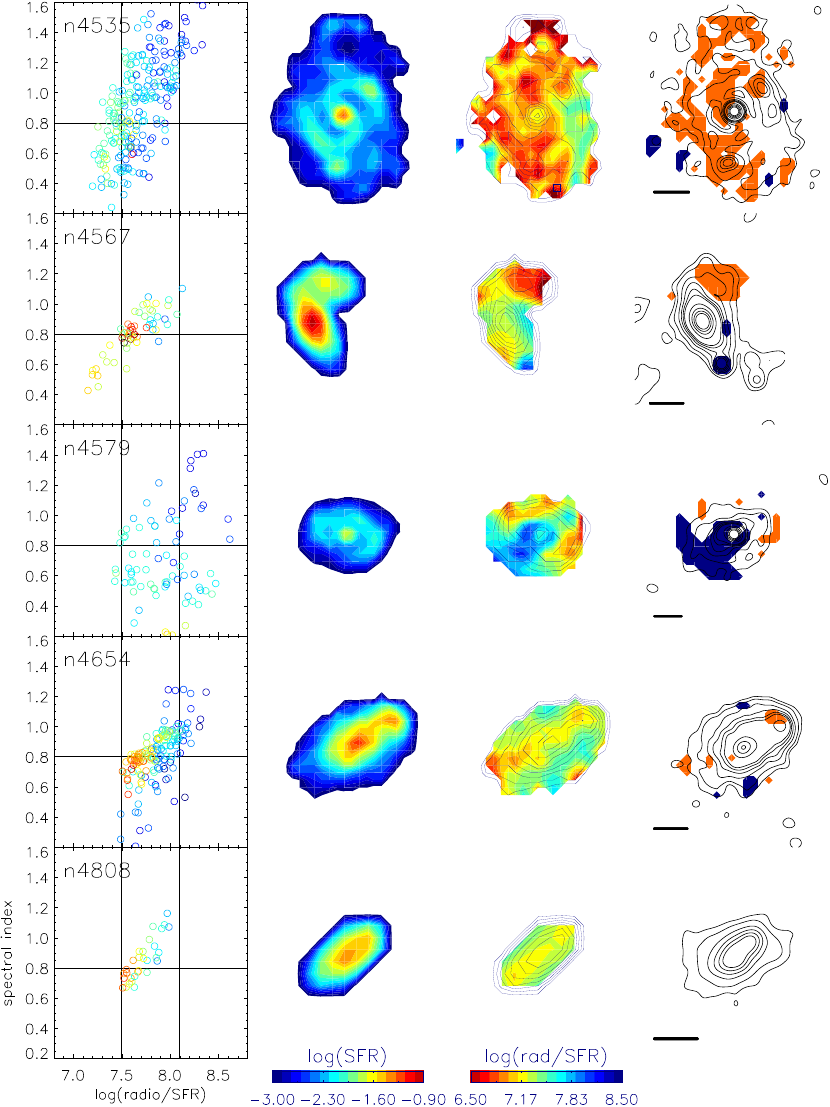}}
  \caption{As Fig.~\ref{fig:rcfir_spixx1c1_nice_5} for the $1.4$~GHz data. The radio continuum contour levels are $(1,2,4,6,8,10,20,30,40,50) \times \xi$ 
    with $\xi=280~\mu$Jy/beam for NGC~4535 and NGC~4654 and $\xi=400~\mu$Jy/beam for NGC~4567/68, $\xi=600~\mu$Jy/beam for NGC~4579 and NGC~4808.
  \label{fig:rcfir_spixx1c1_nice_20_4}}
\end{figure*}

\section{NGC~4579 \label{sec:n4579}}

The radio-bright region in NGC~4579 (Fig.~\ref{fig:rcfir_spixx1c1_nice_4}) is tilted by about $40^{\circ}$ with respect to the
galaxy's major axis, which approximately has an east--west orientation (Chung et al. 2009). 
The outer parts of the radio-bright regions correspond to the outer spiral arms, which are barely visible in the UV and H$\alpha$ line. 
NGC~4579 harbors a gas-rich nuclear spiral (Garcia-Burillo et al. 2005) with a high SFR, which is radio-normal, and an active galactic nucleus (AGN). 
Whereas the major axis of the H{\sc i} 
velocity field coincides with the photometric major axis of the galaxy, the major axis of the ionized gas velocity field within 
$0.5'$ is tilted by $\sim 50^{\circ}$ with respect to the H{\sc i} kinematic major axis (Daigle et al. 2006, Dumas et al. 2007). 
This tilt roughly corresponds to that of the radio-bright region with respect to the galaxy's major axis. On the other hand, the stellar kinematic
major axis within $0.5'$ corresponds to the H{\sc i} kinematic major axis (Dumas et al. 2007). 
Since the galaxy rotates counter-clockwise, the southern half of the disk is far side. If the inner ionized gas velocity field is produced by 
streaming motions within the disk, the gas should have a strong outward radial velocity component, whereas inflow is expected within the inner disk region 
(Garcia-Burillo et al. 2009). 

It is surprising that the magnetic field within $0.5'$ has the same orientation as the kinematic major axis of the ionized gas 
(Vollmer et al. 2013), which might be reminiscent of a vertical outflow. An outflow starting in the galactic disk plane would produce negative radial 
velocities south of the galaxy center. 
Since this is not observed for the ionized thermal gas, this possibility can be discarded. However, if the outflow is driven by the AGN, the
inclination angle of the inner AGN disk can be different from the galactic disk inclination. 
Given that the observed magnetic field lines would be oriented parallel to the putative AGN-driven outflow, we prefer this scenario. 
The relatively flat spectral indices found in the radio-bright region, which are close to the injection index of $-0.5$
(Fig.~\ref{fig:rcfir_spixx1c1_nice_4}), are consistent with this interpretation (see Hummel et al. 1989 for NGC~4258).

\section{Correlations \label{app:correlations}}

Following Leroy et al. (2008) an SFR uncertainty of $0.2$~dex was assumed. We obtained the relations between the SFR and (i) radio continuum, (ii) 
polarized emission (PI), and (iii) fractional polarization (FP) and (iv) the radio continuum emission and the polarized emission for the 10
largest spiral galaxies with low inclination angles: NGC~6946, M~51, NGC~4254, NGC~4303, NGC~4321, NGC~4501, NGC~4535, NGC~4567/68, NGC~4579, and NC~4654.
The disks of NGC~4254 and NGC~4501 were divided into northern and southern halves because the latter contain regions of compressed ISM.
A common Spearman correlation coefficient and slope were calculated for all sample galaxies except NGC~4579, where the
radio/SFR ratio with respect to PI behaves very much differently from those of the other galaxies (Table~\ref{tab:coeffs2}). 
For consistency, the resolution elements of NGC~6946 and M~51 were taken to be $1.5$~kpc as for the Virgo cluster galaxies. 
\begin{table*}
      \caption{Spearman rank correlation coefficients}
         \label{tab:coeffs1}
      \[
       \begin{tabular}{lccccccc}
        \hline
        x-axis -- & log(SFR)-- & log(SFR)-- & log(SFR)-- & log(radio)-- &  log(SFR)-- & log(PI)-- & log(FP)-- \\
        y-axis     & log(radio$^{\rm{(a)}}$)  &  log(PI$^{\rm{(b)}}$) & log(FP$^{\rm{(c)}}$) & log(PI) & log(radio/SFR) & log(radio/SFR) & log(radio/SFR) \\
        \hline
        NGC6946    & 0.74  & (0.16) & -0.64 & 0.41 & -0.68 & (0.21) & (0.30) \\
        M51        & 0.69  & (0.29) & -0.60 & 0.56 & (-0.35) & 0.41 & (-0.04) \\
        NGC4254s   & 0.95  & 0.41 & -0.86 & 0.56 & -0.52 & (0.36) & 0.65 \\
        NGC4254n   & 0.89  & 0.45 & -0.84 & 0.55 & -0.77 & (-0.08) & 0.61 \\
        NGC4303    & 0.94  & 0.32 & -0.85 & 0.48 & (-0.34) & 0.50 & (0.29) \\
        NGC4321    & 0.91  & 0.50 & -0.81 & 0.67 & (-0.17) & 0.41 & (0.11) \\
        NGC4501s   & 0.96  & 0.51 & -0.65 & 0.70 & (-0.04) & 0.62 & (0.28) \\
        NGC4501n   & 0.96  & 0.53 & -0.74 & 0.60 & (-0.04) & (0.32) & (0.14) \\
        NGC4535    & 0.72  & 0.47 & -0.84 & 0.59 & (0.20) & 0.41 & (-0.29) \\
        NGC4567/68    & 0.79  & 0.62 & -0.77 & 0.73 & (0.19) & 0.57 & (-0.37) \\
        NGC4579    & 0.84  & 0.42 & -0.79 & 0.70 & (-0.02) & 0.52 & -0.47 \\
        NGC4654    & 0.96  & (-0.03) & -0.93 & (-0.02) & -0.65 & (0.05) & 0.55 \\ 
        \hline
        total$^{\rm{(d)}}$      & 0.86  & (0.39) & -0.76 & 0.60 & (-0.29) &  0.41 & (0.14) \\
        \hline
        \end{tabular}
      \]
      \begin{list}{}{}
      \item[$^{\rm{(a)}}$] non-thermal (synchrotron) radio continuum emission
      \item[$^{\rm{(b)}}$] surface brightness of polarized radio continuum emission (PI)
      \item[$^{\rm{(c)}}$] fractional polarization (FP)
      \item[$^{\rm{(d)}}$] correlation for data from all galaxies except NGC~4579
      \end{list}
\end{table*}

\begin{table*}
      \caption{Slopes of the linear fits.}
         \label{tab:coeffs2}
      \[
       \begin{tabular}{lccccccc}
        \hline
        x-axis-- & log(SFR)-- & log(SFR)-- & log(SFR)-- & log(radio)-- &  log(SFR)-- & log(PI)-- & log(FP)-- \\
        y-axis     & log(radio)  &  log(PI) & log(FP) & log(PI) & log(radio/SFR) & log(radio/SFR) & log(radio/SFR) \\
        \hline
        NGC6946    & $0.86 \pm 0.03$ & ($ 0.19 \pm 0.03$) &  $ -0.67 \pm 0.04$ &  $ 0.42 \pm 0.03 $ & $-0.49 \pm 0.03$ & ($ 0.22 \pm 0.04 $) & ($ 0.24 \pm 0.03 $) \\
        M51        & $1.06 \pm 0.04$ & $ (0.35 \pm 0.04$) &  $ -0.76 \pm 0.04$ &  $ 0.42 \pm 0.02 $ & ($-0.39 \pm 0.04$) & $ 0.36 \pm 0.04 $ &  ($ -0.10  \pm 0.03 $) \\
        NGC4254s   & $0.93 \pm 0.05$ & $ 0.34 \pm 0.08$ &  $ -0.82 \pm 0.06$ &  $ 0.40 \pm 0.06 $ & $-0.18 \pm 0.07$ & ($ 0.23 \pm 0.10 $) &  $ 0.28 \pm 0.08 $ \\
        NGC4254n   & $0.85 \pm 0.03$ & $ 0.21 \pm 0.05$ &  $ -0.65 \pm 0.05$ &  $ 0.28 \pm 0.05 $ & $-0.34 \pm 0.06$ & ($ -0.14 \pm 0.16 $) &  $ 0.37 \pm 0.08 $ \\
        NGC4303   & $1.02 \pm 0.04$ & $ (0.24 \pm 0.05$) &  $ -0.85 \pm 0.05$ &  $ 0.32 \pm 0.04 $ & ($-0.15 \pm 0.05$) & $ 0.33 \pm 0.07 $ &  ($ 0.07 \pm 0.04 $) \\
        NGC4321   & $1.09 \pm 0.03$ & $ 0.83 \pm 0.07$ &  $ -0.83 \pm 0.04$ &  $ 0.46 \pm 0.03 $ & ($-0.01 \pm 0.05$) & $ 0.38 \pm 0.07 $ &  ($ -0.02 \pm 0.05 $) \\
        NGC4501s   & $1.25 \pm 0.07$ & $ 0.79 \pm 0.20$ &  $ -0.85 \pm 0.16$ &  $ 0.63 \pm 0.09 $ & ($0.06 \pm 0.11$) & $ 0.30 \pm 0.09 $ &  ($ 0.08 \pm 0.10 $) \\
        NGC4501n    & $1.12 \pm 0.04$ & $ 0.55 \pm 0.15$ &  $ -0.78 \pm 0.14$ &  $ 0.48 \pm 0.09 $ & ($-0.34 \pm 0.06$) & ($ 0.12 \pm 0.12 $) &  ($ 0.05 \pm 0.11 $) \\
        NGC4535    & $1.20 \pm 0.07$ & $ 0.05 \pm 0.15$ &  $ -1.12 \pm 0.18$ &  $ 0.15 \pm 0.08 $ & ($0.11 \pm 0.21$) & $ 0.50 \pm 0.56 $ &  $ (-0.14 \pm 0.14 $) \\
        NGC4567/68    & $1.15 \pm 0.10$ & $ 0.64 \pm 0.16$ &  $ -0.85 \pm 0.16$ &  $ 0.44 \pm 0.07 $ & ($0.08 \pm 0.17$) & $ 0.35 \pm 0.22 $ &  ($ -0.23 \pm 0.17 $) \\
        NGC4579    & $1.44 \pm 0.11$ & $ 0.43 \pm 0.09$ &  $ -1.15 \pm 0.13$ &  $ 0.34 \pm 0.03 $ & ($0.02 \pm 0.13$) & $ 1.08 \pm 0.15 $ &  $ -0.48 \pm 0.07 $ \\
        NGC4654   & $0.95 \pm 0.05$ & ($ -0.03 \pm 0.07$) &  $ -0.98  \pm 0.09$ &  ($ -0.02 \pm 0.07 $) & $-0.15 \pm 0.09$ & ($ 0.16 \pm 0.42 $) & $ 0.13 \pm 0.09 $ \\
        \hline
        total$^{\rm{(a)}}$      & $1.11 \pm 0.02$ & $ (0.43 \pm 0.03$) &  $ -0.83 \pm 0.02$ &  $ 0.41 \pm 0.02 $ & ($-0.19 \pm 0.02$) & $ 0.31 \pm 0.03 $ &  ($ 0.05 \pm 0.02 $) \\
        \hline 
        \end{tabular}
      \]
      \begin{list}{}{}
      \item all slopes with an associated Spearman correlation coefficient ${\mid}{\rho}{\mid} < 0.4$ are in brackets
      \item[$^{\rm{(a)}}$] correlation for data from all galaxies except NGC~4579
      \end{list}
\end{table*}

As expected, the tightest overall correlations were found for the SFR--radio relation, followed by the SFR--FP, radio--FP, and radio--PI relations.
The SFR--PI correlation is moderate ($0.4 \leq \rho < 0.6$) for NGC~4254, NGC~4321, NGC~4501, NGC~4535, and NGC~4579.
It is strong ($\rho \geq 0.6$) for NGC~4567/68. On the other hand, NGC~6946, M~51, NGC~4303, and NGC~4654 show no ($\rho < 0.2$) or only a weak 
SFR--PI correlation  ($0.2 \leq \rho < 0.4$). The correlation strength is not linked to perturbations: M~51 and NGC~4567/68
are gravitationally interacting systems with very different SFR--PI Spearman correlation coefficients. Moreover, NGC~4654 and NGC~4501 are both affected 
by ICM ram pressure and show very different SFR--PI Spearman correlation coefficients. NGC~4321 and NGC~4303 have quite symmetric disks
but also have different SFR--PI Spearman correlation coefficients.

All galaxies except NGC~4654 show moderate to strong radio--PI correlations. The associated Spearman correlation coefficients are
significantly higher than those of the SFR--PI correlation. Overall, the radio--PI relations are steeper than the SFR--PI relations
of our sample galaxies. However, there are exceptions: the southern disk halves of NGC~4254 and NGC~4567/68.
A tighter radio--PI than SFR--PI correlation is expected because the same relativistic electron population is responsible for the
PI and radio emission. 
The relatively strong anticorrelations between SFR and FP on the one hand, and radio and FP (not separately presented
because it is not a fundamental relation) on the other hand are driven by the SFR--radio correlation. 
The radio emission most closely related to recent star formation has low polarization because of tangled/irregular magnetic fields.
It is remarkable that the SFR--FP correlations are slightly stronger than the
radio--FP correlations in most of the galaxies. As expected from the shallower SFR--PI relations with respect to the radio--PI relations, 
the SFR--FP relations are generally steeper than the radio--FP relations.

\section{The PI--radio/SFR relation \label{sec:sfrpiapp}}

\begin{figure*}
  \centering
  \resizebox{\hsize}{!}{\includegraphics{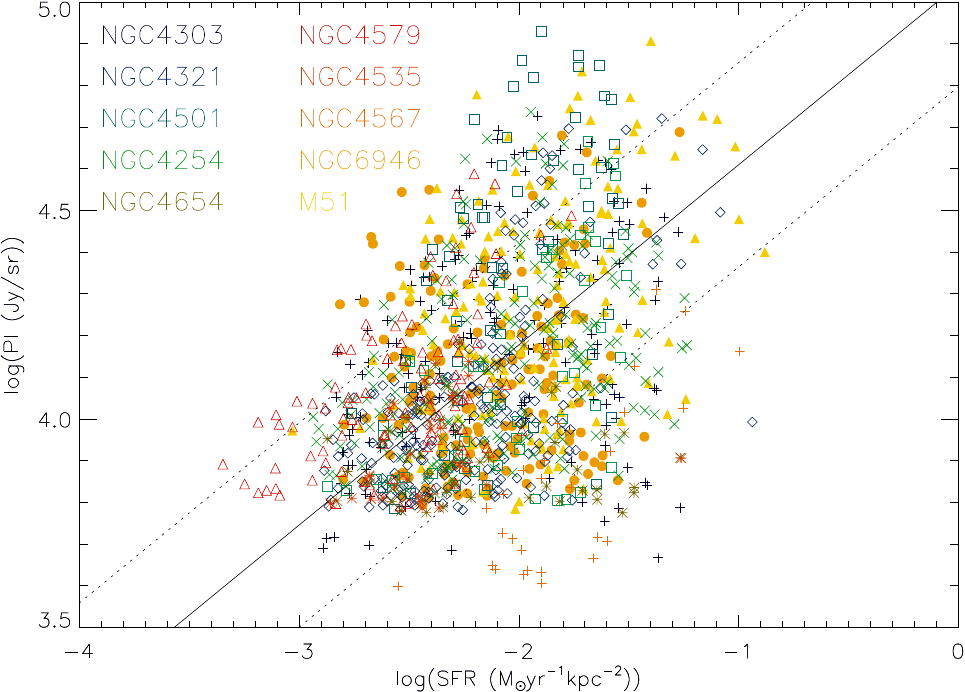}\includegraphics{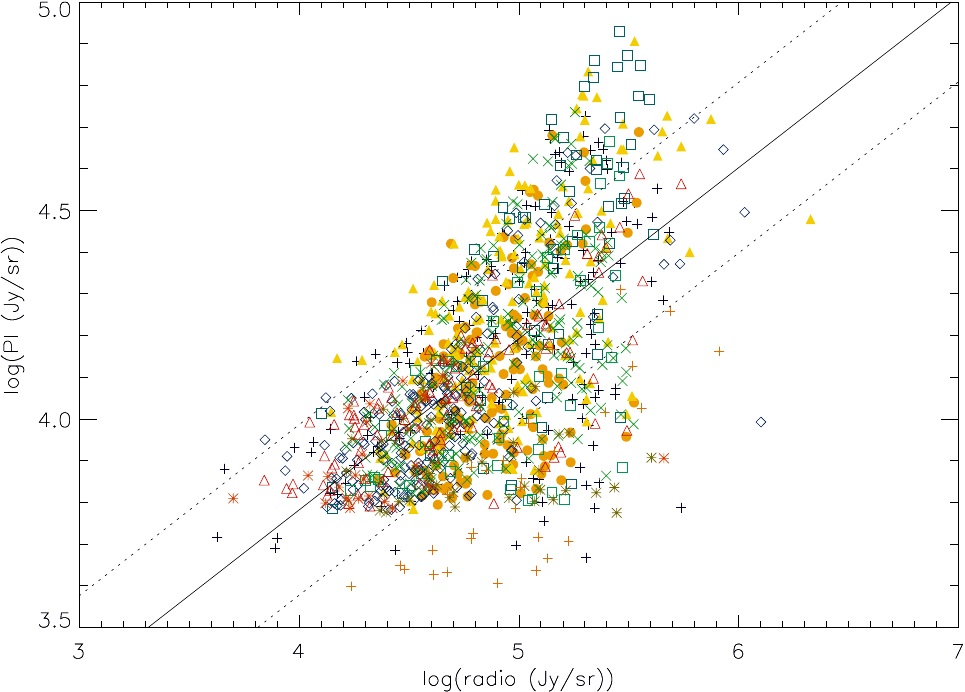}\includegraphics{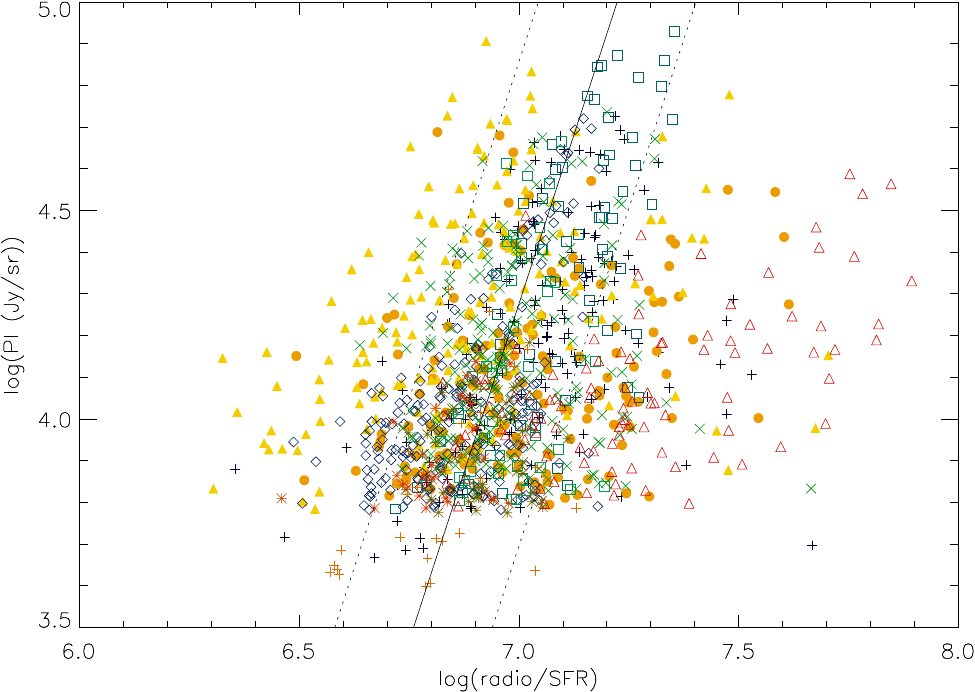}}
  \resizebox{\hsize}{!}{\includegraphics{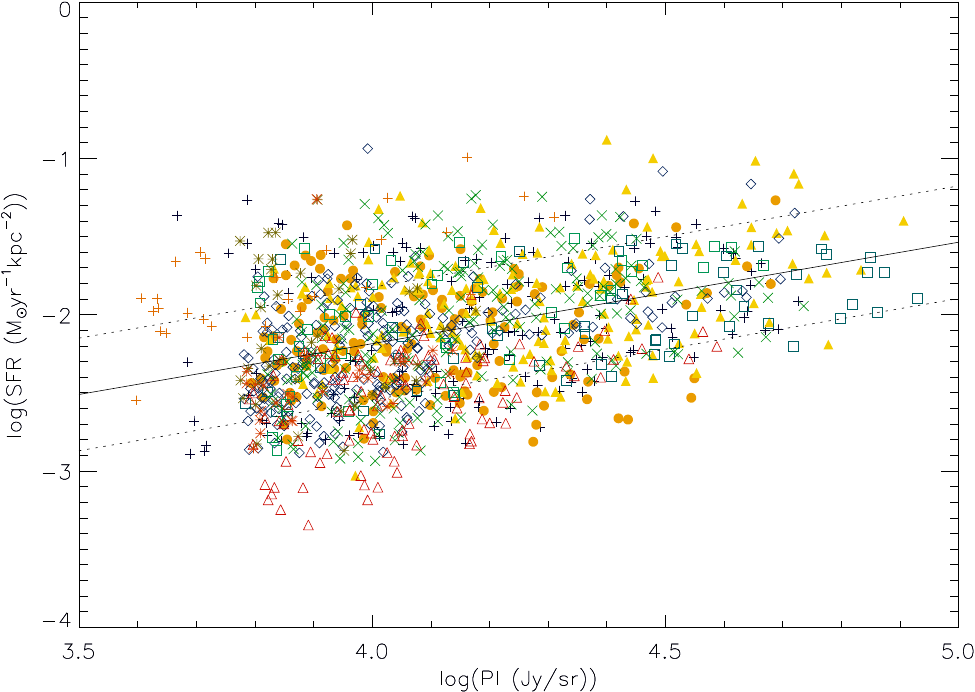}\includegraphics{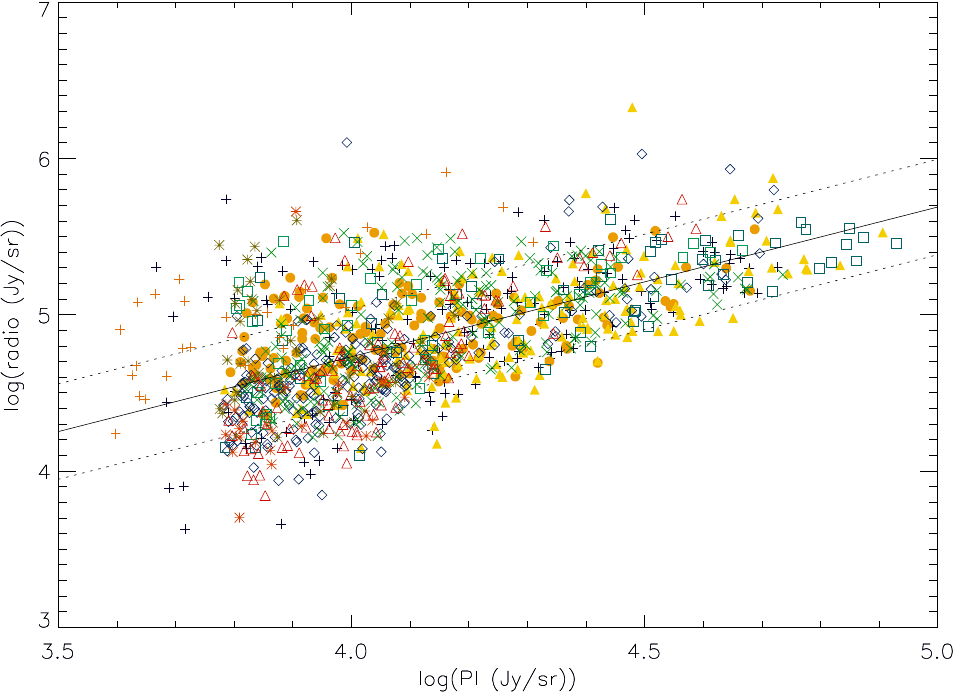}\includegraphics{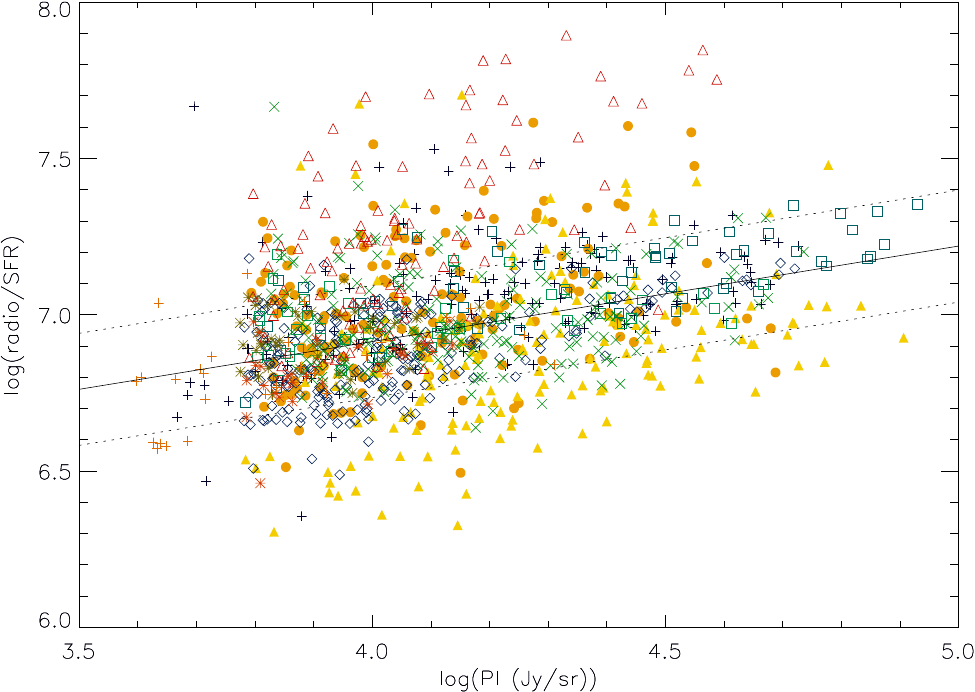}}
  \caption{From upper left to lower right: log(SFR)--log(PI), log(radio)--log(PI), log(radio/SFR)--log(PI), log(PI)--log(SFR),
    log(PI)--log(radio), and log(PI)--log(radio/SFR) relations.  
    The lines show the linear fits (solid) and scatter (dotted). NGC~4579 was excluded from the linear regression.
  \label{fig:radsfr_piall}}
\end{figure*}

Based on the SFR--PI and radio--PI relations, we expect a slope of the log(PI)--log(radio/SFR) relation of $1/0.41-1/0.43=0.11 \pm 0.20$. 
The observationally derived log(PI)--log(radio/SFR) slope of $0.31 \pm 0.03$ is thus consistent within $1\sigma$ with the expected value.
The fact that it is located at the higher end of the expected range can be explained in the following way:
the slopes of the log(PI)--log(SFR) and log(PI)--log(radio) relations calculated with the
Bayesian approach are $0.65 \pm 0.05$ and $0.96 \pm 0.04$, respectively. These are not the inverse of the
log(SFR)--log(PI) and log(radio)--log(PI) relations because of a significant number of resolution elements with high PI and relatively
high SFR or radio continuum emission (Fig.~\ref{fig:radsfr_piall}). These points are more relevant for the fitting of the log(PI)--log(SFR) and log(PI)--log(radio) 
relations than for the fitting of the log(SFR)--log(PI) and log(radio)--log(PI) relations.
The PI--radio/SFR correlation is thus driven by regions of high PI in the interacting galaxies.

The log(PI)--log(radio/SFR) slope is consistent with the log(PI)--log(radio) and log(PI)--log(SFR) slopes $0.96-0.65=0.31 \pm 0.06$.
Remarkably, the resulting  log(radio/SFR)--log(PI) and log(PI)--log(radio/SFR) have consistent slopes of $3.24 \pm 0.24$ and $0.31 \pm 0.02$, respectively 
(Table~\ref{tab:coeffs2} and Fig.~\ref{fig:radsfr_piall}). This makes us confident that the PI--radio/SFR correlation is meaningful. 

\section{Interpretation of the SFR--radio and the SFR--FP correlations \label{app:fpradsfr}}

\begin{figure}
  \centering
  \resizebox{\hsize}{!}{\includegraphics{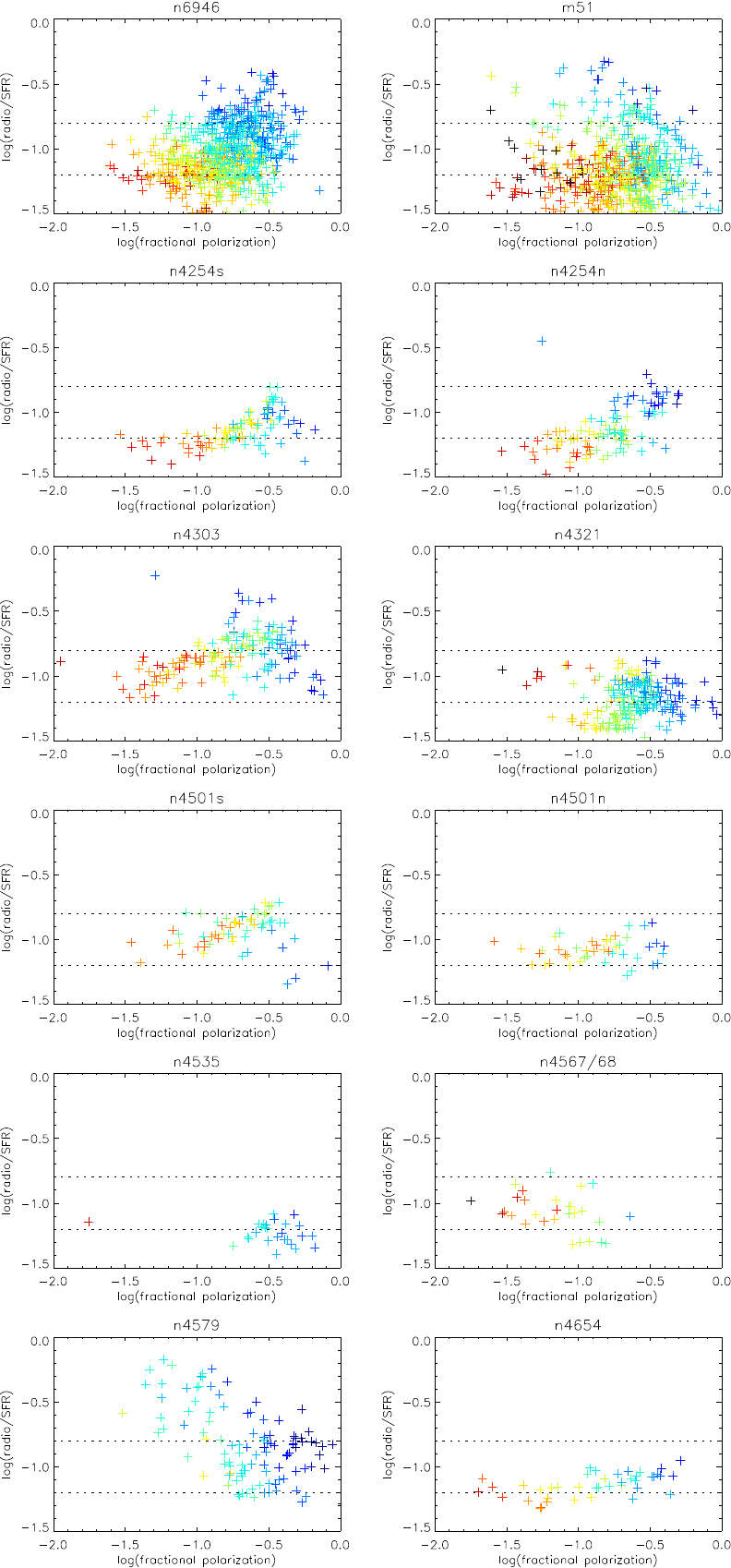}}
  \caption{Ratio between the radio surface brightness and the star formation rate per unit area as a function of the
    fractional polarization. The colors correspond to the star formation rate per unit area; 
    see Fig.~\ref{fig:rcfir_spixx1c1_nice_5} for the color scale.
  \label{fig:radsfr_dp}}
\end{figure}

The mean slope of the log(SFR)--log(radio) correlation is $1.11 \pm 0.02$ (Table~\ref{tab:coeffs2}). In the case of 
CR and magnetic field energy equipartition (Beck \& Krause 2005), a typical synchrotron spectral index of $-0.8$, 
and $B_{\rm tot} \propto SFR^{0.3}$ (Heesen et al. 2014), we obtain for the radio continuum emission $I_{\nu} \propto B_{\rm tot}^{3.8}
\propto SFR^{1.3}$, which is very close to the measured value. The slopes of the individual galaxies significantly vary from $0.85$ to $1.44$.
This behavior might be due to (i) a variation of the spectral index from galaxy to galaxy, (ii) CR--magnetic field equipartition is
not always valid, (iii)  the exponent of the $B_{\rm tot}$--$SFR$ relation varies from galaxy to galaxy, (iv) CR diffusion 
reduces the contrast in radio continuum with respect to the contrast in SFR, or (v) galaxies with high SFR or magnetic field
are closer to the CRe calorimeter condition where the slope should be close to unity. In addition, CR diffusion is less
relevant because the diffusion length becomes shorter with increasing magnetic field strength. Anticipating our model results of Sect.~\ref{sec:symmetricm}, we found the following model SFR--radio slopes: $1.08$ for the calorimeter model $t_{\rm eff}=t_{\rm sync}$, $0.89$ for a model without CR diffusion including bremsstrahlung, IC, ion, and pion losses, and $0.73$ for a model including CR diffusion. Thus both, energy losses (bremsstrahlung, IC, ion, and pion) and CR diffusion lead to shallower SFR--radio slopes. 

\begin{figure*}
  \centering
  \resizebox{\hsize}{!}{\includegraphics{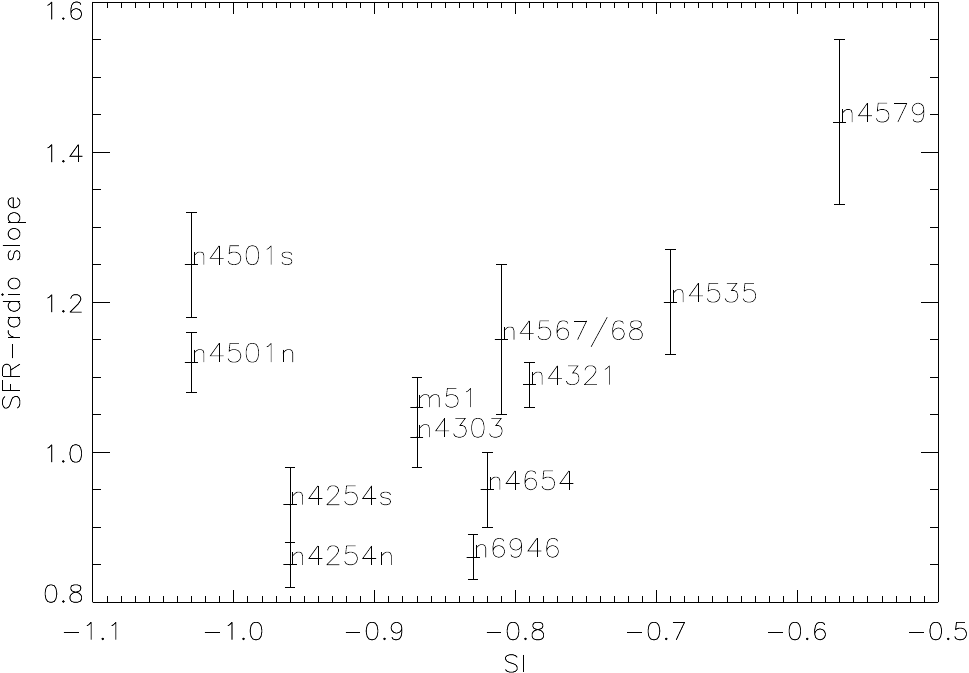}\includegraphics{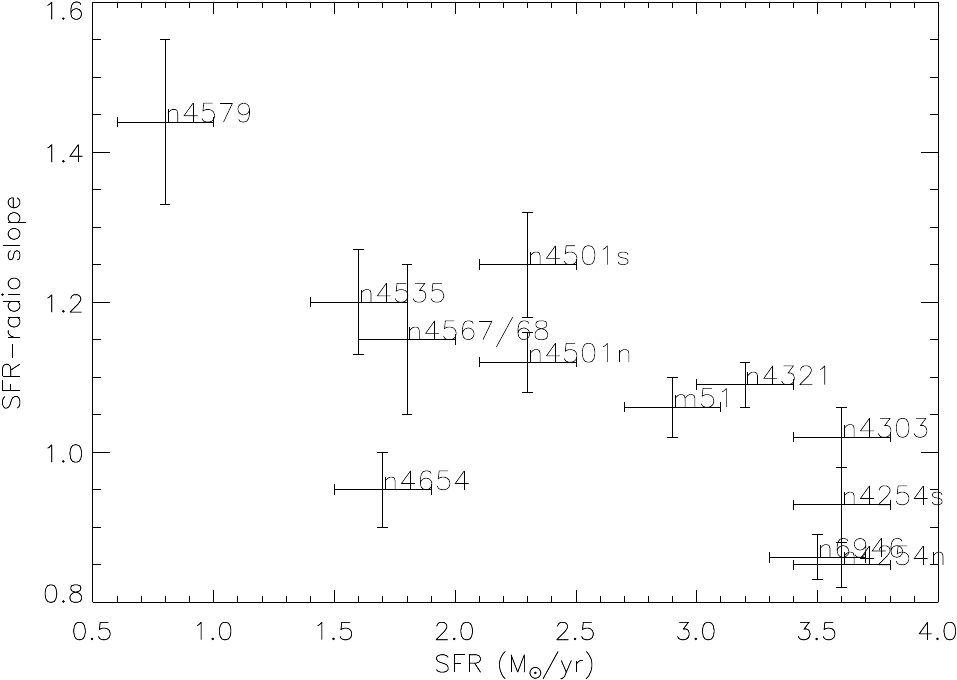}}
  \caption{The log(SFR)--log(radio) slope (from Table~\ref{tab:coeffs2}, column $2$) as a function of the mean spectral index (left) and the SFR (right).
  \label{fig:SFR_TP}}
\end{figure*}

Since the mean spectral index of our sample galaxies varies from $-0.57$ to $-1.03$ and $I_{\nu} \propto \nu^{-\alpha} B^{3+\alpha}
\propto SFR^{0.9+0.3 \alpha}$, the expected slopes range between $1.07$ and $1.21$. 
The inspection of the relation between the log(SFR)--log(radio) slope and the mean SI (left panel of Fig.~\ref{fig:SFR_TP})
shows that only NGC~4501 shows the expected behavior for item (i).
Anticipating the results of our modelling (Sect.~\ref{sec:model}), diffusion can lead to a flattening of the 
log(SFR)--log(radio) slope by $0.2$. Thus, item (iv) is potentially relevant in all galaxies. 
Moreover, the relation between the log(SFR)--log(radio) slope and the SFR
(right panel of Fig.~\ref{fig:SFR_TP}) suggests that item (v) is also relevant in most of the galaxies.
We conclude that items (i), (iv), and (v) contribute to the galaxy-to-galaxy variation of the log(SFR)--log(radio) slope;
items (ii) and (iii) cannot be ruled out.

The SFR--radio correlation is much steeper and tighter than the SFR--PI correlation. This results in an SFR--FP
anticorrelation. We  thus observe a degree of magnetic field anisotropy or field ordering (anisotropy), which decreases with increasing SFR.
This can be interpreted by three mechanisms: (i) the large-scale dynamo operates
preferably in regions with low SFR (see e.g. Beck et al. 2019). In addition, regular fields are tangled and thus destroyed in
regions of higher SFR. (ii) Compression and shear of isotropic random fields, which lead to field ordering (anisotropy), mainly occur in
regions with low SFR where the gas energy density is also low. In regions of high SFR,
magnetic and kinetic energy densities are too high to be significantly affected by compression and shear (see Eqs.~\ref{eq:shearr} and
\ref{eq:compr}).
(iii) Faraday depolarisation is expected to be stronger in regions of high SFR leading to lower PI at high SFR.

\begin{figure}
  \centering
  \resizebox{\hsize}{!}{\includegraphics{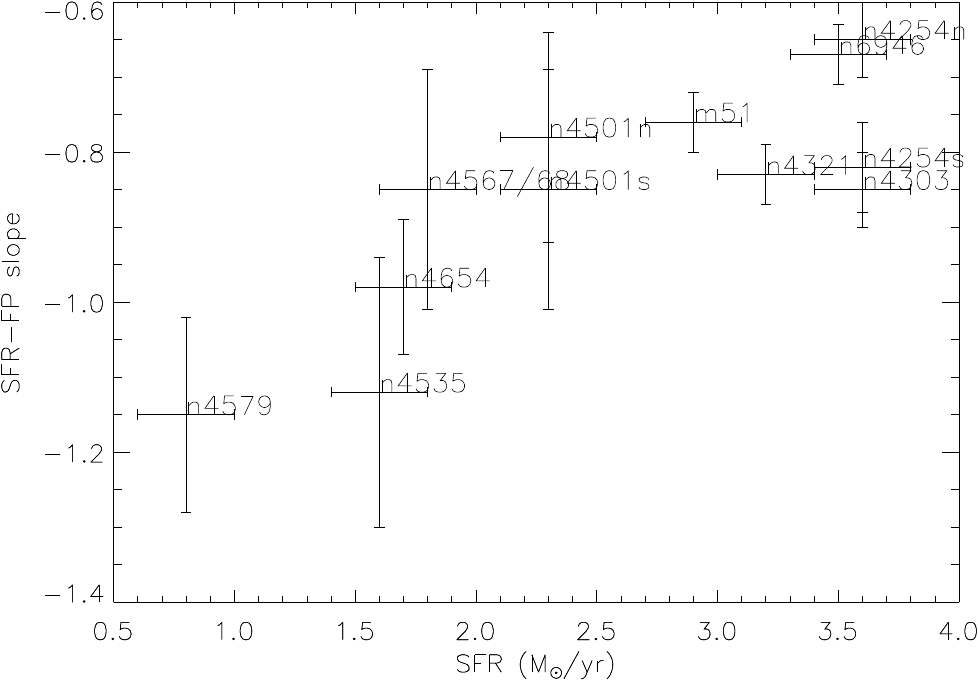}}
  \caption{The log(SFR)--log(FP) slope (from Table~\ref{tab:coeffs2}, column $4$) as a function of the SFR.
  \label{fig:SFR_SFR-DP1}}
\end{figure}

For high dispersions in rotation measure ($\sigma_{\rm RM} \ga 300$~rad\,m$^{-2}$) the degree of polarisation varies with
$\sigma_{\rm RM}^{-2} \sim (<n_{\rm e}> B_{\rm turb})^{-2}$, where $n_{\rm e}$ is the density of thermal electrons along the line of sight
(see, e.g., Eq.~13 of Beck 2015 or Eq.~34 of Sokoloff et al. 1998).
It is thus expected that the log(SFR)--log(FP) slope becomes smaller with increasing SFR. Since this is not observed (Fig.~\ref{fig:SFR_SFR-DP1}),
we conclude that Faraday depolarisation does not play a role for the SFR--FP correlation.
An apparent opposite effect is caused by NGC~4579, NGC~4535, and NGC~4654. The steep SFR--FP correlations of NGC~4535 and NGC~4654 (fourth column of 
Table~\ref{tab:coeffs2}) are due 
to a high FP in regions of low SFR caused by interaction-induced shear motions. In NGC~4579 the steepness of the correlation
is driven by a low FP in the radio-bright region. We speculate that this is due to an intrinsically small FP of the emission from the nuclear disk and the AGN and some beam depolarization (outflow versus disk field).

NGC~6946 is the only isolated galaxy in our sample. Its SFR--FP correlation coefficient is not different from
that of the other galaxies. However, the slope of the correlation is significantly steeper for the interacting galaxies
than for NGC~6946. We suggest that while the SFR--FP correlation in all galaxies is driven by mechanism (i),
the steeper slopes are caused by mechanism (ii).

\section{Additional model calculations \label{sec:additional}}

\begin{figure}
  \centering
  \resizebox{\hsize}{!}{\includegraphics{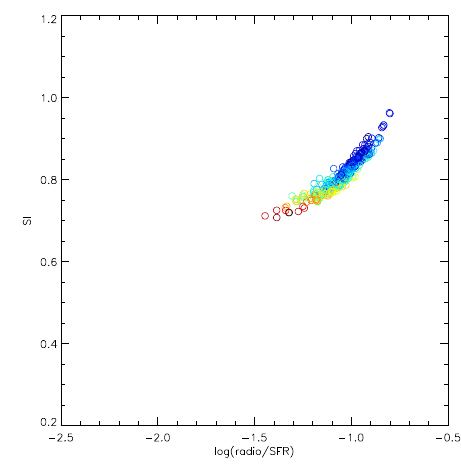}\includegraphics{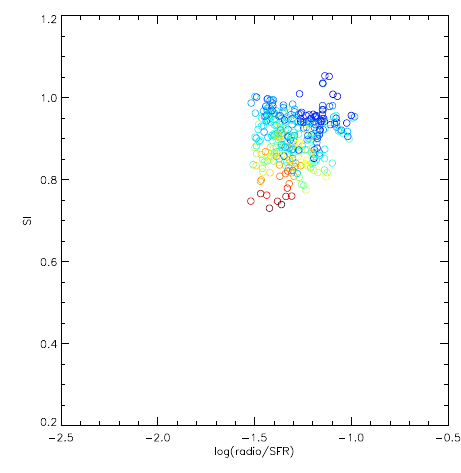}
  \put(-400,50){\Huge $t_{\rm brems}$}
  \put(-180,50){\Huge $t_{\rm brems}$ and $t_{\rm IC}$}}
  \resizebox{\hsize}{!}{\includegraphics{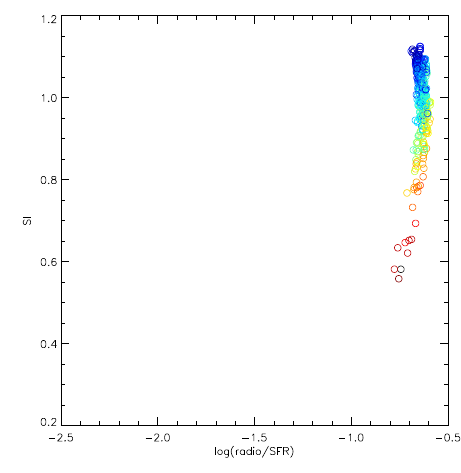}\includegraphics{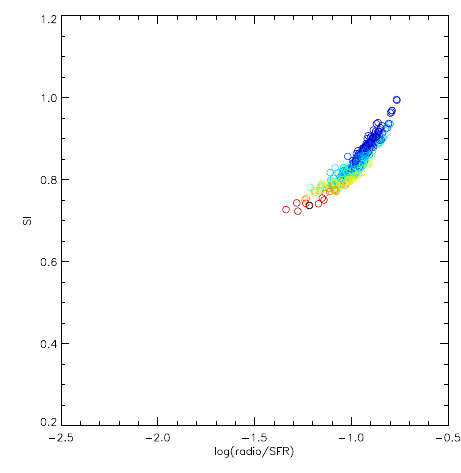}
    \put(-400,50){\Huge $t_{\rm ion}$}
  \put(-180,50){\Huge $t_{\pi}$}}
  \resizebox{5cm}{!}{\includegraphics{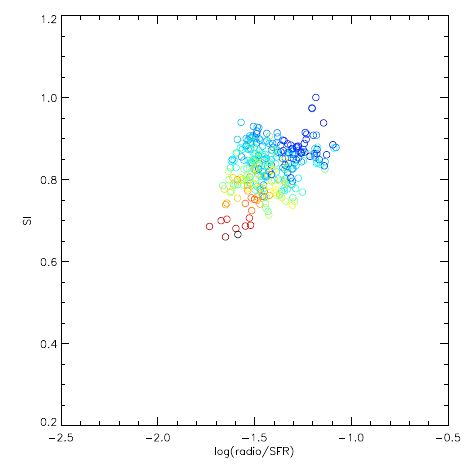}
    \put(-180,50){\Huge all $t$}}
  \caption{Spectral index as a function of the ratio between the radio continuum surface brightness and the star formation rate
    per unit area. The effective timescale contains the timescales that are marked on the plots.
  \label{fig:model3}}
\end{figure}

\begin{figure*}
  \centering
  \resizebox{\hsize}{!}{\includegraphics{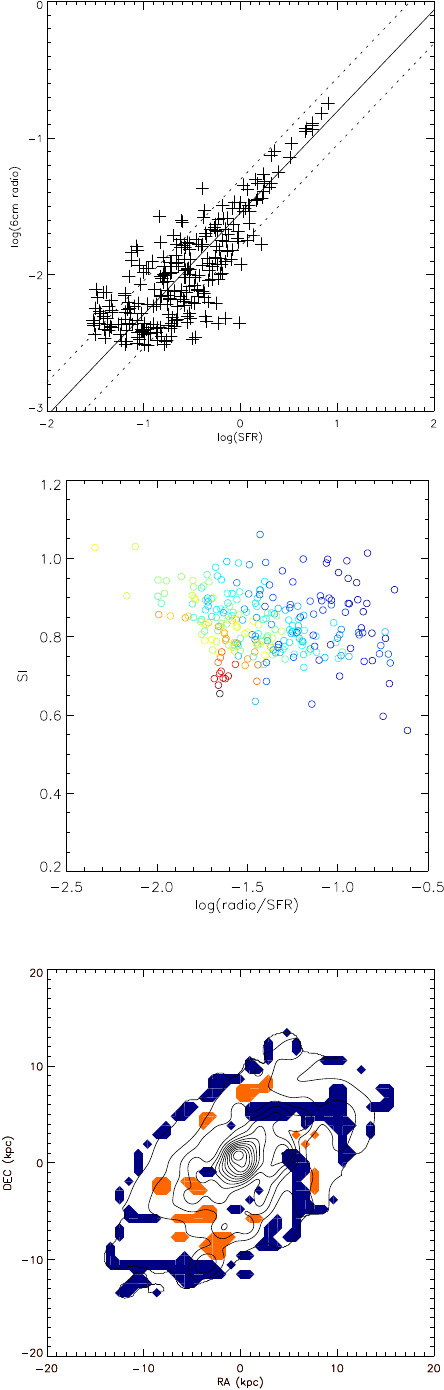}\includegraphics{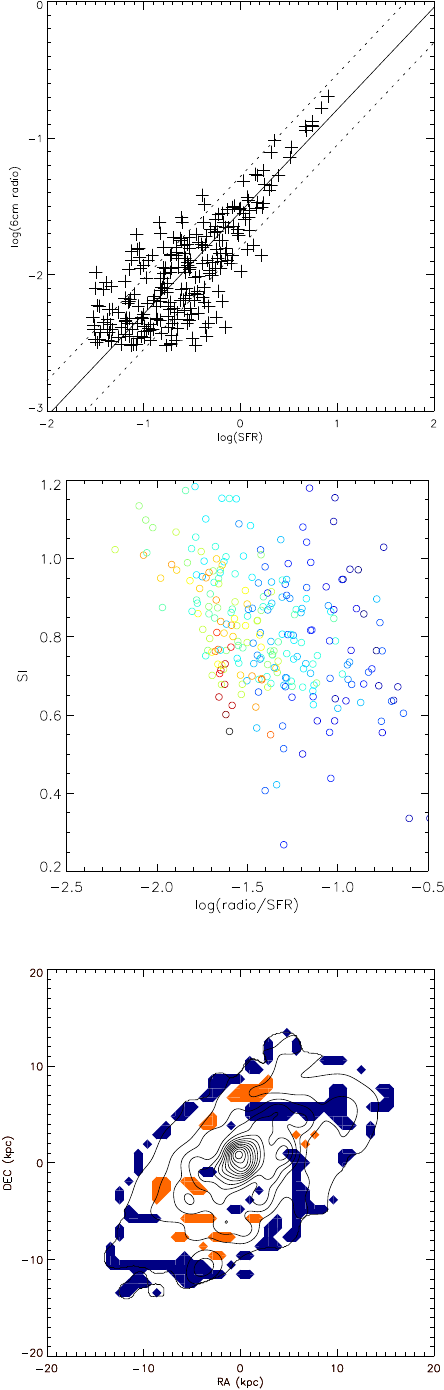}\includegraphics{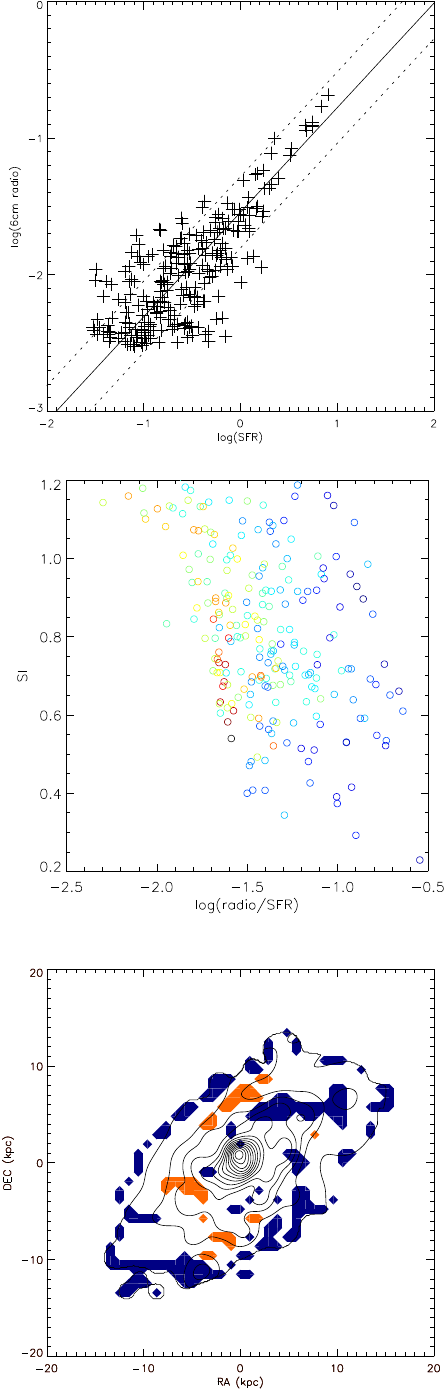}}
\put(-400,400){\Large D(E)}
\put(-230,400){\Large D$_0 \times$2}
\put(-70,400){\Large D(E)$\times$2}
  \caption{As Fig.~\ref{fig:model1}. Left panel: with an energy-dependent diffusion coefficient
  $D=D_0=10^{28}$~cm$^2$s$^{-1}$ for $E \le 3$~GeV and $D=D_0 \big(E/(3~{\rm GeV})\big)^{0.3}$
    for $E > 3$~GeV. Middle panel: with $D=2 \times D_0$ for all energies. Right panel: with 
  $D_0=2 \times 10^{28}$~cm$^2$s$^{-1}$ and an energy-dependent diffusion coefficient.
  \label{fig:modelDE}}
\end{figure*}

\begin{figure}
  \centering
  \resizebox{\hsize}{!}{\includegraphics{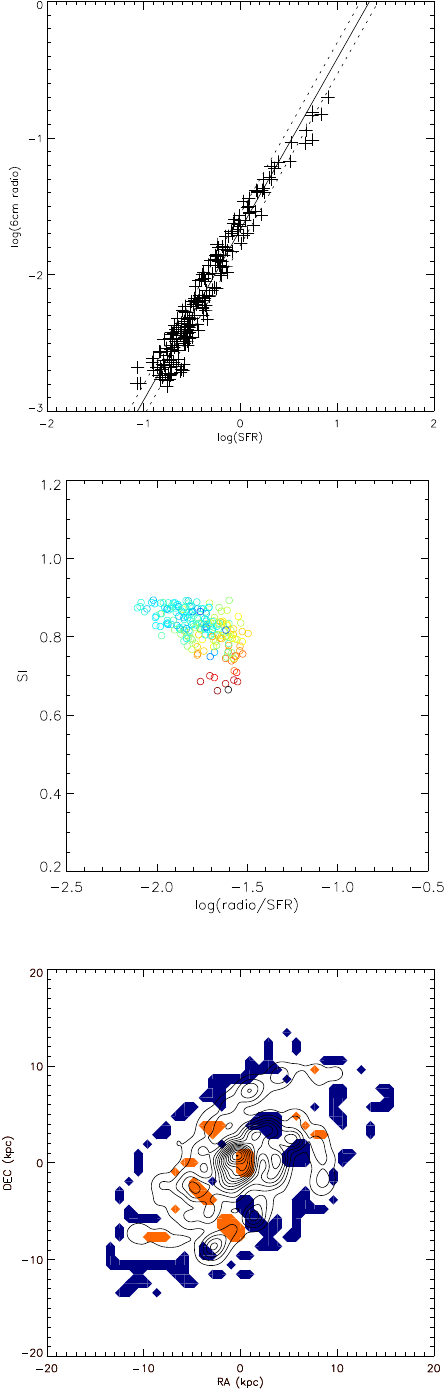}\includegraphics{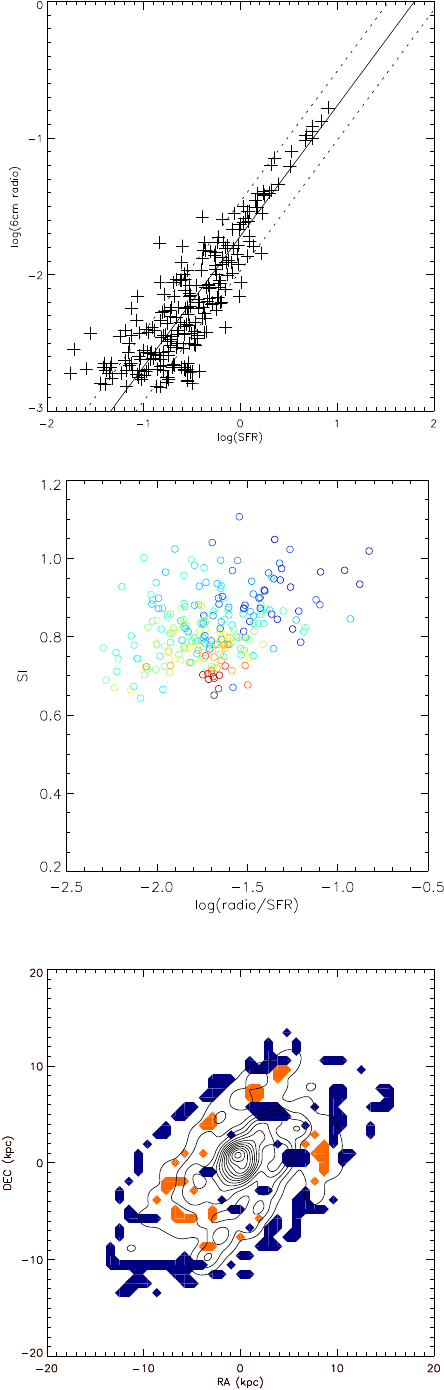}}
  \put(-230,155){\large without diffusion}
  \put(-100,155){\large with diffusion}
  \put(-230,15){\large blue: radio-bright}
  \put(-100,15){\large red: radio-dim}
  \caption{As Fig.~\ref{fig:model1} with additional diffusive escape: $t_{\rm eff}^{-1}=(t_{\rm sync}^{-1}+t_{\rm brems}^{-1}+t_{\rm IC}^{-1}+t_{\rm ion}^{-1}+t_{\pi}^{-1}+t_{\rm escp}^{-1})^{-1}$.
  \label{fig:model1a}}
\end{figure}

\begin{figure}
  \centering
  \resizebox{\hsize}{!}{\includegraphics{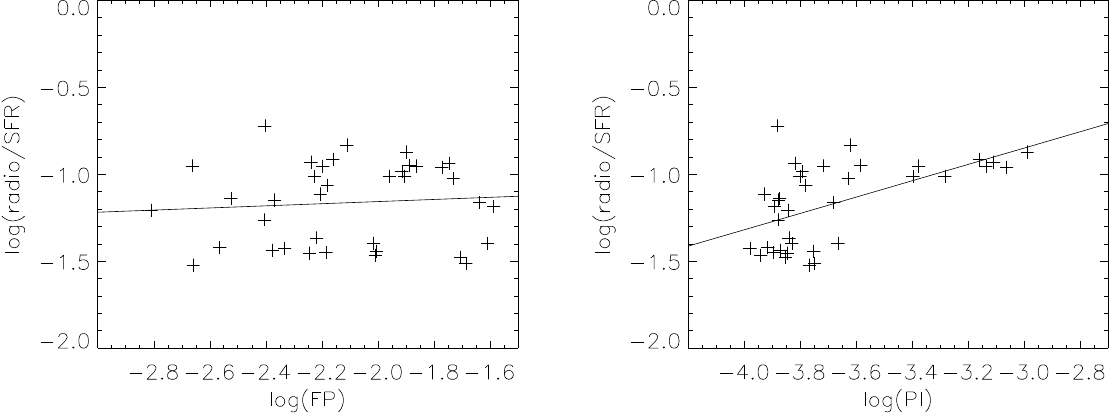}}
  \resizebox{\hsize}{!}{\includegraphics{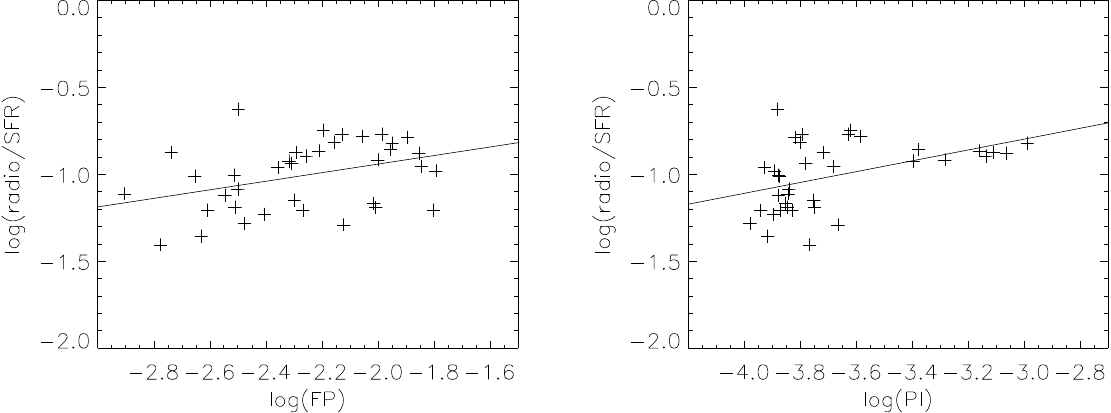}}
  \caption{NGC~4501 model log(radio/SFR)--FP (left panels) and log(radio/SFR)--PI relations (right panels).
    The upper panels correspond to Fig.~\ref{fig:model2}.
    Lower panels: addition of a regular field component $B_{\rm reg}=12$~$\mu$G in regions
    of enhanced model polarized intensity (cf. Fig.~\ref{fig:model2a}).
  \label{fig:radsfr_pi_0}}
\end{figure}

\end{appendix}

\end{document}